\documentclass[a4paper,11pt,authoryear]{elsarticle}

\usepackage[T1]{fontenc}
\usepackage[utf8]{inputenc}
\usepackage[english]{babel}
\usepackage{amsfonts}
\usepackage{amssymb}
\usepackage{amsmath}
\usepackage{amsthm}
\usepackage{mathabx}
\usepackage{bm}
\usepackage{lineno}
\usepackage{graphicx}
\usepackage{subcaption}
\usepackage{float}
\usepackage[authoryear]{natbib}
\usepackage{appendix}
\usepackage{color}
\usepackage{soul}
\usepackage{geometry}
\journal{Journal of the Mechanics and Physics of Solids}

\begin{document}

\newtheorem{remark}{Remark}

\begin{frontmatter}

\title{Effective transient behaviour of heterogeneous media in diffusion problems with a large contrast in the phase diffusivities}

\author[monash]{Laurence Brassart}
\author[ecn]{Laurent Stainier}

\address[monash]{Department of Materials Science and Engineering, Monash University, Clayton, VIC~3800, Australia}
\address[ecn]{Institut de Recherche en G\'enie Civil et M\'ecanique (GeM, UMR 6183 CNRS/ECN/UN)\\{\'E}cole Centrale Nantes, 1 rue de la No\"e, BP 92101, F-44321 Nantes, France}

\begin{abstract}
This paper presents a homogenisation-based constitutive model to describe the effective transient diffusion behaviour in heterogeneous media in which there is a large contrast between the phase diffusivities. In this case mobile species can diffuse over long distances through the fast phase in the time scale of diffusion in the slow phase. At macroscopic scale, contrasted phase diffusivities lead to a memory effect that cannot be properly described by classical Fick's second law. Here we obtain effective governing equations through a two-scale approach for composite materials consisting of a fast matrix and slow inclusions. The micro-macro transition is similar to first-order computational homogenisation, and involves the solution of a transient diffusion boundary-value problem in a Representative Volume Element of the microstructure. Different from computational homogenisation, we propose a semi-analytical mean-field estimate of the composite response based on the exact solution for a single inclusion developed in our previous work [Brassart, L., Stainier, L., 2018. Effective transient behaviour of inclusions in diffusion problems. Z. Angew Math. Mech. 98, 981-998]. A key outcome of the model is that the macroscopic concentration is not one-to-one related to the macroscopic chemical potential, but obeys a local kinetic equation associated with diffusion in the slow phase. The history-dependent macroscopic response admits a representation based on internal variables, enabling efficient time integration. We show that the local chemical kinetics can result in non-Fickian behaviour in macroscale boundary-value problems. 
\end{abstract}


\end{frontmatter}

\section{Introduction}
A number of engineering problems involve the diffusive transport of mobile species in heterogeneous media. Representative examples include atomic transport in polycrystals, fluid and solute transport in geomaterials, and water permeation in porous polymers and gels. When studying diffusion in macroscopic volumes containing a large number of small heterogeneities, direct calculations on a fully-resolved geometry are often impracticable. One then seeks to replace the actual heterogeneous medium by an equivalent homogeneous one with the same "average" behaviour \citep{auriault1991}. This practice is also convenient in view of comparing models to experiments, since many experimental techniques only provide average measures of composition distributions. 

A commonly-adopted hypothesis is that the equivalent medium obeys conservation equations and constitutive relations that have the same structure as the local governing equations. In particular, in linear diffusion problems one often postulates that the effective medium obeys Fick's second law with an effective diffusion coefficient that depends on the phase diffusivities and the microstructure geometry. Analytical expressions for the effective diffusivity are provided by classical bounds and estimates, such as Voigt and Reuss bounds, Hashin-Shtrikman bounds or the self-consistent scheme \citep{hashin1962,budiansky1970} - see also textbooks by \citep{torquato2002} and \citep{auriault2009}\footnote{Many references cited in this Introduction deal with thermal or electrical conduction problems, rather than mass transport. Results obtained in the context of linear conduction can to a large extent be transposed to the mass transport problem.}. 

Yet Fick's second law is not always appropriate to describe the effective diffusion behaviour. In polycrystals, the presence of fast diffusion paths (grain boundaries, free surfaces, dislocations) leads to distinct, possibly non-Fickian limiting regimes, depending on the diffusivity contrast, macro- and microscopic length scales and the observation time scale \citep{harrison1961,balluffi2005}. For example, diffusion over long distances may take place rapidly through the grain boundaries, while diffusion in the grain interior is much slower. This situation cannot be described by a single effective diffusion coefficient. Anomalous diffusion has also been reported in double-porosity media, where the non-Fickian behaviour manifests itself in the long tail observed in solute concentration distributions \citep{gist1990,sternberg1996,levy2003,ngoc2014}. The long-tail effect has been attributed to local non-equilibrium effects associated with the mass exchange between the low and high diffusivity regions. 

A simple Fickian description may also not be suited to describe mass transport in electrodes of Li-ion batteries at macroscopic scale. Typical electrodes present a porous architecture consisting of active particles (e.g. LiCoO$_2$ for the cathode, graphite for the anode), conductive fillers and a polymer binder \citep{shearing2010,stephenson2011}. The diffusion coefficient of lithium ions in electrolyte-filled pores can be orders of magnitude larger than the diffusion coefficient of lithium atoms in the active particles ($D \sim 10^{-14}-10^{-16}$ m$^2$/s for lithium in LiCoO$_2$ \citep{xie2008} compared to $D \sim 10^{-11}$ m$^2$/s for lithium ions in commonly-used organic carbonates electrolyte with LiPF$_6$ salt \citep{danilov2008}. As a result, transient lithiation of active particles can occur concurrently to large scale transport through the electrolyte in sufficiently thick electrodes, with direct implication for the battery capacity and rate capability.

The general objective of this work is to develop governing equations for the effective diffusion behaviour of heterogeneous media in which there is a high contrast between the diffusivities of the constituents. This paper focuses on two-phase particulate composites consisting of "slow" inclusions dispersed in a “fast" percolating matrix. In this case distinct limiting regimes can be anticipated for the relaxation by diffusion of a compositional heterogeneity. Let $\tau$ be the characteristic time for diffusion in a typical inclusion with size $a$: $\tau=a^2/D_1$, where $D_1$ is the diffusion coefficient of the mobile species in the inclusion. The relaxation time $\tau$ and the diffusion coefficient in the matrix, $D_2$, together define a length, $\Lambda=\sqrt{D_2 \tau}$, which characterises  the diffusion length of the mobile species through the matrix over a time $\tau$. This length only depends on the diffusion properties and inclusion size, and is thus an intrinsic property of the material. For a boundary-value problem with typical size $L$ much larger than $\Lambda$, the system relaxation is limited by the long-range transport through the fast matrix, while short-range diffusion within the inclusions has relaxed. For a boundary-value problem with size $L$ smaller than $\Lambda$, relaxation is limited by the short-range diffusion in the inclusions, while diffusion in the matrix has reached steady state. Effective governing equations should include these two regimes as limiting cases.   
  
These two limiting regimes were previously predicted by phenomenological theories for coupled diffusion and viscous flow \citep{li2014,brassart2018}. These theories rest on the postulate that long-range transport of species and local concentration changes are mediated by distinct molecular processes with different kinetics. In the mathematical formulation, this is accounted for by relaxing the assumption of local chemical equilibrium, hence introducing a kinetic relation relating the concentration rate to the chemical potential. In response to a jump in chemical potential, the local concentration does not immediately adjust, but rather evolves towards its equilibrium value according to a kinetic model with relaxation time $\tau$. The latter could represent for example the kinetics of breaking and reforming chemical bonds or creep relaxation associated with volume change due to species insertion \citep{brassart2016}. In the context of supercooled liquids \citep{li2014}, the theory assumes fast diffusion through regions of high mobility and creep-limited species insertion dominated by regions of low mobility. In the present case, the local kinetics of species insertion will be associated to diffusion in inclusions with low diffusivity. One of the aims of this paper is to provide a micromechanically-based motivation for the relaxation of the local chemical equilibrium assumption at macroscopic scale in the context of double-diffusivity media.    

In this paper we develop effective diffusion equations using a micro-macro approach. To simplify the treatment and focus on the essential ideas, we consider linear diffusion problems and do not introduce any coupling with mechanics. The proposed upscaling strategy involves the solution of a transient diffusion problem on a Representative Volume Element (RVE) of the microstructure subject to boundary conditions in terms of a macroscopic chemical potential and its gradient. Corresponding macroscopic concentration rate and flux are obtained by averaging of the microscopic fields in the RVE, in accordance to Hill-Mandel condition. Different from previously-proposed computational homogenisation methods, e.g. \citep{ozdemir2008a,larsson2010,salvadori2015}, here we propose a semi-analytical mean-field estimate of the effective transient behaviour, assuming steady-state matrix and transient inclusions, allowing us to derive the effective governing equations in closed-form. The model relies on the exact solution for a single inclusion subject to time-varying, uniform chemical potential on its boundary, as presented in our previous work \citep{brassart2017}. The mean-field estimate gives the macroscopic concentration rate and flux as functions of the macroscopic chemical potential, its gradient, and the loading history. The model can be written in terms of internal variables, enabling efficient numerical implementation for solving macroscale boundary-value problems. Fickian to non-Fickian transition is predicted at macroscopic scale, in qualitative agreement with the phenomenological model proposed by \cite{brassart2018}.     

\subsection{Review of existing approaches}
Macroscopic models addressing the problem of diffusion in the presence of fast-diffusion paths have been formulated by postulating separate conservation equations for each family of diffusion path \citep{aifantis1979,aifantis1980}. The conservation equations are coupled through phenomenological source terms representing the mass exchange between the slow and fast regions. The total concentration obeys a fourth-order differential equation, showing the non-Fickian character of the average diffusion behaviour \citep{aifantis1980}. Similar two-equation models have been obtained by using the Method of Volume Averaging \citep{whitaker1999}, which introduces a spatial smoothing of the governing equations in each phase through volume averaging. The phase-exchange term is identified by solving a closure problem at the microscopic scale. Both steady-state and unsteady-state closure problems were considered \citep{quintard1993,moyne1997}. In the latter case, the exchange term between the phases takes the form of a linear relaxation process. 

Volume-averaged conservation equations are also considered in the phenomenological Porous Electrode Theory, pioneered by Newman and coworkers, see e.g. \citep{thomas2002}. In this approach, the conservation equation in the electrolyte is averaged over a volume assumed small relative to the electrode dimensions, but much larger than the typical size of active particles. The effect of tortuosity on the effective transport behaviour is typically described using analytical mean-field estimates or percolation theory. Active particles are distributed through the simulation volume and interact with the electrolyte through a reactive source term, which depends on the potential difference between the phases. Particles are usually treated as spheres and subjected to prescribed flux directly related to the source term in the electrolyte conservation equation. An extension of the theory to account for elastic and inelastic deformations of the particles was proposed by \cite{golmon2009}. In recent years, the theory was reformulated in a rigorous irreversible thermodynamics framework by Bazant and coworkers, see \citep{ferguson2012,smith2017}. Notably, these authors also included a description of phase transforming electrode materials.   

The determination of the effective diffusion behaviour of heterogeneous media can be rigorously addressed using asymptotic homogenization theory \citep{bensoussan1978,sanchez1980}. Conduction problems in periodic composites have been considered by \cite{auriault1983}, and recently revisited by \cite{matine2013,matine2015} to include a description of short-time and edge effects. In the presence of a high contrast between the phase conductivities, \cite{auriault1983} showed that, to the first-order, the temperature field obeys a heat conservation equation with pulsation-dependent heat capacity, bringing about a memory effect. \cite{auriault1995} developed the parallel theory for mass transport in double-diffusivity media, leading to similar conclusions. \cite{dureisseix2015} recently proposed a computational approach to estimate the memory function by direct finite element calculation on a unit cell of the microstructure. Alternatively, \cite{curto2015,curto2016} used the multiscale convergence approach \citep{allaire1992,allaire1996} to determine the effective diffusion behaviour of ions in microstructured electrolytes driven by concentration gradients and electric fields. However, the authors did not specifically investigate the effect of high diffusivity contrast on the overall behaviour. A drawback of asymptotic homogenisation approaches - in addition to their relative complexity - is that they require a priori judgement as to the magnitude of the diffusivity ratio in relation to the scaling parameter \citep{auriault1995,moyne1997}. 

Computational homogenisation is another, increasingly popular upscaling technique, according to which the effective behaviour is calculated numerically by solving a boundary value problem on a RVE of the microstructure at each integration point of a macroscale analysis \citep{kouznetsova2001,geers2010}. A computational homogenisation procedure was developed by \cite{ozdemir2008a} for heat conduction, assuming steady-state within the RVE. Microscopically-transient conduction problems were considered by \cite{monteiro2008} and \cite{larsson2010}. Numerical frameworks for heat conduction coupled to thermomechanical problems were also proposed, still relying of the assumption of steady-state heat transfer at microscale \citep{ozdemir2008b,temizer2011,berthelsen2017}. A multiphysics computational homogenisation framework for coupled electrochemo-mechanics was recently developed by \cite{salvadori2014,salvadori2015} in the context of Li-ion batteries. The advantage of computational homogenisation is that it can handle nonlinear constitutive models and general microstructures. Its drawback is the high computational cost associated with solving large-scale boundary value problems, as the method does not provide expressions of the effective behaviour in closed-form. The two-scale framework adopted in the present work is similar to that of \cite{larsson2010} and \cite{salvadori2015}. The key contribution of the present work is the formulation of a closed-form mean-field estimate for the RVE behaviour, which enables two-scale simulations at a much lower computational cost than computational homogenisation techniques.       

The paper is organised as follows. Section \ref{sec:prob-descr} presents the local governing equations and defines the effective behaviour. A general two-scale approach is presented in Section \ref{sec:two-scale}. The mean-field estimate is presented in Section \ref{sec-model}, and a numerical strategy for upscaling based on internal variables is proposed in Section \ref{sec:upscaling}. The mean-field model is validated in Section \ref{sec:validation} by comparing its predictions to reference, full-field predictions on unit cells with random microstructures. Finally, two-scale simulations are presented in Section \ref{sec:macro_bvp} and confirm the accuracy of the proposed model.

\section{Problem description}\label{sec:prob-descr}

\subsection{Diffusion boundary value problem}\label{sec:local_bvp}
We consider the transient diffusion problem of a mobile species in a continuous, heterogeneous medium $\Omega$. The heterogeneous medium is taken as a two-phase composite consisting of inclusions (phase 1)  distributed in a continuous matrix (phase 2).  The local state at a point $\bm x \in \Omega$ is described by the concentration $c$ of mobile species (number of molecules per unit volume). The free energy density of the medium (energy per unit volume) is written $G(\bm x,c)$, where the dependence in $\bm x$ indicates the spatially-varying chemical properties of the medium. The chemical potential of the mobile species at a point is derived from the free energy according to: 
\begin{equation}\label{state_law}
\mu = \frac{\partial G}{\partial c}. 
\end{equation}
We assume that the chemical potential in each phase is a linear function of concentration:  
\begin{equation} \label{chempot_lin}
\mu(c) = \mu_r + K_r (c-c_r),
\end{equation}
where $K_r$ is the chemical modulus of phase $r$ ($r=1,2$), $c_r$ a reference concentration for the phase, and $\mu_r$ the corresponding chemical potential. The free energy density function is thus quadratic in each phase:
\begin{equation}\label{gibbs_lin}
G(\bm x,c) = \sum_{r=1}^2 \chi_r(\bm x) G_r(c) \quad\text{with}\quad
G_r(c) = \mu_r c + \frac{K_r}{2}(c-c_r)^2,
\end{equation}
where $\chi_r(\bm x)$ is the indicator function of the domain occupied by phase $r$ ($\chi_r(\bm x)=1$ if $\bm x$ is in phase $r$, $\chi_r(\bm x)=0$ otherwise).
In the following, we will set $c_r=0$ and $\mu_r=0$, without loss of generality. Concentration and chemical potential $(c,\mu)$ can then be interpreted as perturbations about the reference state $(c_r,\mu_r)$. 

\begin{remark} The constitutive model (\ref{chempot_lin})-(\ref{gibbs_lin}) corresponds to the linearisation of a commonly-adopted chemical constitutive model about the reference concentration $c_r$:
\begin{equation} \label{chempot_nonlin}
\mu(c) = \mu_r + k_bT \ln\left( \frac{c}{c_r} \right), 
\end{equation} 
where $k_b$ is Boltzmann's constant and $T$ is the absolute temperature. Expression (\ref{chempot_nonlin}) is a reasonable approximation for a dilute solution of interstitial atoms in a host \citep{balluffi2005}. This model corresponds to the following free energy density function:
\begin{equation}\label{gibbs_nonlin}
G_r(c) = \mu_r c + k_bT \left[ c\ln\left( \frac{c}{c_r} \right) - c \right] + k_bT c_r.
\end{equation}
The chemical modulus $K_r$ in Eq. (\ref{chempot_lin}) is thus identified as $K_r=k_bT/c_r$. The model (\ref{chempot_lin}) is a valid approximation of the nonlinear model (\ref{chempot_nonlin}) provided that the concentration does not deviate too much from the reference concentration.  
\end{remark}

Diffusion of mobile species is driven by the gradient of chemical potential $\bm g\equiv \bm{\nabla} \mu$. The simplest isotropic model of diffusion assumes a quasi-linear relationship between the flux of mobile species, $\bm j$, and the gradient of chemical potential:
\begin{equation}\label{kinetic_model}
\bm j = -k(\bm x)\bm g, 
\end{equation}
where $k$ is the molecular conductivity, which we assume to be uniform in each phase:
\begin{equation}\label{conductivity}
k(\bm x) = \sum_{r=1}^2 \chi_r(\bm x)k_r, 
\end{equation}
with $k_r$ the molecular conductivity of phase $r$. The evolution of the concentration field $c(\bm x,t)$ in space and time is governed by the species conservation equation:   
\begin{equation} \label{mass_conservation}
\dot c = - \bm{\nabla}\cdot \bm j.
\end{equation}
\begin{remark}
Combining Eqs (\ref{state_law}), (\ref{chempot_lin}), (\ref{kinetic_model}) and (\ref{mass_conservation}), we recover Fick's second law within each phase: 
\begin{equation} \label{Fick2ndLaw}
\dot c = D(\bm x) \bm{\nabla}^2 c,
\end{equation}
where the diffusion coefficient is given by:
\begin{equation}\label{diffusion_coeff}
D(\bm x) = \sum_{r=1}^2 \chi_r(\bm x) D_r, \quad D_r = k_r K_r.
\end{equation}
\end{remark}

\begin{figure}
\centering
\includegraphics[width=0.7\textwidth]{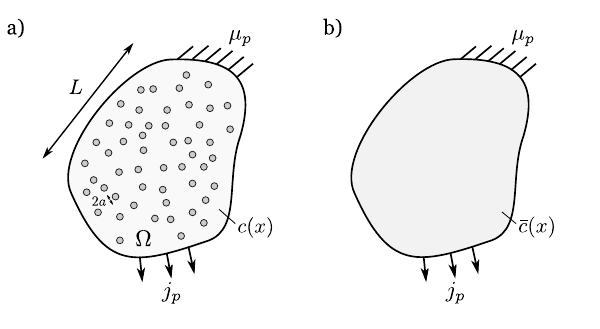}
\caption{a) Diffusion boundary-value problem in a heterogeneous medium. (b) Equivalent homogeneous medium subjected to the same boundary conditions.}
\label{fig-bvp}
\end{figure}

Boundary conditions are written either in terms of prescribed chemical potential or prescribed flux on the boundary $\partial \Omega$:
\begin{eqnarray}
-\bm j(\bm x,t)\cdot \bm n(\bm x) &=& j_p(\bm x,t) \quad \textnormal{on} \quad \partial \Omega_j,\label{bc1}\\
\mu(\bm x,t) &=& \mu_p(\bm x,t) \quad \textnormal{on} \quad \partial \Omega_{\mu}, \label{bc2}
\end{eqnarray}   
where $\bm n$ is the outward unit normal to the external surface, and $j_p$ and $\mu_p$ are prescribed functions respectively defined on portions $\partial \Omega_j$ and $ \partial \Omega_{\mu}$ of the boundary, with $\partial \Omega_j \cup \partial \Omega_{\mu} = \partial \Omega$ and $\partial \Omega_j \cap \partial \Omega_{\mu} = \emptyset$. At the interface between an inclusion and the matrix, the chemical potential and normal fluxes should be continuous. Finally, the initial condition $c(\bm x,0)=c_0(\bm x)$ must also be specified. In this work we assume $c_0(\bm x) = 0$. The diffusion boundary-value problem in the heterogeneous medium is represented in Fig. \ref{fig-bvp}a. 

The boundary value problem (\ref{bc1})-(\ref{bc2}) can be rewritten in an equivalent, weak form as follows. First introduce the set
\begin{equation}
\mathcal K(\mu_p) = \left\{ \hat{\mu} | \hat{\mu} \;  \textnormal{"sufficiently smooth" and } \hat{\mu} = \mu_p \ \textnormal{on} \ \partial \Omega_{\mu} \right\}.
\end{equation}
For all admissible fields $\hat{\mu} \in \mathcal K(\mu_p)$, the following virtual power principle is satisfied:
\begin{equation}\label{weak_form}
\int_{\Omega} \left( \hat{\mu}\dot c - \bm j \cdot \hat{\bm g}  \right) dV = -\int_{\partial \Omega_{\mu}} \mu_p \bm j \cdot \bm n dS - \int_{\partial \Omega_j} \hat{\mu} j_p dS,
\end{equation} 
where $\hat{\bm g}=\bm{\nabla}\hat{\mu}$. Note that the physical field $\mu$ belongs to the set of admissible fields, yielding a power conservation principle: the power of applied chemical potential and flux at the boundary is equal to the sum of powers of stored and dissipated energy in the bulk.

\subsection{Effective behaviour}\label{sec:macro_bvp}
We are interested in boundary-value problems with characteristic length $L$ much larger than the characteristic size $a$ of an inclusion (Fig. \ref{fig-bvp}a). In such cases, it is desirable to replace the actual heterogeneous medium with a fictitious homogeneous medium whose behaviour under applied external chemical loads is identical to the actual behaviour (Fig. \ref{fig-bvp}b). In the equivalent medium, the local state is described by an effective concentration $\bar{c}$ (representing a volume average of the concentration in the underlying microstructure, to be formally defined later), which satisfies a conservation equation of a form similar to (\ref{mass_conservation}):
\begin{equation}\label{macro_mass_cons}
\dot{\bar c} = - \bm{\nabla} \cdot \bar{\bm j}.
\end{equation}
This equation can be seen as the definition of the effective diffusion flux $\bar{\bm j}$. We also introduce an effective chemical potential $\bar{\mu}$, as well as its gradient, $\bar{\bm g}$. By analogy with the weak form of the local problem (\ref{weak_form}), the effective chemical potential is formally defined as the solution to the effective diffusion problem (\ref{macro_mass_cons}) written in a weak form. Thus, the effective chemical potential field is such that:
\begin{equation}\label{weak_form_macro}
\int_{\Omega} \left( \bar{\mu}\dot{\bar c} - \bar{\bm j} \cdot \bar{\bm g}  \right) dV = -\int_{\partial \Omega_{\mu}} \mu_p \bar{\bm j} \cdot \bm n dS - \int_{\partial \Omega_j} \bar{\mu} j_p dS.
\end{equation} 
The statement (\ref{weak_form_macro}) identifies the effective chemical potential as power-conjugate to the effective concentration rate, and its gradient as power-conjugate to the effective flux. 

The weak form (\ref{weak_form_macro}) must be supplemented by effective constitutive relations relating the generalised forces ($\bar{\mu}$,$\bar{\bm g}$) to the generalised fluxes ($\dot{\bar c}$,$\bar{\bm j}$). In general, the effective constitutive relations can be written under the form: 
\begin{eqnarray}
\dot{\bar c} &=& \mathcal F(\bar{\mu},\bar{\bm g}, \dot{\bar{\mu}},\dot{\bar{\bm g}},\textnormal{loading history}) \label{macro_kin1}, \\ 
\bar{\bm j} &=& \mathcal G (\bar{\mu},\bar{\bm g}, \dot{\bar{\mu}},\dot{\bar{\bm g}},\textnormal{loading history}) \label{macro_kin2}.
\end{eqnarray}
Our aim is to identify these effective constitutive relationships through a micro-macro approach, as described in the following sections.

\section{Two-scale approach}\label{sec:two-scale}
\begin{figure}
\centering
\includegraphics[width=0.5\textwidth]{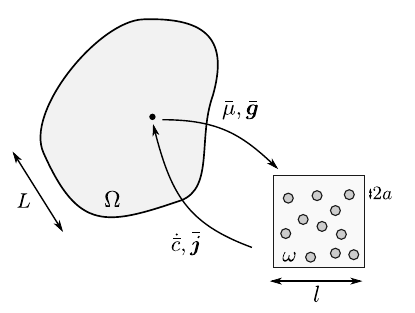}
\caption{Two-scale procedure: a RVE of the microstructure is associated to every material point in the effective medium. Effective concentration rate and flux are obtained from an analysis of the transient diffusion problem in the RVE domain.}
\label{fig-2scale}
\end{figure}

We adopt a two-scale approach according to which a Representative Volume Element (RVE) of the microstructure is associated to each material point in the effective medium (Fig.  \ref{fig-2scale}). For a random microstructure, the RVE should be chosen sufficiently large so that it is statistically representative of the actual microstructure. Let $l$ be the characteristic size of the RVE. We assume separation of scales:
\begin{equation}\label{separation_scales}
a \ll l \ll L.
\end{equation}  

At a given loading step, the effective chemical potential and gradient ($\bar{\mu}$,$\bar{\bm g}$) are viewed as given loading parameters, and we seek to identify the corresponding effective concentration rate and diffusion flux for the RVE, $(\dot{\bar c},\bar{\bm j})$. As is standard in two-scale approaches, two steps are successively considered: 
\begin{enumerate}
\item The macroscopic loading parameters are translated into boundary conditions on the RVE boundary, allowing for the microscopic concentration and diffusion fields to be calculated in the RVE by solving a transient diffusion problem (localisation step). 
\item The effective concentration and diffusion flux are calculated from local field through suitable averaging conditions (homogenisation step).   
\end{enumerate}

\subsection{Localisation}
Consider a macroscopic material point at which the values of macroscopic chemical potential and chemical potential gradient are $\bar{\mu}$ and $\bar{\bm g}$. In general, $\bar{\mu}$ and $\bar{\bm g}$ can take arbitrary values. We associate to this macroscopic point a RVE of domain $\omega$, and write $\bm x_0$ the centre of volume of the RVE in a microscale coordinate system. The centre of volume of the RVE is defined such that $\int_{\omega} (\bm x - \bm x_0) dV = \bm 0$. At any point $\bm x$ in $\omega$, the field of chemical potential is expressed as:   
\begin{equation} \label{expansion}
\mu(\bm x,t) = \bar{\mu}(t) + \bar{\bm g}(t) \cdot (\bm{x} - \bm{x}_0) + \tilde{\mu}(\bm x,t),
\end{equation}
where $\tilde{\mu}$ represents the fluctuation relative to a development of the chemical potential field up to the first order around the macroscopic value. We adopt a first-order homogenisation scheme, and require that the volume average of the chemical potential gradient over the RVE should equal the macroscopic value: 
\begin{equation}\label{average_g}
\frac{1}{V}\int_{\omega} \bm g \ dV = \bar{\bm g},
\end{equation}
where $V$ is the RVE volume. This condition in turn implies that: 
\begin{equation} \label{av_fluc}
\frac{1}{V} \int_{\omega} \bm{\nabla} \tilde{\mu} \ dV =\bm 0.
\end{equation}
Using the divergence theorem, the latter condition can be rewritten as a surface integral: 
\begin{equation} \label{boundary_constraint}
\int_{\partial \omega} \tilde{\mu} \bm n \ dS = \bm 0,
\end{equation}
where $\bm n$ is the outward unit normal to the RVE boundary. Condition (\ref{boundary_constraint}) is satisfied by setting $\tilde{\mu}=0$ on $\partial \omega$. Eq. (\ref{expansion}) then leads to affine chemical potential boundary conditions:  
\begin{equation} \label{affineBC}
\mu = \bar{\mu} + \bar{\bm{g}} \cdot (\bm x - \bm x_0) \quad \textnormal{on} \ \partial \omega. 
\end{equation} 
Note that, while affine boundary conditions satisfy condition (\ref{average_g}), in general the volume average of the chemical potential in the RVE does not necessarily coincide with the macroscopic value:  
\begin{equation}
\frac{1}{V}\int_{\omega} \mu dV \neq \bar{\mu}. 
\end{equation}
For heat conduction problems, similar affine boundary conditions can be written in terms of prescribed temperature (\cite{larsson2010}). 

\begin{remark} 
Alternatively, condition (\ref{av_fluc}) is  satisfied by requiring the micro-fluctuation field $\tilde{\mu}$ to be periodic:
\begin{equation}\label{periodicity}
\tilde{\mu}(\bm x^+) = \tilde{\mu}(\bm x^-), 
\end{equation}
where $\bm x^+$ and $\bm x^-$ are position vectors of corresponding points on the boundary. The boundary conditions then write: 
\begin{equation} \label{periodicBC}
\mu(\bm x^+) - \mu(\bm x^-) = \bar{\bm{g}} \cdot (\bm x^+ - \bm x^-)  \quad \textnormal{on} \ \partial \omega.
\end{equation}
Periodic boundary conditions only specify the field of chemical potential up to an undetermined constant. This is similar to periodic displacement boundary conditions in mechanics problems, which only determine the displacement field up to a rigid body motion. Similar to mechanics problems, the indeterminacy can be suppressed by prescribing the chemical potential value at one point on the RVE boundary. Alternatively, one can require that the average chemical potential should equal the macroscopic chemical potential: 
\begin{equation}\label{av_mu}
\frac{1}{V}\int_{\omega} \mu dV = \bar{\mu}.
\end{equation}
Introducing expansion (\ref{expansion}) into (\ref{av_mu}), it amounts to requiring that the volume average of the microscopic fluctuation field vanishes:
\begin{equation}\label{mean_fluc}
\frac{1}{V}\int_{\omega} \tilde{\mu} dV = 0.
\end{equation}
For heat conduction problems, a requirement similar to (\ref{av_mu}) was proposed by \cite{ozdemir2008a}, where it amounts to enforcing consistency of stored heat at macroscopic and microscopic levels.  
\end{remark}

\subsection{Averaging}\label{sec:averaging}
Expressions for the effective quantities ($\dot{\bar c}$,$\bar{\bm j}$) in terms of the local fields in the RVE are obtained by requiring that the macroscopic power should equal the power supplied to the RVE through its boundaries: 
\begin{eqnarray}
\bar{\mu} \dot{\bar c} - \bar{\bm j} \cdot \bar{\bm g} &=& - \frac{1}{V}\int_{\partial \omega} \mu \bm j \cdot \bm n dS \label{Hill1}\\
  &=& \frac{1}{V}\int_{\omega} \mu \dot c \ dV - \frac{1}{V} \int_{\omega} \bm j \cdot \bm{\nabla}\mu \ dV,  \label{Hill2}  
\end{eqnarray} 
where the second equality follows from application of the divergence theorem and the local species conservation equation (\ref{mass_conservation}). Eqs. (\ref{Hill1})-(\ref{Hill2}) are equivalent to the Hill-Mandel condition in (quasi-static) mechanical homogenisation problems. In its original derivation \citep{hill1967}), Hill's lemma follows from the a-priori definition of effective fields as volume averages. In contrast, here we postulate the equality (\ref{Hill1}), and determine expressions for the effective concentration rate and diffusion flux which ensures that the equality is satisfied. The generalisation of the Hill-Mandel condition as a principle of multiscale virtual power for a broad class of RVE-based methods is discussed in \cite{blanco2016}.  

Using the affine boundary condition (\ref{affineBC}) together with the divergence theorem and the conservation equation (\ref{mass_conservation}), the right-hand side of Eq. (\ref{Hill1}) becomes:  
\begin{equation}
- \frac{1}{V}\int_{\partial \omega} \mu \bm j \cdot \bm n dS = \bar{\mu} \frac{1}{V}\int_{\omega} \dot c \ dV - \frac{1}{V} \bar{\bm{g}} \cdot \int_{\partial \omega} (\bm x - \bm x_0) (\bm j \cdot \bm n) \ dS. 
\end{equation}
Comparing this expression to Eq. (\ref{Hill1}), and recalling that $\bar{\mu}$ and $\bar{\bm g}$ can be varied independently, the effective concentration rate and flux are identified:
\begin{equation} \label{macro_c}
\dot{\bar c} = \frac{1}{V} \int_{\omega} \dot c \ dV,   
\end{equation}
and 
\begin{equation} \label{macro_j1}
\bar {\bm j}  = \frac{1}{V} \int_{\partial \omega}(\bm x - \bm x_0) (\bm j \cdot \bm n) \ dS. 
\end{equation}     
The effective flux can alternatively be rewritten in terms of volume averages: 
\begin{equation} \label{macro_j2}
\bar {\bm j} = \frac{1}{V}\int_{\omega} \bm j\ dV - \frac{1}{V} \int_{\omega} \dot c (\bm x - \bm x_0) \ dV.
\end{equation}

\begin{remark}
In the case of periodic boundary conditions instead of affine boundary conditions, the same expressions (\ref{macro_c}) and (\ref{macro_j2}) can be obtained if $\int_{\partial \omega} \tilde{\mu} \bm j \cdot \bm n dS=0$. This condition can then be used, instead of \eqref{av_mu} or \eqref{mean_fluc}, to control the indeterminacy on the field $\mu$.
\end{remark}

The result (\ref{macro_c}) shows that the effective concentration $\bar{c}$ coincides with the volume average of the local concentration in the RVE, as one would have expected. On the other hand, Eq. (\ref{macro_j2}) shows that the effective diffusion flux does not coincide with the volume average of the local flux. The second term of the right-hand side of (\ref{macro_j2}) accounts for microscale inertia through the first moment of the rate of concentration in the RVE. This term depends on the RVE size, and therefore introduces a size effect which vanishes when the RVE size tends to zero. A similar size-dependent term was previously identified by \cite{larsson2010} for transient heat conduction, where it corresponds to the "moment of heat content". In the context of diffusion, a similar relation was given by \cite{salvadori2015}, \cite{nilenius2015}, and \cite{kaessmair2016}. It also appears in elastodynamics, where it corresponds to a "moment of momentum" \citep{pham2013}.  

\subsection{Orientation for the rest of the study}
In the rest of this study, we will consider composites in which the diffusivity in the matrix phase is much larger than in the inclusion phase, $D_2 \gg D_1$. We define $\tau_1$ the characteristic time for diffusion in an inclusion, $\tau_1=\frac{a^2}{D_1}$, and $\tau_2$ the characteristic time for diffusion in the RVE through the fast percolating matrix: $\tau_2 = \frac{l^2}{D_2}$. We assume that the macroscopic excitation time scale $T$ is much larger than the characteristic time for diffusion in the matrix, $T \gg \tau_2$, so that the assumption of quasi steady-state holds everywhere in the matrix:
\begin{equation}
0 = - \bm{\nabla} \cdot \bm j \quad \textnormal{in} \ \omega_2,
\end{equation}
where $\omega_2$ is the matrix domain. On the other hand, transient diffusion (Eq. (\ref{mass_conservation})) is considered in the inclusions collectively occupying domain $\omega_1$, i.e. $T\leq \tau_1$.  
 
The RVE problem under affine chemical potential boundary conditions can be solved very accurately for arbitrary geometries and material properties using a computational approach, such as the finite element method. The relationship between the loading parameters ($\bar{\mu}$,$\bar{\bm g}$) and the average concentration rate and flux $(\dot{\bar c},\bar{\bm j})$ identified in Eqs (\ref{macro_c}) and (\ref{macro_j2}) can then be obtained numerically. However, we are mostly interested in identifying the structure of the constitutive relationships (\ref{macro_kin1})-(\ref{macro_kin2}) in transient diffusion problems. Therefore, a semi-analytical mean-field model will be developed and validated by comparing its predictions to reference results obtained from full-field simulations.  

\section{Mean-field model for two-phase composites} \label{sec-model}
In this section we develop a mean-field model for transient diffusion in two-phase composites with steady-state matrix. The inclusions are assumed spherical (3D problems) or circular (2D problems) with radius $a$. For a cubic or square RVE of size $l$ containing $N$ inclusions, the inclusion volume fraction $f$ is thus given by $f = \frac{4N\pi a^3}{3l^3}$ (3D) or $f = \frac{N\pi a^2}{l^2}$ (2D). Our aim is to establish a relationship between the generalised forces ($\bar{\mu}$,$\bar{\bm g}$) and the generalised fluxes $(\dot{\bar c},\bar{\bm j})$ identified as (\ref{macro_c})-(\ref{macro_j2}) in a semi-analytical form. We introduce the following usual notations for volume averages:
\begin{equation}
\langle \cdot \rangle = \frac{1}{V} \int_{\omega} \cdot \ dV,\quad \langle \cdot \rangle_r = \frac{1}{V_r} \int_{\omega_r} \cdot \ dV,
\end{equation} 
where $V_r$ is the volume of phase $r$, such that $V_1 + V_2 = V$. Then,
\begin{equation}
\langle \cdot \rangle = f \langle \cdot \rangle_1 + (1-f)\langle \cdot \rangle_2.
\end{equation}
For later use, we introduce the following first- and second-order tensors: 
\begin{eqnarray}
\bm s_r &\equiv& \frac{1}{l}\langle (\bm x - \bm x_0) \rangle_{r}, \label{sr} \\
\bm S_r &\equiv& \frac{1}{l^2}\langle (\bm x - \bm x_0)\otimes (\bm x - \bm x_0) \rangle_{r}. \label{Sr}
\end{eqnarray}
Since  $\langle (\bm x - \bm x_0) \rangle = 0$ by definition of the centre of volume, the vectors $\bm s_1$ and $\bm s_2$ are related by: 
\begin{equation}\label{rel_s1_s2}
f \bm s_1 + (1-f) \bm s_2 = \bm 0. 
\end{equation}
For a cubic or square RVE of size $l$, one can readily verify that: $\langle (\bm x - \bm x_0)\otimes (\bm x - \bm x_0) \rangle = \frac{l^2}{12}\bm 1$. Therefore, $\bm S_1$ and $\bm S_2$ are related by:
\begin{equation}
f \bm S_1 + (1-f) \bm S_2 = \frac{1}{12}\bm 1. 
\end{equation}
Taking advantage of the spherical or circular symmetry of the individual inclusions, $\bm s_1$ and $\bm S_1$ can be expressed in terms of the position vectors of the inclusions centres, $\bm x_k$ ($k=1,N$):
\begin{eqnarray}
\bm s_1 &=& \frac{1}{Nl} \sum_{k=1}^N (\bm x_k-\bm x_0), \label{s1}\\
\bm S_1 &=& \frac{1}{Nl^2} \sum_{k=1}^N (\bm x_k-\bm x_0) \otimes (\bm x_k-\bm x_0). \label{S1}
\end{eqnarray}
The first- and second-order tensors $\bm s_1$ and $\bm S_1$ thus represent the first and second moment of the inclusion distribution about the RVE centre of volume.  

\subsection{Exact solution for a single inclusion subject to a uniform, time-varying chemical potential}
Our model relies on the exact solution for the transient diffusion problem in a single inclusion subject to a uniform, time-varying chemical potential at its boundaries developed in \citep{brassart2017}. Consider an inclusion with chemical modulus $K$, molecular conductivity $k$ and diffusivity $D=Kk$. The inclusion is subject to a time-varying chemical potential, $\mu_0(t)$ prescribed on its boundaries. The average concentration in the inclusion, $\langle c \rangle_0$, is given by \citep{brassart2017}: 
\begin{equation}\label{c_1inclusion}
\langle c \rangle_0(t) =  \frac{1}{K} \int_0^t J(t-t') \frac{d\mu_0}{dt'} dt',
\end{equation}
where $J(t)$ is the chemical creep function, which depends on the inclusion size, geometry and diffusion coefficient\footnote{We adopted the expression "chemical creep" because of the similarity of expression (\ref{c_1inclusion}) with the description of the creep response of a linear viscoelastic material.}. In writing (\ref{c_1inclusion}), we have assumed that the concentration in the inclusion is zero at $t=0$. The chemical creep function has the following properties: $J(0)=0$ (initial condition) and $\lim_{t\rightarrow \infty}J(t) = 1$ (chemical equilibrium). In general, the chemical creep function can be written as a series expansion:    
\begin{equation}\label{creep_fun}
J(t) = 1 - \sum_{m=1}^{\infty} A_m \exp(-t/\tau_m),
\end{equation}  
where the coefficients $A_m$ are mode amplitudes and $\tau_m$ the associated relaxation times, with the following property: $\sum_{m=1}^{\infty} A_m=1$, which ensures that the initial condition is satisfied. For the radial diffusion in a circular inclusion with radius $a$, these coefficients are given by: 
\begin{equation}\label{modes_cylinder}
A_m = \frac{4}{z^2_m},\quad \tau_m = \frac{a^2}{D}\frac{1}{z^2_m},
\end{equation}
where $z_m$ is the m$^{th}$ root of the zero-order Bessel function of the first kind, $J_0(z)$. For the radial diffusion in a spherical inclusion with radius $a$, these coefficients are given by:  
\begin{equation}\label{modes_sphere}
A_m = \frac{6}{m^2 \pi^2},\quad \tau_m = \frac{a^2}{D}\frac{1}{m^2 \pi^2}.
\end{equation}

\begin{remark}
When the inclusion is subject to a non-zero chemical potential applied in $t=0^+$, the solution (\ref{c_1inclusion}) should be rewritten in the more general form:
\begin{equation}
\langle c \rangle_0(t) = \frac{\mu_0(0^+)}{K}J(t) + \frac{1}{K} \int_0^t J(t-t') \frac{d\mu_0}{dt'} dt'.
\end{equation}
In particular, for a step load $\mu_0(t) = \alpha H(t)$, with $\alpha$ an arbitrary non-zero value and $H(t)$ the Heaviside step function, the inclusion response simply reduces to:
\begin{equation}
\langle c \rangle_0(t) = \frac{\alpha}{K}J(t). 
\end{equation}
In the following, we write the history-dependent response under the form (\ref{c_1inclusion}) for simplicity.
\end{remark}

\subsection{Estimate for the transient composite response}
Our mean-field model relies on the assumption that the chemical potential can be considered as uniform on the boundary of each inclusion in the RVE. This approximation is reasonable provided that the length scale associated with the gradient of effective chemical potential is much larger than the inclusion size, $\bar{\mu}/|\bar{\bm g}| \gg a$. It follows that the average flux in the inclusions is negligible, $\langle \bm j \rangle_1 \approx 0$. The average flux in the RVE is then obtained by solving the steady-state diffusion problem in the RVE, assuming non-conducting inclusions:   
\begin{equation}\label{av_flux_macro}
\langle \bm j \rangle = -\bar{\bm k}\cdot \bar{\bm g},
\end{equation}
where $\bar{\bm k}$ is the effective conductivity tensor for a composite with non-conducting inclusions. The corresponding average chemical potential gradient in the matrix phase is given by
\begin{equation}\label{g_matrix}
\langle \bm g \rangle_2 = \frac{1}{k_2(1-f)}\bar{\bm k}\cdot \bar{\bm g}. 
\end{equation}
To the first order, the field of chemical potential in the matrix is approximated by an affine relation:
\begin{equation}\label{mu_matrix}
\mu(\bm x,t) = \bar{\mu}(t) + \langle \bm g \rangle_2(t) \cdot(\bm x - \bm x_0).
\end{equation}
From the latter expression, a mean-field estimate of the concentration in the steady-state matrix is proposed: 
\begin{equation}\label{c_matrix}
\langle c \rangle_2(t) = \frac{1}{K_2}\left(\bar{\mu}(t) + l \bm s_2 \cdot \langle \bm g \rangle_2(t)   \right),
\end{equation} 
where $\bm s_2$ was defined in Eq. (\ref{sr}). On the other hand, the contribution $\langle \dot c (\bm x - \bm x_0) \rangle_2$ to the macroscopic flux vanishes, by virtue of the quasi steady-state assumption in the matrix phase.  

The uniform chemical potential $\mu_k$ on the boundary of the $k^{\textnormal{th}}$ inclusion ($k=1,..,N$) centred at $\bm x_k$ is estimated from an affine relation similar to Eq. (\ref{mu_matrix}):   
\begin{equation} \label{mu_boundary_inclusions}
\mu_k(t) = \bar{\mu}(t) + \langle \bm g \rangle_2(t) \cdot (\bm x_k - \bm x_0).
\end{equation}
Let $\langle c \rangle_{1,k}$ be the average concentration in the $k^{\textnormal{th}}$ inclusion. The solution $\langle c \rangle_{1,k}(t)$ is of the form (\ref{c_1inclusion}), with $\mu_0(t)=\mu_k(t)$. Averaging over all inclusions in the RVE then gives: 
\begin{eqnarray} 
\langle c \rangle_{1}(t) &=& \frac{1}{N} \sum_{k=1}^N \langle c \rangle_{1,k}(t) \\
&=& \frac{1}{N}\sum_{k=1}^N \frac{1}{K_1}\int_0^t J(t-t') \frac{d\mu_k}{dt'}dt' \label{c_allinclusions1}\\
 &=& \frac{1}{K_1}\int_0^t J(t-t') \frac{d\hat{\mu}}{dt'} dt' \label{c_allinclusions2}
\end{eqnarray}
where: 
\begin{equation}\label{hat_mu}
\hat{\mu}(t) = \bar{\mu}(t) + l\bm s_1 \cdot \langle \bm g \rangle_2(t). 
\end{equation}
Expression (\ref{c_allinclusions2}) shows that the average concentration response in the inclusion phase can be obtained from the solution for a single inclusion subject to an effective chemical potential $\hat{\mu}(t)$ on its boundary. 

The first moment of the concentration $\langle c (\bm x-\bm x_0)\rangle_1$ in the inclusions is obtained as follows:  
\begin{eqnarray}
\langle c (\bm x - \bm x_0) \rangle_{1} &=& \frac{1}{N}\sum_{k=1}^N \langle c (\bm x-\bm x_0) \rangle_{1,k} \label{cmom_icl0}\\
&=& \frac{1}{N}\sum_{k=1}^N \langle c \rangle_{1,k} (\bm x_k - \bm x_0) \label{cmom_icl1}\\
&=&\frac{l}{K_1}\int_0^t J(t-t') \frac{d\check{\bm{\mu}}}{dt'}  dt'. \label{cmom_icl2}
\end{eqnarray}
where:
\begin{equation}\label{check_mu}
\check{\bm{\mu}}(t) = \bm s_1 \bar{\mu}(t) + l\bm S_{1}\cdot \langle \bm g \rangle_2(t).
\end{equation}
Eq. (\ref{cmom_icl1}) follows from the spherical or circular symmetry, and Eq. (\ref{cmom_icl2}) from the single inclusion solution (\ref{c_1inclusion}) together with the prescription   (\ref{mu_boundary_inclusions}). Expression (\ref{cmom_icl2}) shows that each component $\langle c (x_i - x_{0,i}) \rangle_{1}$ can be obtained from the solution for a single inclusion subject to an effective chemical potential $\check{\mu}_i$ on its boundary. The contribution $\langle \dot c (\bm x - \bm x_0) \rangle_1$ to the effective flux in Eq. (\ref{macro_j2}) then directly follows from (\ref{cmom_icl2}) by time differentiation. 

\begin{remark}
The assumption of uniform chemical potential on the boundary of spherical or circular inclusions implies that $\langle \bm g \rangle_1=\bm 0$, which is not true for a composite with non-conducting inclusions at steady-state, as in that case:
\begin{equation}
\langle \bm g \rangle_1 = \frac{1}{k_2 f}(k_2\bm 1 - \bar{\bm k})\cdot \bar{\bm g}.
\end{equation}
Nonetheless, this inconsistency is expected to have negligible influence on the results as long as the macroscopic chemical potential gradient is not too large.
\end{remark}

Combining Eqs (\ref{macro_c}), (\ref{c_matrix}) and (\ref{c_allinclusions2}), the macroscopic concentration is obtained:   
\begin{equation}\label{macro_c_integral}
\bar c(t) =\frac{(1-f)}{K_2}\left(\bar \mu(t) + l \bm s_2 \cdot \langle \bm g \rangle_2(t)  \right) + \frac{f}{K_1}\int_0^t J(t-t') \frac{d\hat{\mu}}{dt'} dt', 
\end{equation}
where $\langle \bm g\rangle_2$ is given by Eq. (\ref{g_matrix}) and $\hat{\mu}$ by Eq. (\ref{hat_mu}). The macroscopic concentration has an instantaneous component associated with the steady-state matrix, and a transient, history-dependent component associated with the inclusions, bringing about a memory effect. The macroscopic flux is obtained by combining Eqs (\ref{macro_j2}), (\ref{av_flux_macro}) and (\ref{cmom_icl2}):
\begin{equation}\label{macro_flux_integral}
\bar{\bm j} = -\bar{\bm k}\cdot \bar{\bm g} - \frac{fl}{K_1} \int_0^t J'(t-t') \frac{d\check{\bm{\mu}}}{dt'} dt', 
\end{equation}
where $\check{\bm{\mu}}$ was defined in Eq. (\ref{check_mu}). In Expression (\ref{macro_flux_integral}), the first term is the steady-state flux contribution due to fast diffusion through the matrix, and the second term represents the transient contribution to the flux due to the inclusions. The latter represents the microscale inertia, and includes the size-effect mentioned in Section \ref{sec:averaging}. 

\subsection{Isotropic estimate}
When the distribution of inclusions is isotropic, the structure tensors can be simplified as follows (see \ref{sec:app_A}):
\begin{equation}
\bm s_1 = \bm 0,\quad \bm S_1 = \frac{1}{12}\bm{1}.
\end{equation}
The mean-field model then reduces to:
\begin{eqnarray}
\bar c &=&\frac{(1-f)\bar{\mu}}{K_2} + \frac{f}{K_1}\int_0^t J(t-t') \frac{d\bar{\mu}}{dt'} dt' \label{macro_c_iso}\\
\bar{\bm j} &=& -\bar{k} \bar{\bm g} - \frac{fl^2}{12K_1} \int_0^t J'(t-t') \frac{d\bm{g}}{dt'} dt', \label{macro_flux_iso}
\end{eqnarray}
where $\bar k$ is the effective conductivity of an isotropic composite with spherical, non-conducting inclusions. A closed-form estimate of the latter for spherical inclusions is for example provided by the Hashin-Shtrikman upper bound \citep{hashin1962}, see also \citep{benveniste1986} and \citep{torquato2002}:
\begin{equation}\label{HS_3D}
\bar k = \frac{1-f}{1+\frac{f}{2}} k_2.
\end{equation}
For 2D problems with circular inclusions, the estimate becomes:
\begin{equation}\label{HS_2D}
\bar k = \frac{1-f}{1+f} k_2.
\end{equation} 
 
\section{Strategy for upscaling}\label{sec:upscaling}
\subsection{Internal variable representation} 
Eqs (\ref{macro_c_integral})-(\ref{macro_flux_integral}) (or (\ref{macro_c_iso})-(\ref{macro_flux_iso}) for isotropic composites) completely specify the RVE transient response in terms of the past history of loading. However, the integral representation of the loading history is not practical for numerical implementation. In this section, we reformulate the model in terms of a finite number of internal variables, and also briefly discuss time discretisation. The model is then illustrated in the case of an isotropic distribution of inclusions.  

We start by developing an estimate $\tilde J(t)$ of the chemical creep relaxation function (\ref{creep_fun}) using a finite number $M$ of relaxation modes: 
\begin{eqnarray}
\tilde J(t) &=& 1 - \sum_{m=1}^{M} \tilde A_m \exp(-t/\tilde{\tau}_m) \label{creep_estimate1} \\
     &=& \sum_{m=1}^{M} \tilde A_m (1- \exp(-t/\tilde{\tau}_m)) + \tilde A_{M+1}  \label{creep_estimate2}
\end{eqnarray}
where $\tilde{A}_{M+1} = 1-\sum_{m=1}^M \tilde{A}_m$. Note that this estimate in general predicts a non-physical instantaneous concentration response to a step load in the single inclusion problem, since it does not a priori satisfy the condition $\tilde J(0)=0$, unless the coefficients $\tilde A_m$ have been chosen such that $\sum_{m=1}^M \tilde A_m = 1$. The error on the initial response increases as $\sum_{m=1}^M \tilde A_m$ deviates from one. However the estimate is exact in the long time limit by construction. In our previous work \citep{brassart2017} we have proposed and discussed several strategies to identify mode amplitudes and relaxation times for estimates of the form (\ref{creep_estimate1})-(\ref{creep_estimate2}) with a limited number of modes, including a collocation method and a FE-based modal analysis. The proposed methods are applicable to arbitrary inclusion geometries, thus also in cases where the chemical creep function is not available in closed form. In this work, we consider only inclusions with a circular shape for simplicity, for which the coefficients $A_m$ and $\tau_m$ have an analytical expression, cf. Eq. (\ref{modes_cylinder})-(\ref{modes_sphere}). In this case, the simplest method for identifying the coefficients in the estimate (\ref{creep_estimate1})-(\ref{creep_estimate2}) is to truncate the infinite series up to the M$^{\textnormal{th}}$ term, that is:
\begin{equation}
\tilde{A}_m = A_m,\ \tilde{\tau}_m = \tau_m,\quad (m=1,M).  \label{chosen_estimate}
\end{equation}
The minimum number of modes for an accurate estimate of the chemical creep function depends on the excitation time scale, see \citep{brassart2017}. For the loading conditions considered in the following, we used $M=20$ , which is  more than sufficient for the single inclusions estimate (\ref{bm}) to be virtually identical to the reference solution obtained when considering $M\rightarrow \infty$.

Based on the estimate (\ref{creep_estimate2}), it is then possible to express the single inclusion solution in terms of $M+1$ internal variables \citep{ricaud2009,brassart2017}. For the inclusion phase response, we write:
\begin{equation} \label{bm}
\langle c \rangle_1 = \sum_{m=1}^{M+1} b_m,
\end{equation}
where the internal variable $b_m$ obey the following evolution laws \citep{brassart2017}:
\begin{eqnarray}
\dot{b}_{m} &=& \frac{1}{\tilde{\tau}_m}\left( \frac{\tilde A_m \hat{\mu}}{K_1} - b_{m} \right), \quad (m=1,M)  \label{ode1} \\
b_{M+1} &=& \frac{\tilde A_{M+1}}{K_1}\hat{\mu}. \label{ode2}
\end{eqnarray}
Eq. (\ref{ode2}) expresses the instantaneous response that follows from the prescription (\ref{chosen_estimate}), which does not satisfy $\sum_{m=1}^M \tilde A_m = 1$. Note that the internal variable representation and associated evolution laws remain valid even in the presence of a step load applied in $t=0$. Similarly, the first moment of the concentration in the inclusion phase is decomposed as:
\begin{equation} \label{int_var2}
\langle c (\bm x - \bm x_0) \rangle_{1} = \sum_{m=1}^{M+1} \bm d_{m} ,  
\end{equation}
where the pseudo-vectors $\bm d_{m}$ obey:
\begin{eqnarray}
\dot{\bm d}_{m} &=& \frac{l}{\tilde{\tau}_m}\left( \frac{\tilde A_m \check{\bm{\mu}}}{K_1} - \bm d_{m} \right), \quad (m=1,M)  \label{ode3} \\
\bm d_{M+1} &=& \frac{l\tilde A_{M+1}}{K_1} \check{\bm{\mu}}. \label{ode4}
\end{eqnarray}

Using the internal variable representation, the mean-field model (\ref{macro_c_integral})-(\ref{macro_flux_integral}) rewrites as:
\begin{equation}\label{macro_c_intvar}
\dot{\bar c} =  \frac{(1-f)}{K_2} \left( \dot{\bar{\mu}} + l\bm s_2 \cdot \langle \dot{\bm g}\rangle_2 \right) + f\sum_{m=1}^{M} \frac{1}{\tilde{\tau}_m} \left(\frac{\tilde A_m}{K_1} \hat{\mu} - b_m \right) +f\frac{\tilde A_{M+1}}{K_1}\dot{\hat{\mu}}. 
\end{equation}
and
\begin{equation}\label{macro_flux_intvar}
\bar{\bm j} = - \bar{\bm k} \cdot \bar{\bm{g}} - fl \sum_{m=1}^{M} \frac{1}{\tilde{\tau}_m} \left(\frac{\tilde A_m}{K_1} \check{\bm{\mu}} - \bm d_m\right) - fl \frac{\tilde A_{M+1}}{K_1} \dot{\check{\bm{\mu}}}. 
\end{equation}
These two relations completely specify the sought-after constitutive relation (\ref{macro_kin1})-(\ref{macro_kin2}), where the quantities $b_m$ and $\bm d_m$ ($m=1,M$) are internal variables accounting for the loading history. 

\subsection{Time discretisation} \label{sec-update}
In a time-discretised setting, the mean-field model (\ref{macro_c_intvar})-(\ref{macro_flux_intvar}) can be readily integrated in time using a fully implicit Euler scheme. Suppose that all internal variables are known at a simulation time $t_n$ $(b^{(n)}_m,\bm d^{(n)}_m)$ ($m=1,M+1$), where the superscript indicates the time step. For given values of the macroscopic chemical potential and its gradient at time $t_{n+1}$, $\bar{\mu}^{(n+1)}$ and $\bar{\bm g}^{(n+1)}$, the corresponding values of $\hat{\mu}^{(n+1)}$ and $\check{\bm{\mu}}^{(n+1)}$ are calculated from their definition (\ref{hat_mu}) and (\ref{check_mu}). The updates $(b^{(n+1)}_m,\bm d^{(n+1)}_m)$ can then be calculated as:
 \begin{eqnarray}
b^{(n+1)}_{m} &=& b^{(n)}_m + \frac{\Delta t}{\tilde{\tau}_m}\left( \frac{\tilde A_m \hat{\mu}^{(n+1)}}{K_1} - b^{(n+1)}_{m} \right), \quad (m=1,M)  \label{euler1}\\
b^{(n+1)}_{M+1} &=& \frac{\tilde A_{M+1}}{K_1}\hat{\mu}^{(n+1)}. 
\end{eqnarray}
and:
\begin{eqnarray}
\bm d^{(n+1)}_{m} &=& \frac{\Delta t l}{\tilde{\tau}_m}\left( \frac{\tilde A_m \check{\bm{\mu}}^{(n+1)}}{K_1} - \bm d^{(n+1)}_{m} \right), \quad (m=1,M)  \label{euler2} \\
\bm d^{(n+1)}_{M+1} &=& \frac{l\tilde A_{M+1}}{K_1} \check{\bm{\mu}}^{(n+1)}. 
\end{eqnarray}
Eqs (\ref{euler1}) and (\ref{euler2}) are linear and can thus be solved analytically for $b^{(n+1)}_m$ and $\bm d^{(n+1)}_m$. The updates of macroscopic concentration and macroscopic fluxes are finally calculated:
\begin{eqnarray}
\bar c^{(n+1)} &=& \frac{(1-f)}{K_2} \left( \bar{\mu}^{(n+1)} + l\bm s_2 \cdot \langle \bm g \rangle_2^{(n+1)}  \right) + f \sum_{m=1}^{M+1} b^{(n+1)}_m \\
\bar{\bm j}^{(n+1)} &=&  - \bar{\bm k} \cdot \bar{\bm{g}}^{(n+1)} - f \sum_{m=1}^{M+1} \bm d^{(n+1)}_m 
\end{eqnarray}
where $\langle \bm g \rangle_2^{(n+1)}$ is calculated from $\bar{\bm g}^{(n+1)}$ using Eq. (\ref{g_matrix}). Algorithmic tangent operators $\frac{\partial \bar c^{(n+1)}}{\partial \bar{\mu}^{(n+1)}}$, $\frac{\partial \bar c^{(n+1)}}{\partial \bar{\bm g}^{(n+1)}}$, $\frac{\partial \bar{\bm j}^{(n+1)}}{\partial \bar c^{(n+1)}}$ and $\frac{\partial \bar{\bm j}^{(n+1)}}{\partial \bar{\bm g}^{(n+1)}}$ can be obtained by straightforward differentiation in view of two-scale simulations. 

The complete model comprises respectively $3(M+1)$ and $4(M+1)$ internal variables in 2D and 3D. For the chosen value $M=20$, this leads to respectively 63 and 84 internal variables. However, given that internal variables obey linear evolution laws, the computational cost of the mean-field model is actually very small. Note that the number $M$ could probably be further reduced without impacting much the quality of the prediction.  

\subsection{Numerical example}
\begin{figure}
\begin{center}
\begin{subfigure}[b]{0.46\textwidth}
\includegraphics[width=\textwidth]{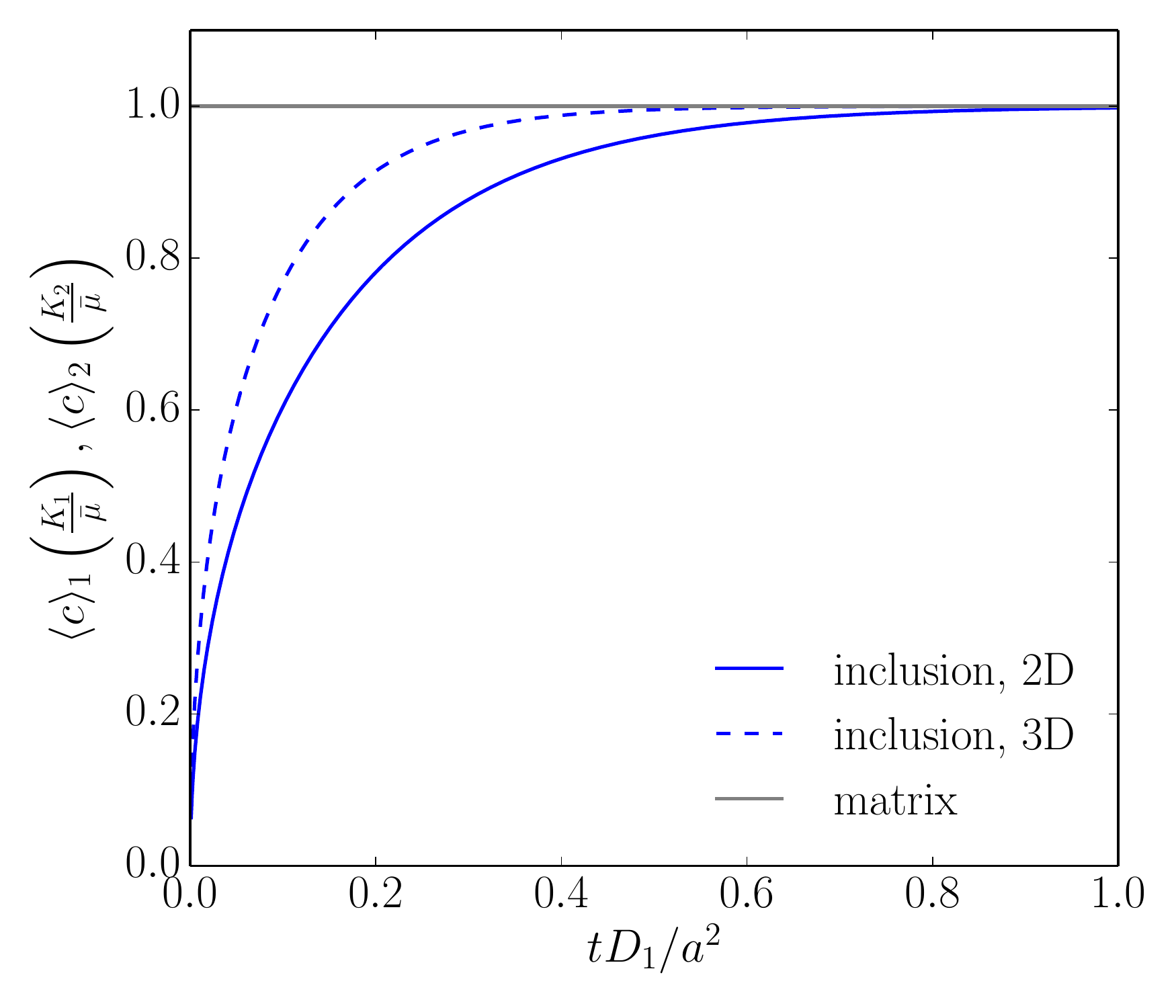}
\caption{}
\label{step_mu}
\end{subfigure}
\begin{subfigure}[b]{0.46\textwidth}
\includegraphics[width=\textwidth]{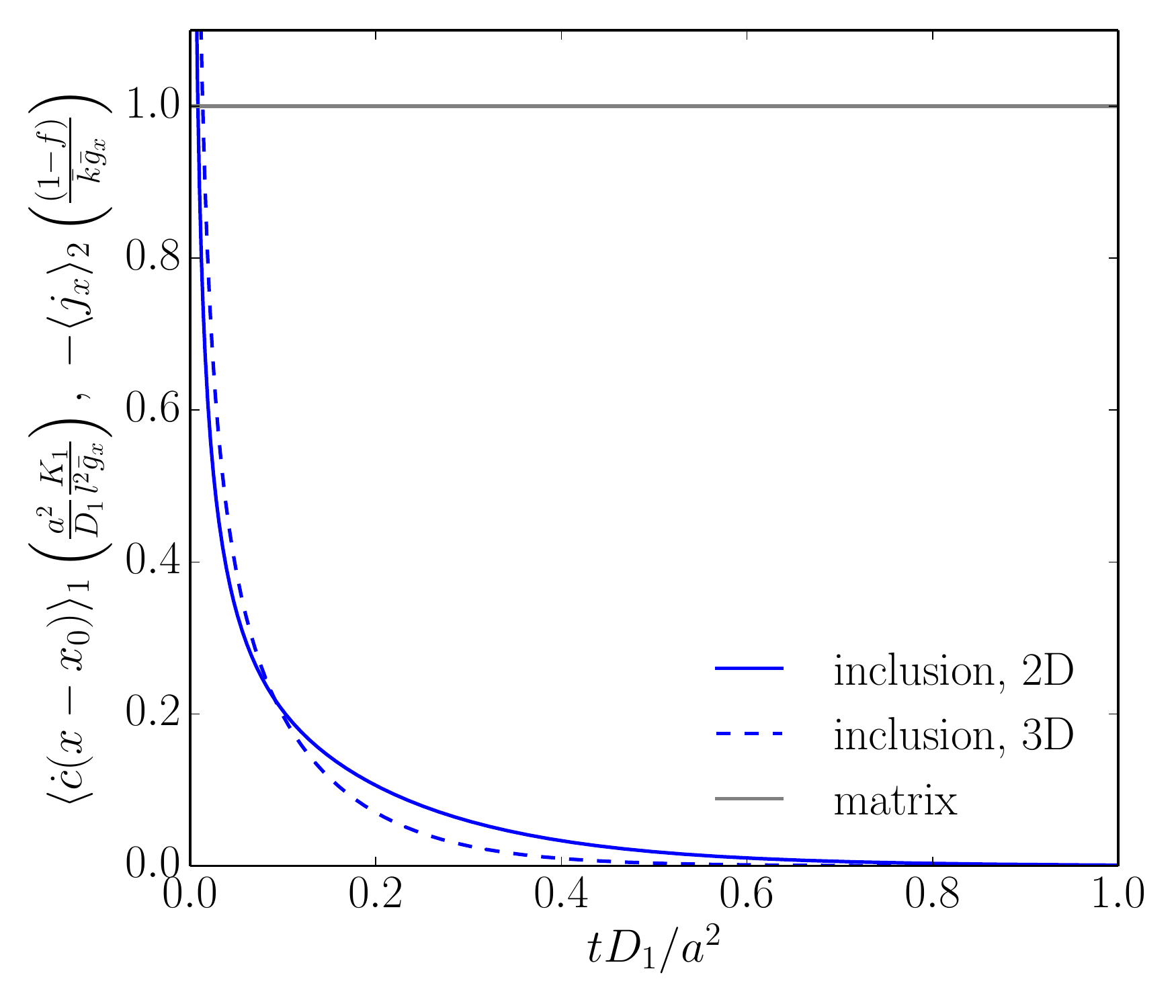}
\caption{}
\label{step_g}
\end{subfigure}
\caption{Predictions of the isotropic mean-field model in response to (a) a unit step loading $\bar{\mu}(t) = H(t)$ and (b) a unit step loading $\bar g_x = H(t)$.}
\end{center}
\end{figure}

We illustrate the model in the case of a composite with an isotropic distribution of inclusions in 2D (circular inclusions) and 3D (spherical inclusions). Per-phase concentration responses in a RVE subjected to a step load $\bar{\mu}H(t)$ are represented in Fig. \ref{step_mu}. The average concentration of each phase is normalised by its equilibrium value, $\bar{\mu}/K_r$. Time is normalised by a characteristic time for diffusion in the inclusion, $a^2/D_1$. Since the matrix is assumed infinitely fast in the RVE, the equilibrium concentration is reached instantaneously in the matrix. In contrast, the average concentration in the slow inclusion phase gradually evolves towards its equilibrium value. The kinetics also depends on the chosen geometry, circular or spherical. According to the model, the per-phase concentration response is independent of any applied macroscopic gradient $\bar{\bm g}$ (because of isotropy), and the results in the figure thus also hold in the presence of an applied gradient. 
 
Per-phase flux contributions in a RVE subject to a step load in terms of macroscopic chemical potential gradient in the x-direction, $\bar g_x H(t)$ (with $\bar g_y=\bar g_z=0$), are represented in Fig. \ref{step_g}. The inclusion phase contributes to the macroscopic flux through the term $\langle \dot c (x-x_0) \rangle_1$, and the matrix phase through the average flux $\langle j_x \rangle_2 = \langle j_x\rangle/(1-f)$. These terms are normalised in such a way that the curves are independent of specific values of volume fraction, RVE size and material parameters. The inclusion contribution tends to infinity at small times but quickly decays with time, while the average flux in the steady-state matrix is constant. According to the model, both contributions are independent of any applied macroscopic chemical potential $\bar{\mu}$, and the results in the figure also hold for a non-zero applied macroscopic chemical potential. 
 
\section{Model validation}\label{sec:validation}
\begin{figure}
\begin{center}
\centering
\begin{subfigure}[b]{0.3\textwidth}
\includegraphics[width=\textwidth]{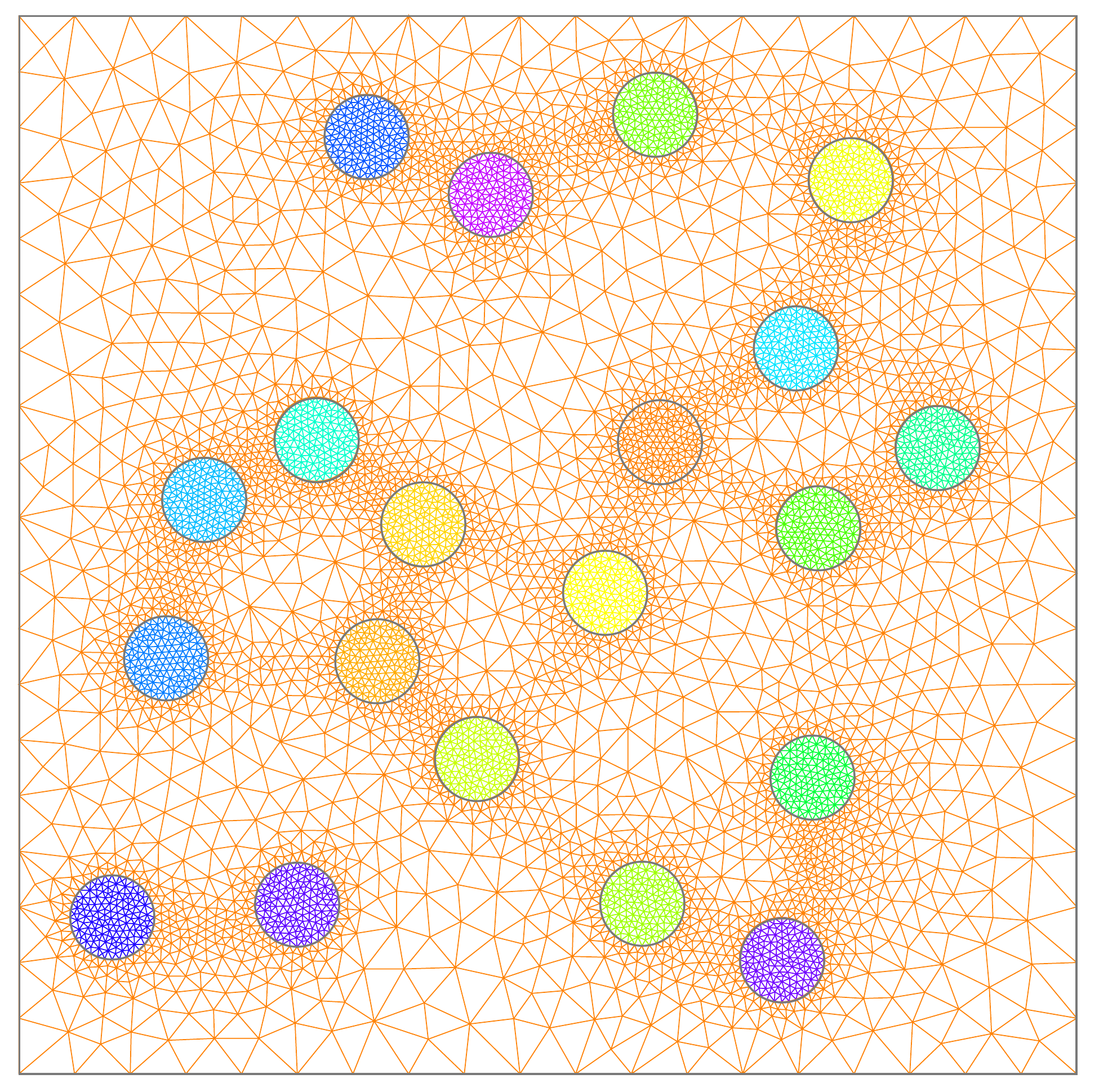}
\begin{center}
\scalebox{0.6}{$\bm s_1=\left(\begin{array}{c} -0.008\\-0.004 \end{array} \right)$}\\
\vspace{0.25cm}

\scalebox{0.6}{$\bm S_1=\left( \begin{array}{cc} 0.053&0.012\\0.012&0.062\end{array} \right) $}
\end{center}
\caption{}
\end{subfigure}
\hspace{0.25cm}
\begin{subfigure}[b]{0.3\textwidth}
\includegraphics[width=\textwidth]{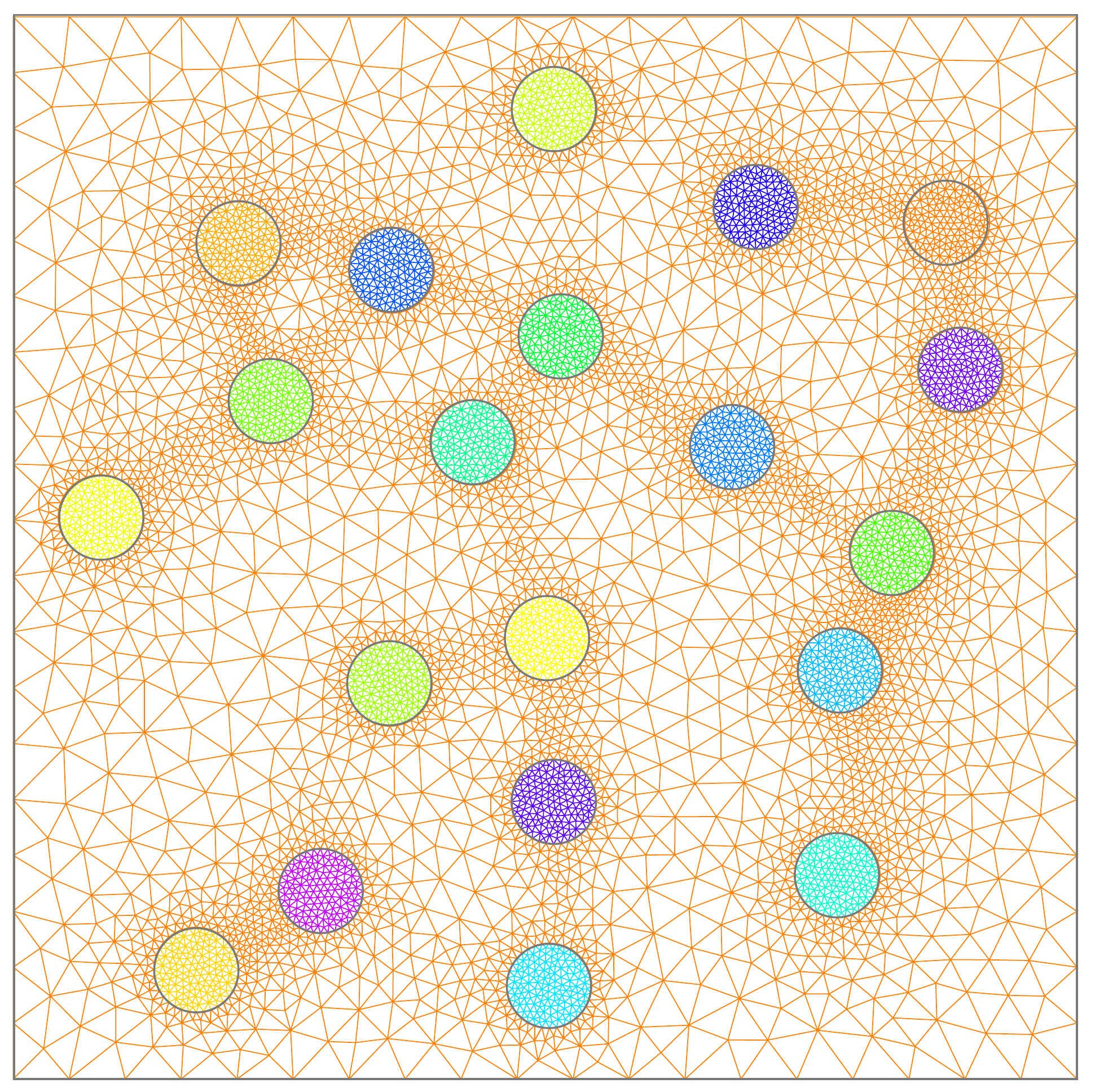}
\begin{center}
\scalebox{0.6}{$\bm s_1=\left(\begin{array}{c} 0.010\\0.015 \end{array} \right)$}\\
\vspace{0.25cm}

\scalebox{0.6}{$\bm S_1=\left( \begin{array}{cc} 0.057&0.009\\0.009&0.062\end{array} \right) $}
\end{center}
\caption{}
\end{subfigure}
\hspace{0.25cm}
\begin{subfigure}[b]{0.3\textwidth}
\includegraphics[width=\textwidth]{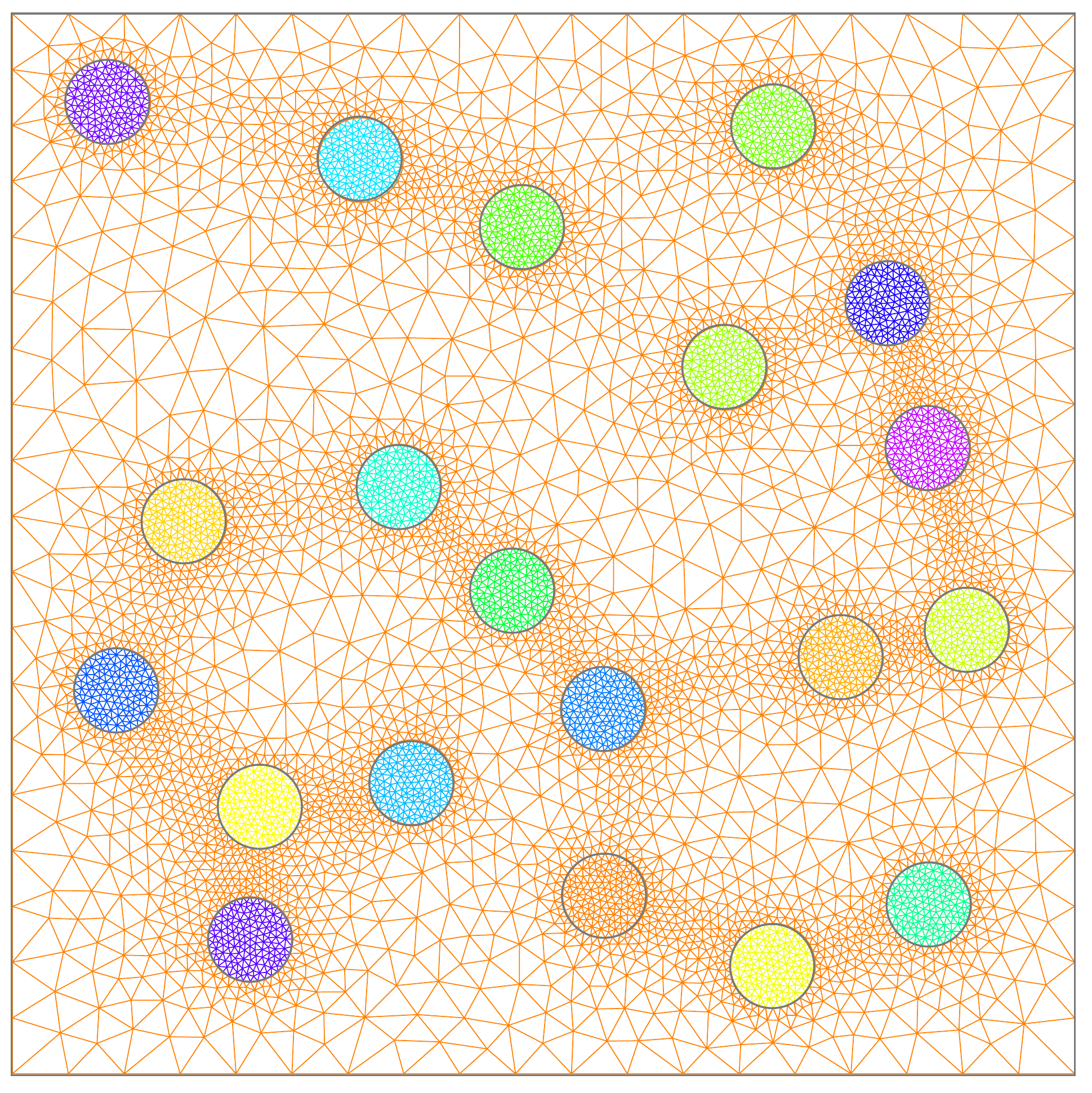}
\begin{center}
\scalebox{0.6}{$\bm s_1=\left(\begin{array}{c} 0.014\\-0.021 \end{array} \right)$}\\
\vspace{0.15cm}

\scalebox{0.6}{$\bm S_1=\left( \begin{array}{cc} 0.070&-0.004\\-0.004&0.067\end{array} \right) $}
\end{center}
\caption{}
\end{subfigure}
\caption{The three considered random distributions with $N=20$ and $f=0.1$.}
\label{fig-geoms}
\end{center}
\end{figure}

We validate the proposed mean-field model by comparing its predictions to reference results obtained by solving the transient diffusion problem on 2D RVEs using the Finite Element Method (FEM). Three random realisations of RVEs with 20 inclusions and volume fraction $f=0.1$ were considered (Fig. \ref{fig-geoms}). Morphology tensors $\bm s_1$ and $\bm S_1$ for each of these inclusion distributions were calculated numerically from the inclusion positions and are reported in the figure. One can see from the calculated values of $\bm s_1$ and $\bm S_1$ that the considered inclusion distributions are not perfectly isotropic, which directly results from the limited number of inclusions, as well as from the fact that inclusions were not allowed to intersect RVE boundaries in the random sequential algorithm used to generate the random microstructures. Therefore, the general anisotropic model (\ref{macro_c_integral})-(\ref{macro_flux_integral}) was used. Effective anisotropic conductivity tensors $\bar{\bm k}$ were calculated numerically for each geometry by subjecting each unit cell to unit chemical potential gradients in each direction, assuming non-conducting inclusions.

The material properties and external loading time scale $T$ were selected in such a way that the assumptions of steady-state matrix and transient inclusions hold simultaneously. These two conditions are met provided that $\tau_2 \ll T < \tau_1$, with $\tau_1=a^2/D_1$ and $\tau_2=l^2/D_2$. For a fixed $a/l$ ratio, the range of relevant time scales thus increases with the diffusivity contrast $D_2/D_1 = K_2 k_2/K_1 k_1$. In the following, we used $K_2/K_1=6$ and $k_2/k_1 = 10^5$, in combination with excitation periods $T=0.1\tau_1$. If the conductivity contrast is reduced at constant loading time scale, the steady-state matrix assumption breaks down and the mean-field model is no longer accurate. The effect of decreasing diffusivity contrast on the accuracy of the model predictions is examined in \ref{sec:app_B}. If one simultaneously increases the loading time scale to maintain the matrix in a steady state, then one may reach the point where the inclusions are also in a steady state. In this case the model also looses its interest, and one should instead rely on available mean-field estimates derived under steady-state RVE assumption.

The geometries were meshed with $\sim 12500$ second-order triangular elements using the software Gmsh \citep{gmsh}. Fully-implicit, finite element simulations were carried out using an in-house finite element code. Reference FE predictions of the macroscopic concentration and flux were obtained from their definition (\ref{macro_c}) and (\ref{macro_j2}), and volume averages were numerically-calculated as weighted averages over integration points. All simulations (mean-field and full-field) were carried out using a time step $\Delta t = 10^{-3}\tau_1$. The number of internal variables in each single inclusion companion problem was set to $M=20$, as previously mentioned. 

In the following, we also compare mean-field predictions of concentrations to a simpler model that assumes that concentrations are at equilibrium with the macroscopic chemical potential at all times: 
\begin{equation}\label{ss}
\bar c_{eq} = fc_{1,eq} + (1-f)c_{2,eq} = \frac{\bar{\mu}}{\bar K},
\end{equation}     
with $c_{1,eq} = \bar{\mu}/K_1$ and $c_{2,eq} = \bar{\mu}/K_2$. In Eq. (\ref{ss}), the effective chemical modulus $\bar K$ is given by:  
\begin{equation}\label{eq_capa}
\bar K = \left(\frac{f}{K_1} + \frac{(1-f)}{K_2} \right)^{-1}.
\end{equation}    

\subsection{Macroscopically-uniform chemical potential}
We first consider macroscopic loading conditions where the macroscopic chemical potential varies harmonically in time, and the macroscopic chemical potential gradient is zero:  
\begin{equation}\label{loading1}
\begin{array}{rcl}
\bar{\mu}(t) &=& \mu_0 \sin (\omega t) \\
\bar{\bm{g}} &=& \bm 0
\end{array}
\end{equation}
where $\omega = 2\pi/T$ is the angular frequency, $T=0.1\tau_1$ the excitation period and $\mu_0$ a constant. In the absence of a macroscopic chemical potential gradient, the chemical potential in the matrix is uniform at all times and equal to the macroscopic chemical potential, and the average flux in the matrix vanishes. Each inclusion is subject to a uniform, time-varying macroscopic chemical potential $\bar{\mu}(t)$ on its boundaries. We thus expect the mean-field model to be exact in this case, up to the discretisation error introduced by considering a finite number of internal variables and the time-discretisation error. We found these two sources of error to be negligible with the chosen values for $M$ and $\Delta t$.    

\begin{figure}
\begin{center}
\begin{subfigure}[b]{0.4\textwidth}
\includegraphics[width=\textwidth]{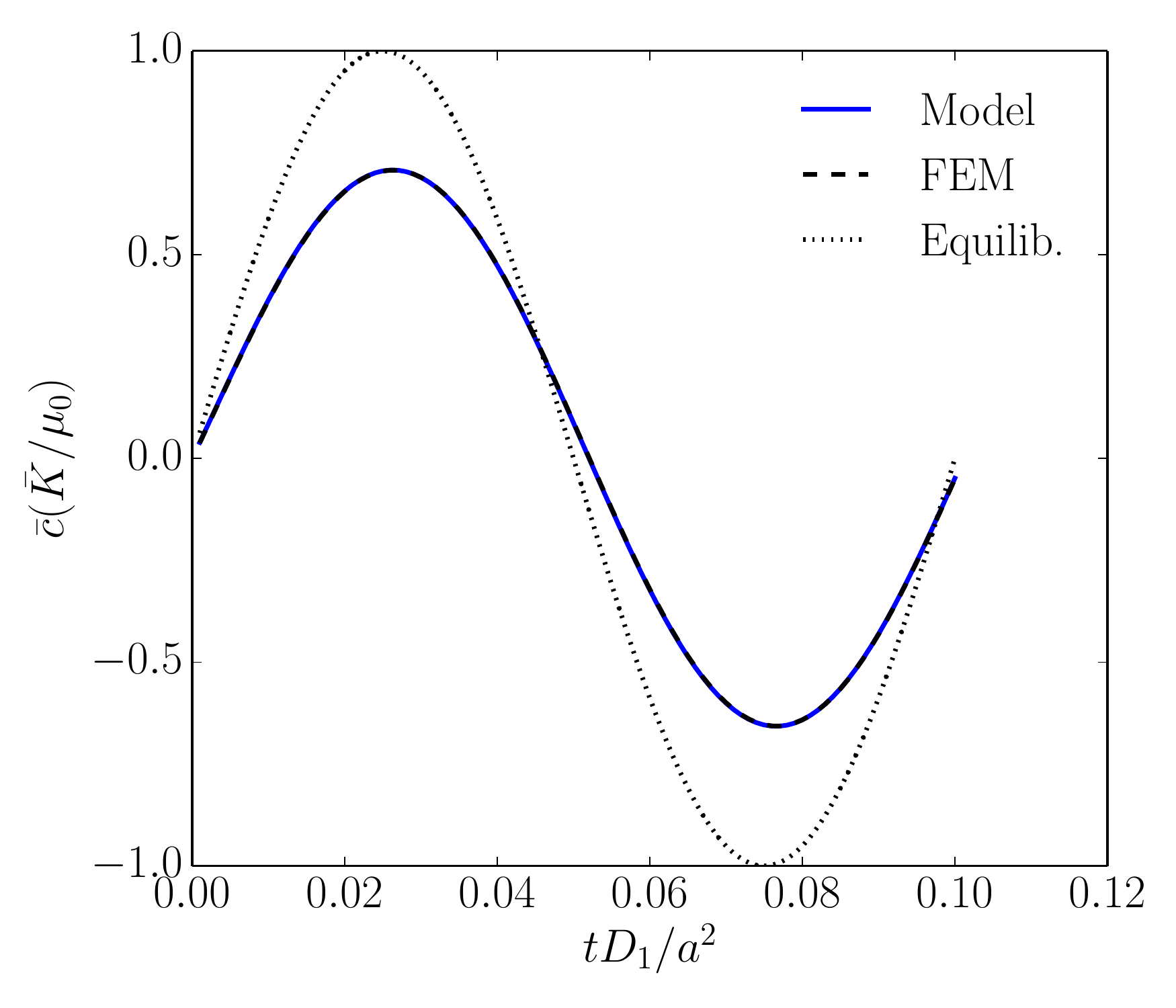}
\caption{}
\end{subfigure}
\begin{subfigure}[b]{0.4\textwidth}
\includegraphics[width=\textwidth]{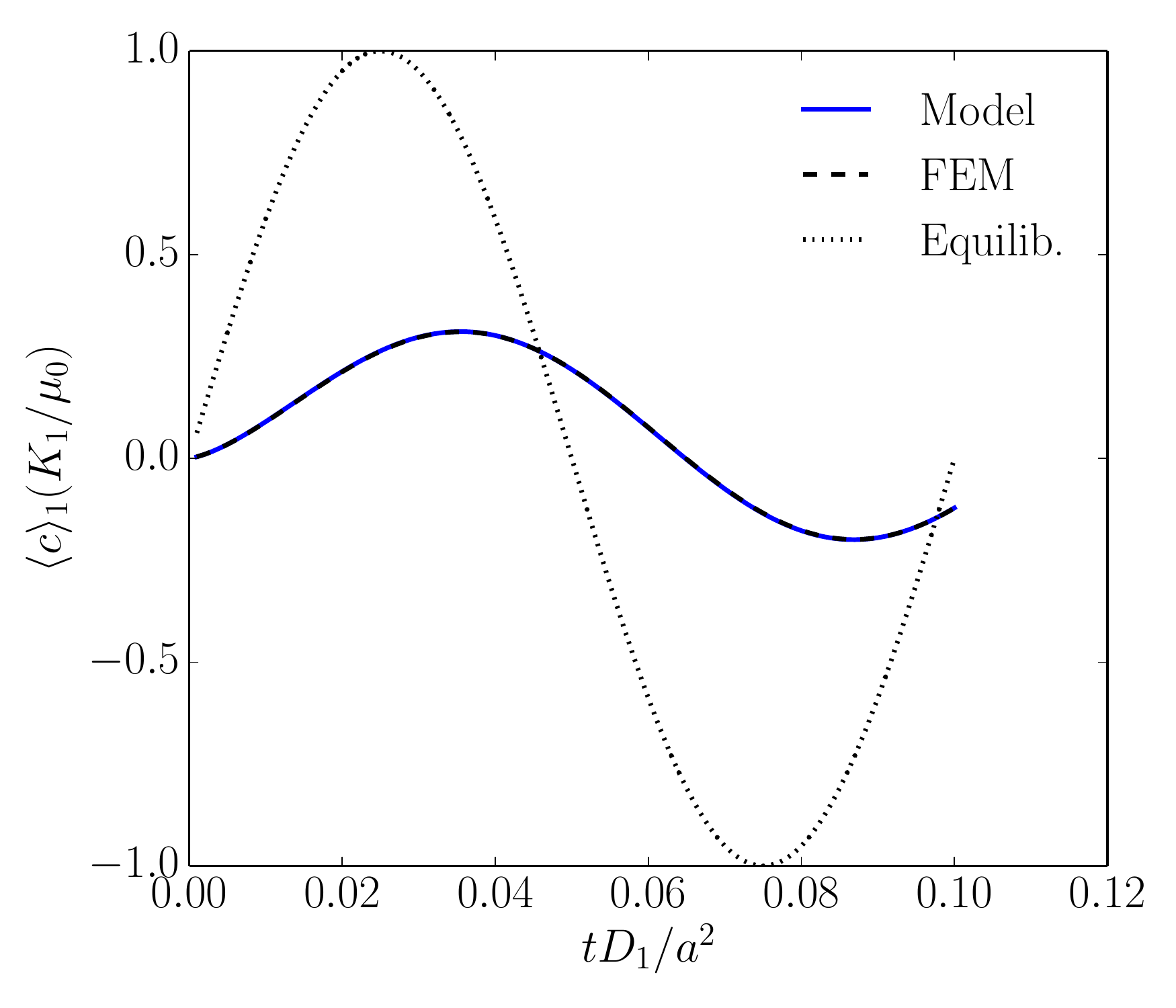}
\caption{}
\end{subfigure}
\begin{subfigure}[b]{0.4\textwidth}
\includegraphics[width=\textwidth]{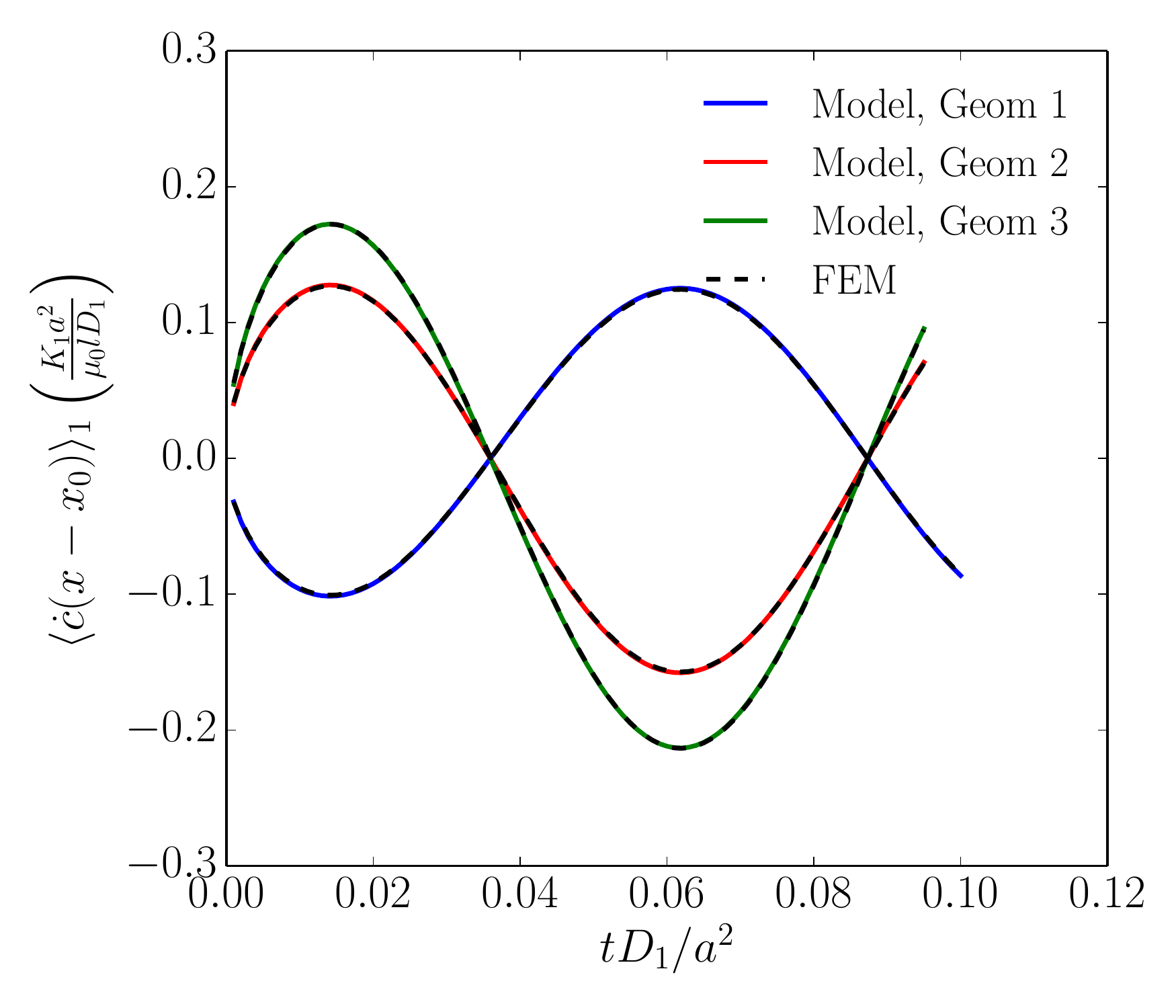}
\caption{}
\end{subfigure}
\begin{subfigure}[b]{0.4\textwidth}
\includegraphics[width=\textwidth]{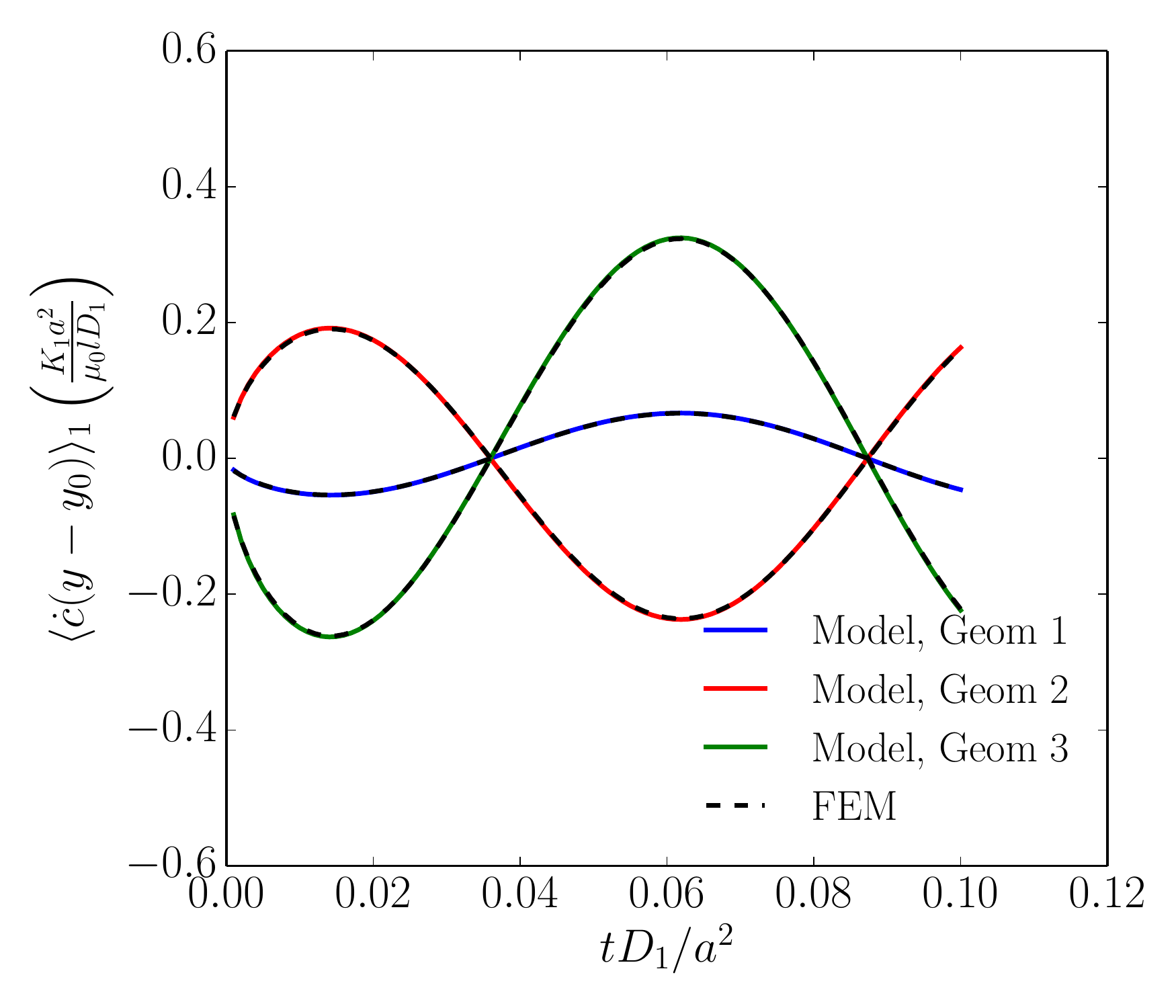}
\caption{}
\end{subfigure}
\caption{Effective behaviour corresponding to the loading conditions (\ref{loading1}). (a)-(b): Macroscopic concentration and inclusion average concentration in the first geometry. (c)-(d): Moments of the inclusion concentration rates for the three geometries represented in Fig. \ref{fig-geoms}.}
\label{fig-valid1}
\end{center}
\end{figure}

The average concentration response is shown in Fig. \ref{fig-valid1}(a)-(b) for the first geometry (Fig. \ref{fig-geoms}(a)). Identical results are found for the other two geometries. The concentration response is independent of the particular arrangement of inclusions, and is perfectly predicted by the mean-field model at both macro and phase levels, as expected. The actual transient response is markedly different from the response obtained from a the simple equilibrium model (\ref{ss}), which directly follows from the inclusion transient response. The first moments of the concentration rate in the x- and y-directions are shown in Fig. \ref{fig-valid1}(c)-(d) for the three geometries, also showing excellent agreement. In contrast to the average inclusion concentration, the first moment of the concentration does depend on the specific arrangement of the inclusions in the RVE, and induces a non-zero macroscopic flux. These contributions to the flux would vanish if the inclusion distribution was perfectly isotropic.  

\subsection{Macroscopically non-uniform chemical potential}\label{sec-nonunif-mu}
Next, we consider loading conditions that involve a combination of time-varying macroscopic chemical potential and macroscopic chemical potential gradient. As a first example, consider the following loading conditions: 
\begin{equation} \label{loading2}
\begin{array}{rcl}
\bar{\mu}(t) &=& \mu_0 \frac{t}{\tau_1} \\
\bar g_x(t) &=&  g_0 \sin(\omega t) \\
\bar g_y(t) &=& 0
\end{array}
\end{equation}
where $\omega = 2\pi/T$, $T=0.1 \tau_1$, and $\mu_0$ and $g_0$ are constants with $g_0=10 \mu_0/l$. 

\begin{figure}
\begin{center}
\begin{subfigure}[b]{0.45\textwidth}
\includegraphics[width=\textwidth]{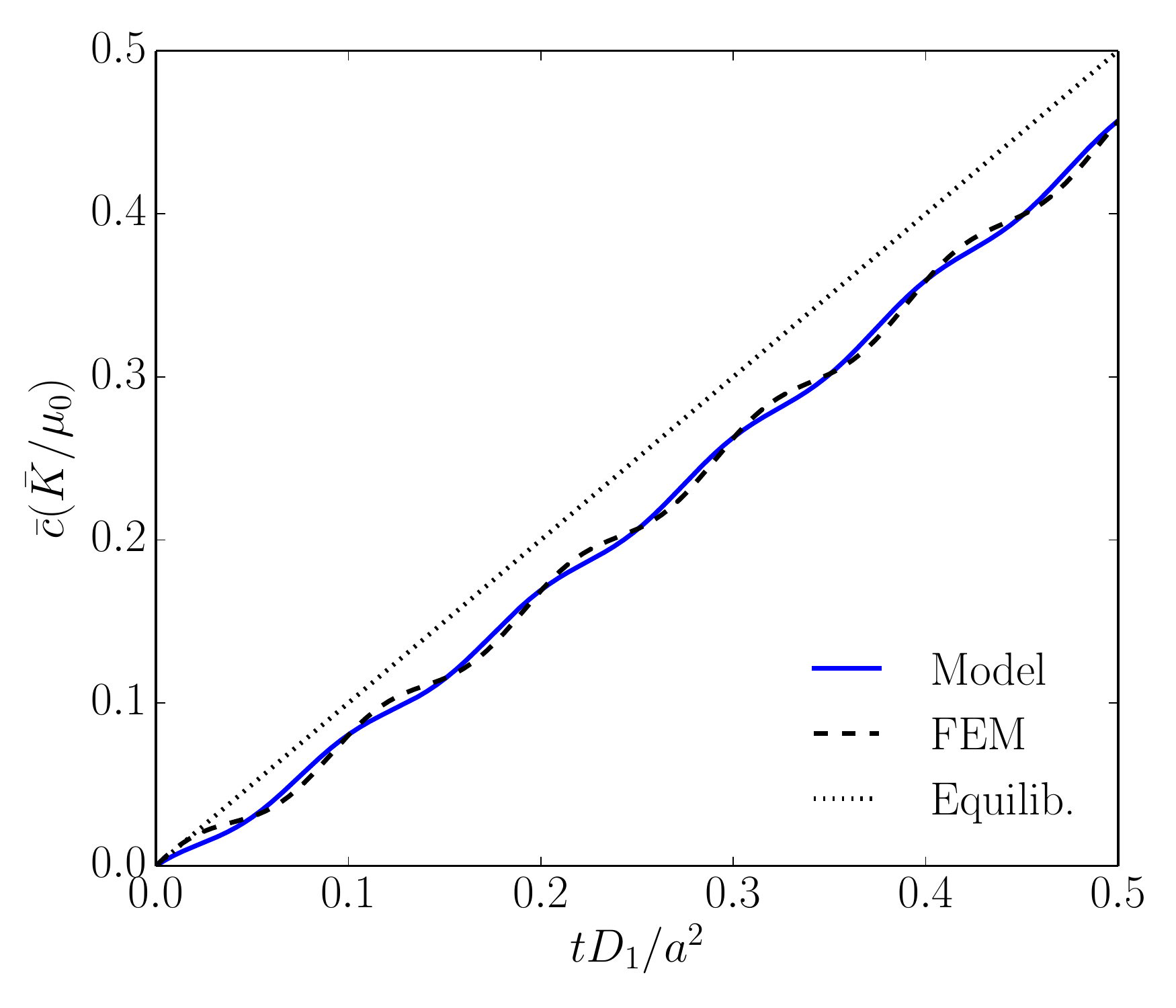}
\caption{}
\end{subfigure}
\begin{subfigure}[b]{0.45\textwidth}
\includegraphics[width=\textwidth]{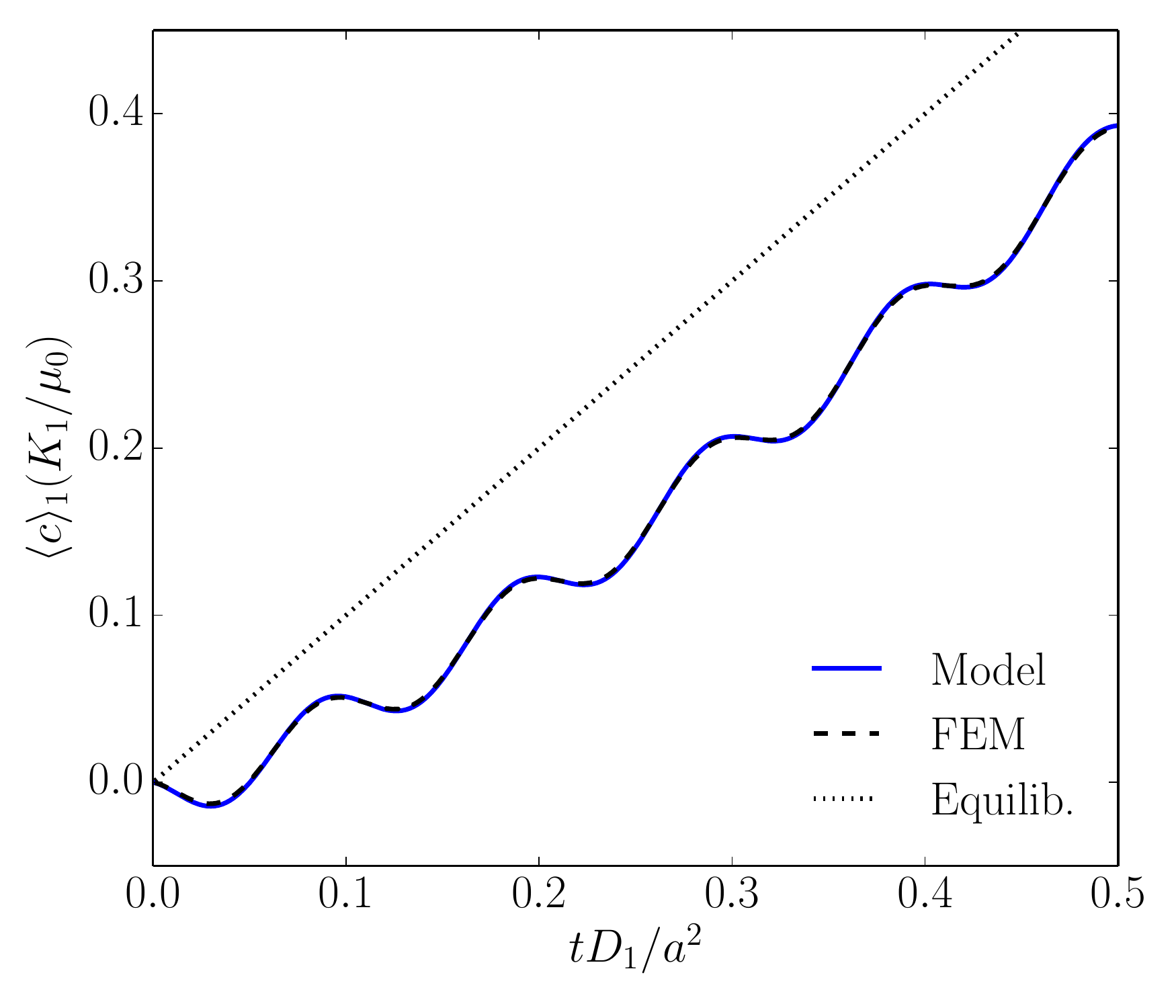}
\caption{}
\end{subfigure}
\begin{subfigure}[b]{0.45\textwidth}
\includegraphics[width=\textwidth]{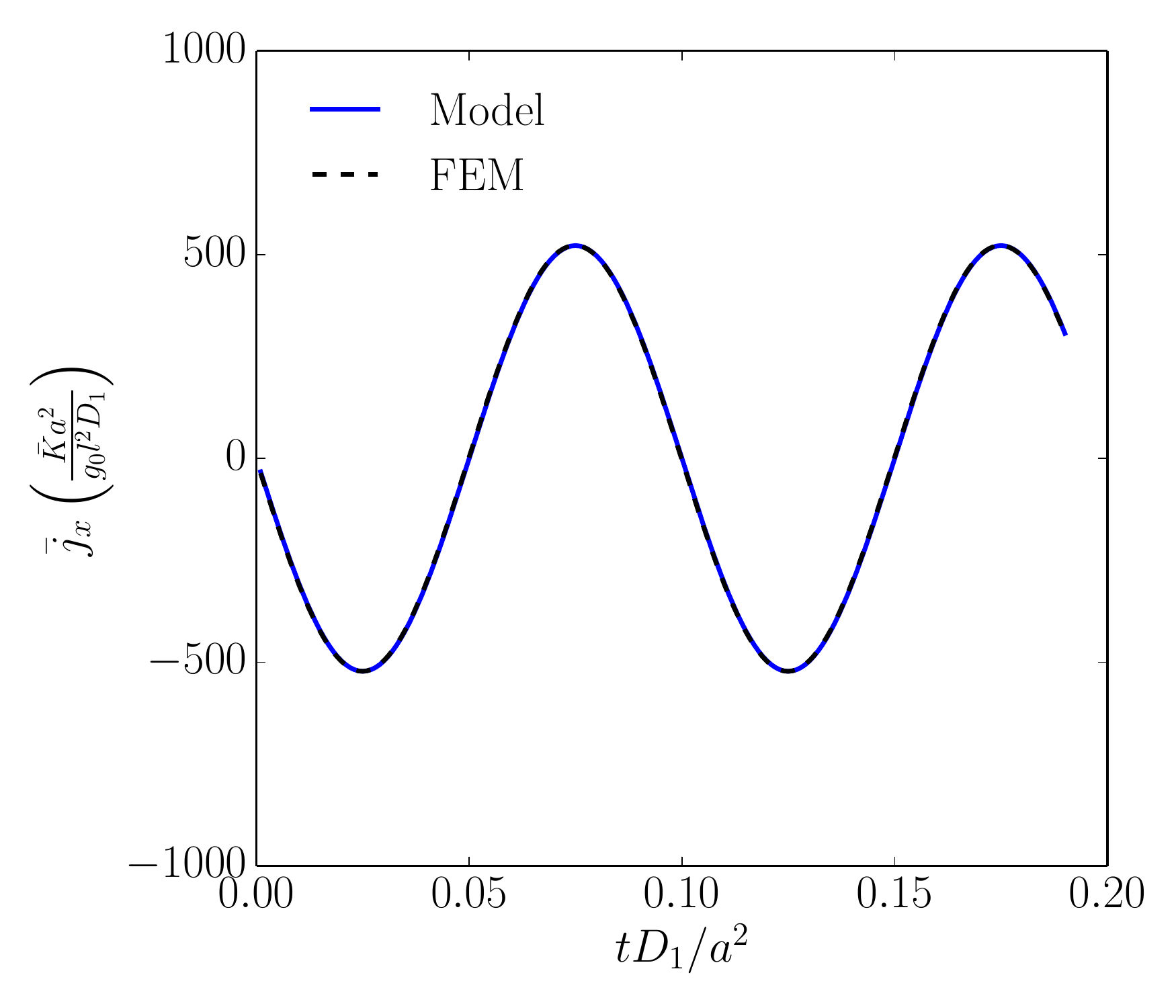}
\caption{}
\end{subfigure}
\begin{subfigure}[b]{0.45\textwidth}
\includegraphics[width=\textwidth]{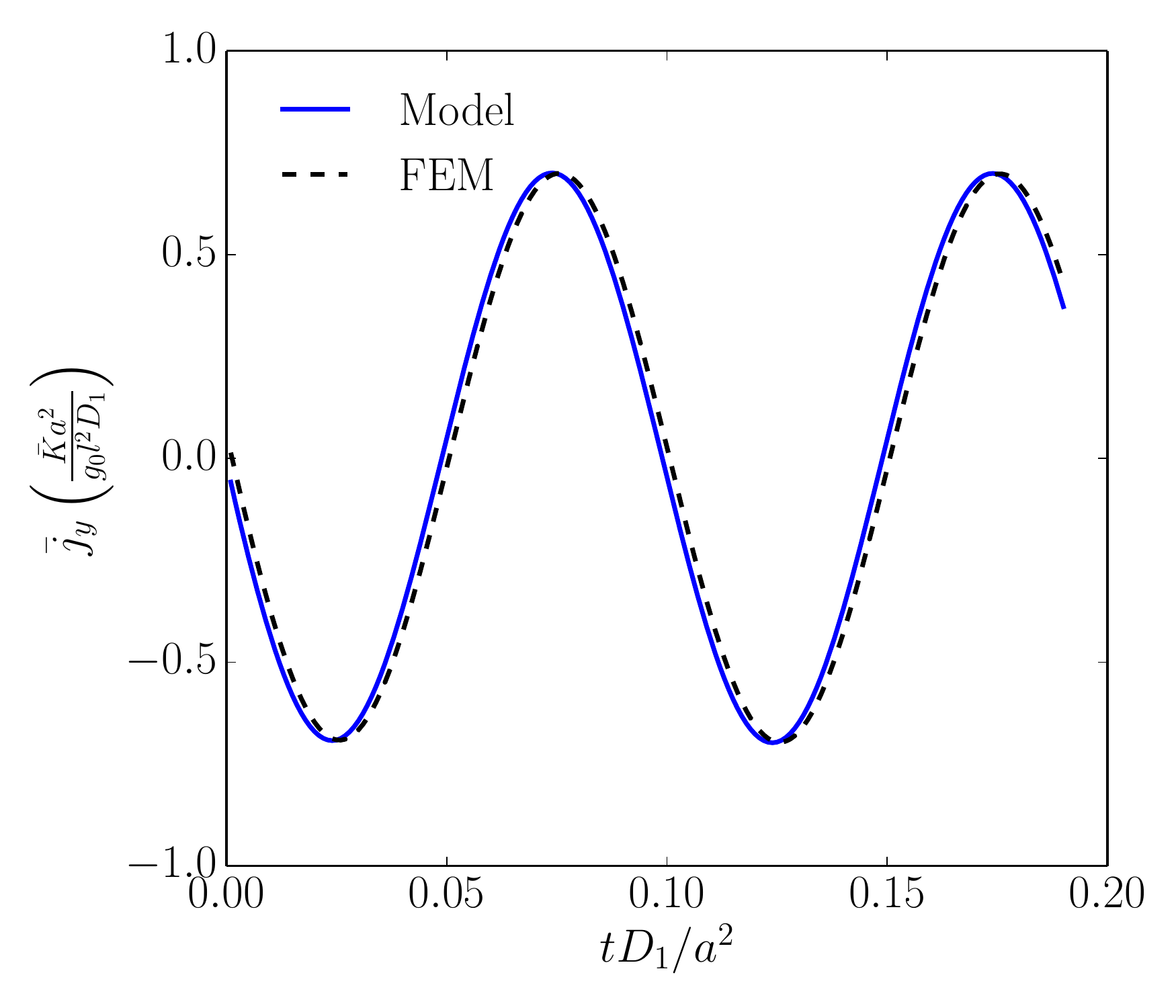}
\caption{}
\end{subfigure}
\begin{subfigure}[b]{0.45\textwidth}
\includegraphics[width=\textwidth]{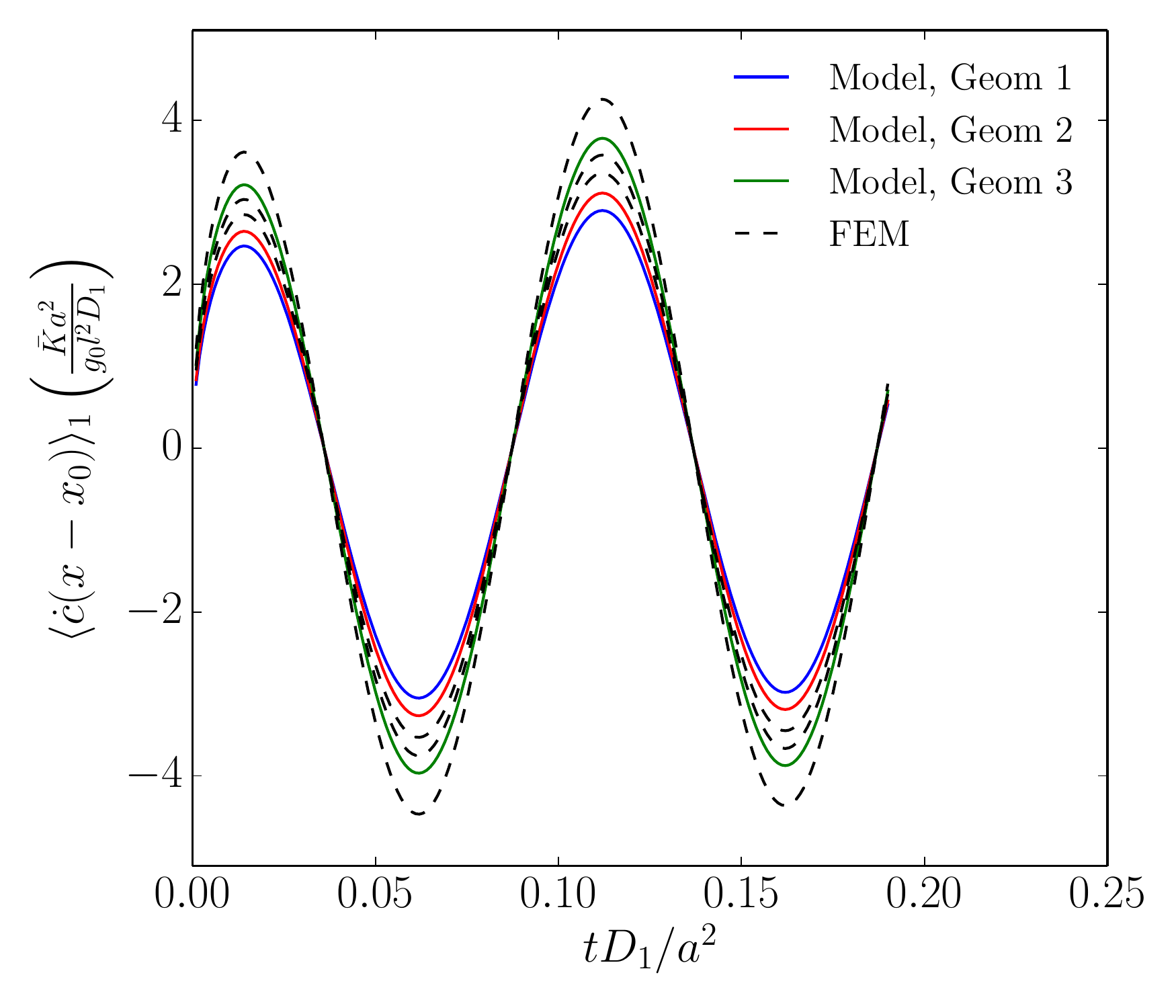}
\caption{}
\end{subfigure}
\begin{subfigure}[b]{0.45\textwidth}
\includegraphics[width=\textwidth]{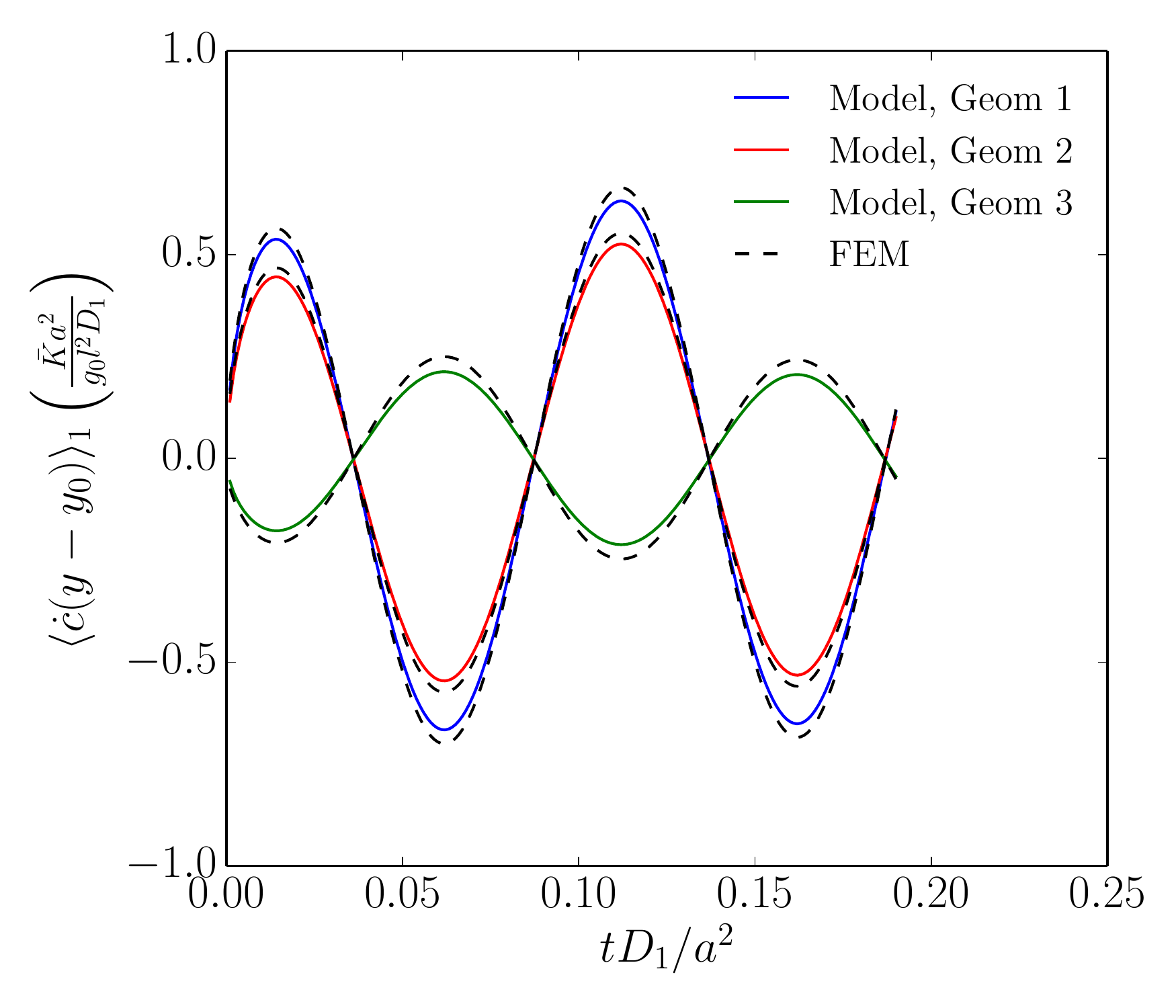}
\caption{}
\end{subfigure}
\caption{Effective behaviour corresponding to the loading conditions (\ref{loading2}). (a) Effective concentration, (b) inclusion average concentration, and (c)-(d) macroscopic flux components for the first geometry. (e)-(f) Moments of the inclusion concentration rates for the three considered geometries.}
\label{fig-valid2}
\end{center}
\end{figure}

Macroscopic and average inclusion concentration responses for the first geometry are shown in Figs \ref{fig-valid2}(a)-(b). A similar degree of accuracy was found for the other two geometries. The mean-field model provides a very good estimate of the inclusion average concentration (Fig. \ref{fig-valid2}(b)). The prediction of the macroscopic concentration is less accurate, which is due to a less accurate prediction of the matrix average concentration in the presence of a macroscopic chemical potential gradient, Eq. (\ref{c_matrix}). The oscillations in the predicted concentration response at both inclusion and macroscopic levels result from anisotropy through the vectors $\bm s_1$ and $\bm s_2$ in Eqs (\ref{c_matrix}) and (\ref{c_allinclusions2})-(\ref{hat_mu}), respectively. In comparison, the simple equilibrium model (\ref{ss}) does not capture the effect of anisotropy, and significantly overestimates the concentration at both inclusion and macroscopic levels.  

Macroscopic flux components are shown in Figs \ref{fig-valid2}(c)-(d) for the first geometry. Both components are very well predicted by the mean-field model, which means that the approximation of non-conducting inclusions in a steady-state matrix is realistic in order to predict the macroscopic flux. The non-zero flux component in the y-direction results from the slight anisotropy in the inclusion distribution, and is much smaller in magnitude than the flux in the x-direction. In the mean-field model, anisotropy is accounted for through the anisotropic effective conductivity tensor (here computed numerically) and through the moment of the rate of concentration in the inclusion, see below. 

Moments of concentration rate in the inclusions are represented in Figs \ref{fig-valid2}(e)-(f). The agreement between the mean-field model and the FE results is reasonably good. In particular, the mean-field model captures the anisotropic response. A comparison of Figs \ref{fig-valid2}(c) and (e) shows that the transient inclusion contribution to the macroscopic flux in the x-direction is negligible. Comparing Figs \ref{fig-valid2}(d) and (f), it appears that the inclusion contribution to the macroscopic flux becomes more significant. However, the inclusion contribution is weighted by the inclusion volume fraction in Eq. (\ref{macro_flux_integral}) and therefore the net effect of the macroscopic flux remains small. This effect probably explains the slight discrepancy between FE and mean-field results in Fig. \ref{fig-valid2}(d).

As a second example, consider loading conditions in which the macroscopic chemical potential ramps up before reaching a plateau, from where a rotating chemical potential gradient of unit norm is superimposed:
\begin{equation}\label{loading3-1}
\bar{\mu}(t) = \left\{ \begin{array}{ll}
\mu_0 \frac{5t}{\tau_1} & \textnormal{if} \ t\leq 0.2\tau_1 \\
1 & \textnormal{otherwise}
\end{array} \right.
\end{equation}
\begin{equation}\label{loading3-2}
\bar g_x(t) = \left\{ \begin{array}{ll}
0 & \textnormal{if} \ t\leq 0.2\tau_1 \\
g_0 \sin(\omega (t-0.2\tau_1)) & \textnormal{otherwise}
\end{array} \right.
\end{equation}
\begin{equation}\label{loading3-3}
\bar g_y(t) = \left\{ \begin{array}{ll}
0 & \textnormal{if} \ t\leq 0.225\tau_1 \\
g_0 \sin(\omega (t-0.225\tau_1)) & \textnormal{otherwise}
\end{array} \right.
\end{equation}
where $\omega = 2\pi/T$, $T=0.1 \tau_1$, and $\mu_0$ and $g_0$ are constants such that $g_0=\mu_0/l$. 

\begin{figure} 
\begin{center}
\begin{subfigure}[b]{0.45\textwidth}
\includegraphics[width=\textwidth]{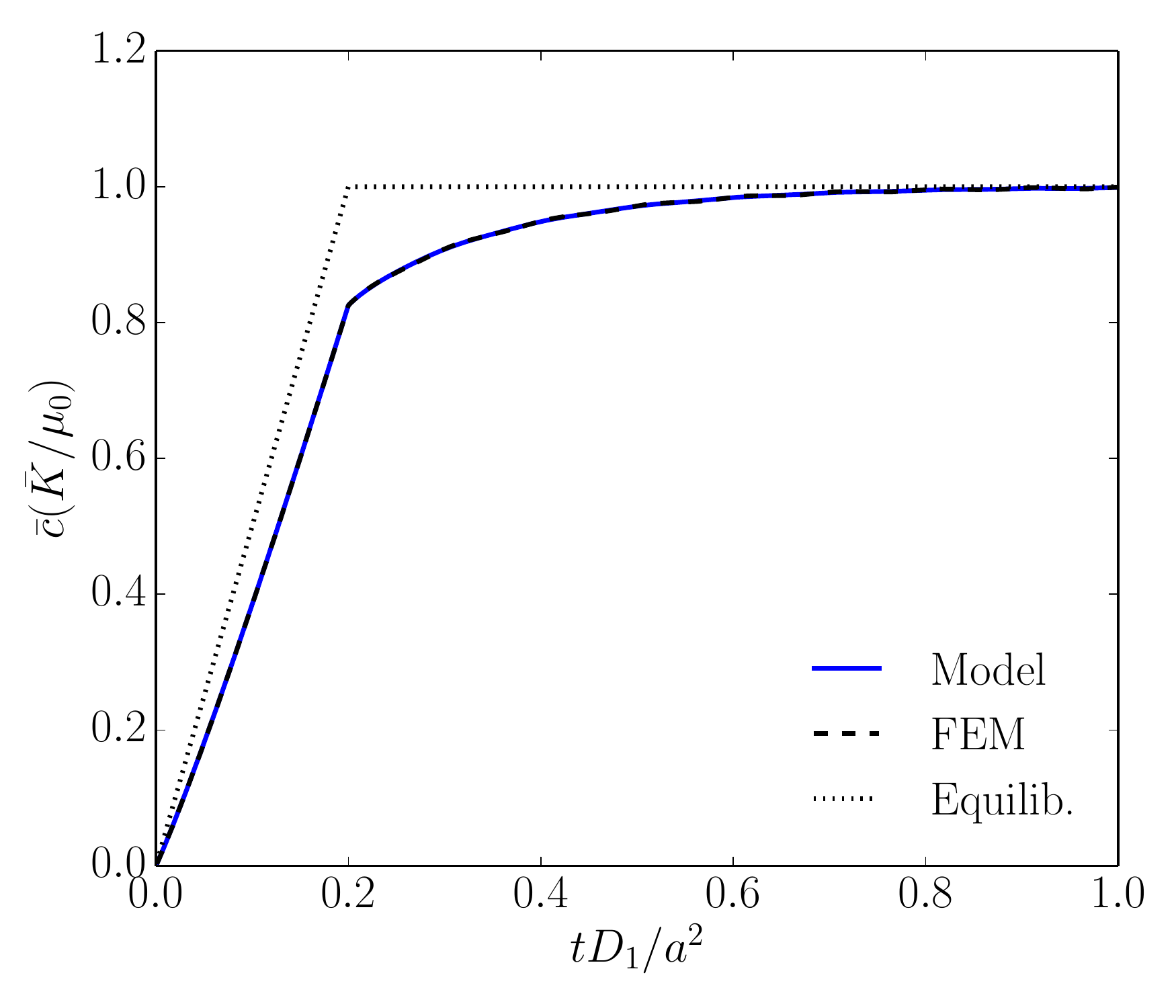}
\caption{}
\end{subfigure}
\begin{subfigure}[b]{0.45\textwidth}
\includegraphics[width=\textwidth]{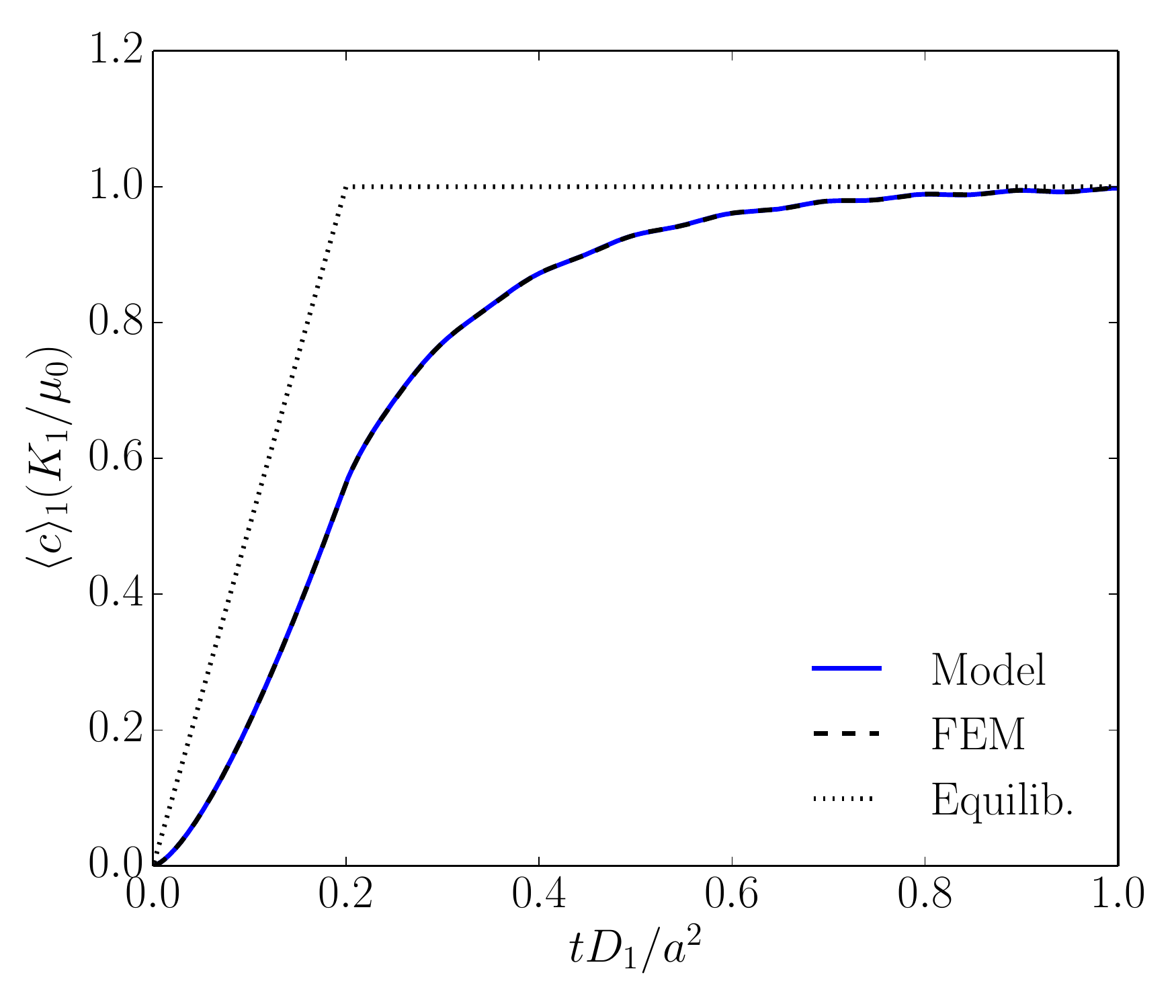}
\caption{}
\end{subfigure}
\begin{subfigure}[b]{0.45\textwidth}
\includegraphics[width=\textwidth]{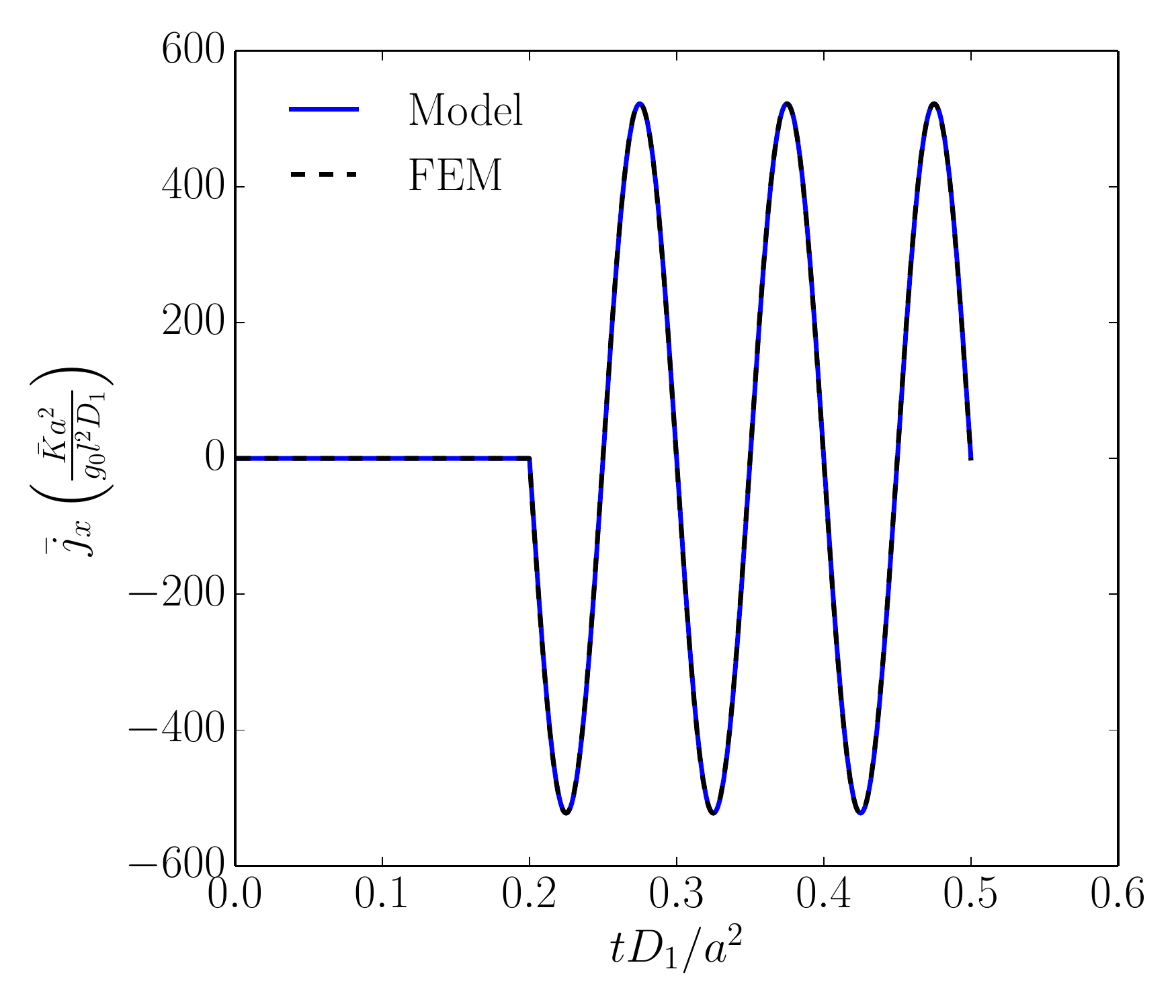}
\caption{}
\end{subfigure}
\begin{subfigure}[b]{0.45\textwidth}
\includegraphics[width=\textwidth]{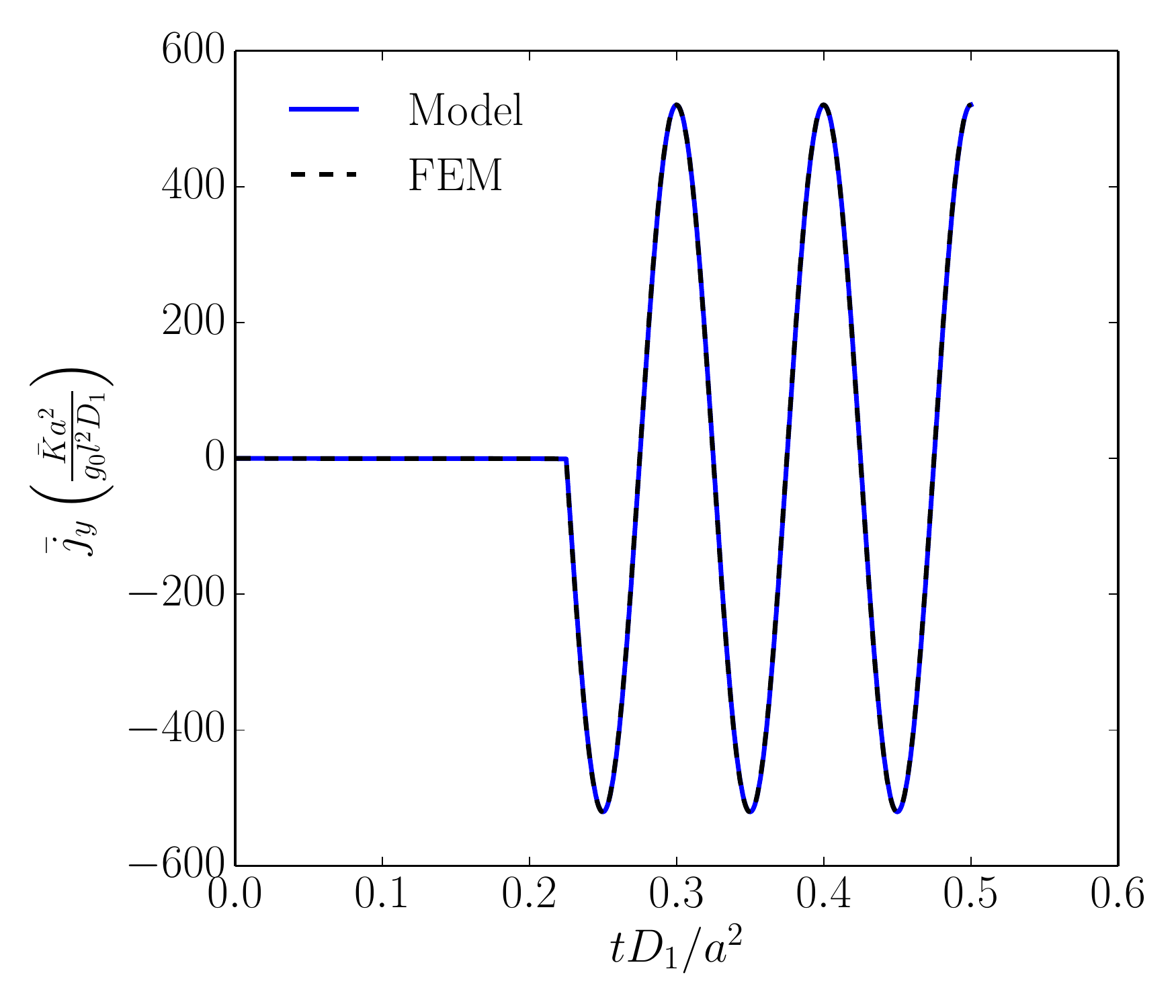}
\caption{}
\end{subfigure}
\begin{subfigure}[b]{0.45\textwidth}
\includegraphics[width=\textwidth]{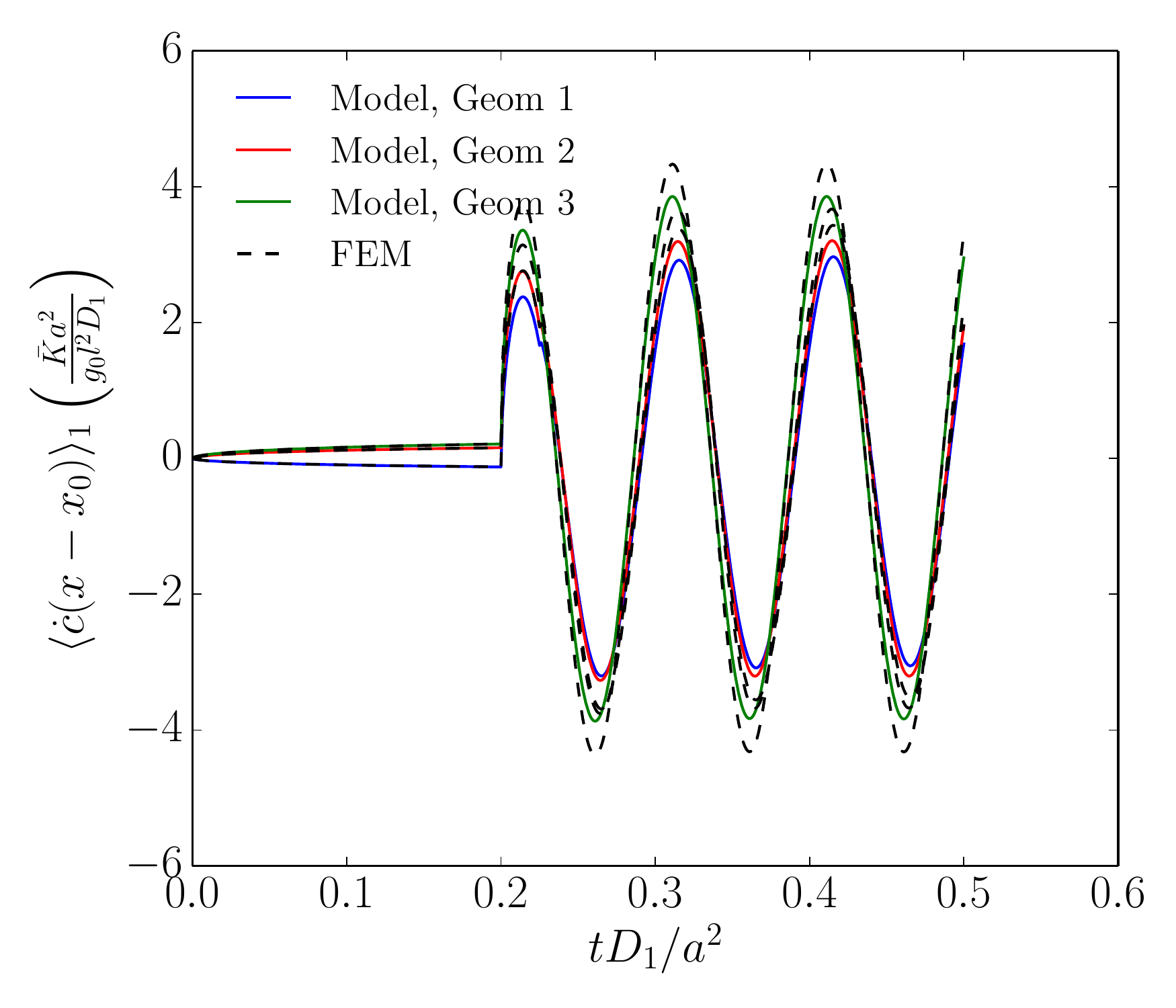}
\caption{}
\end{subfigure}
\begin{subfigure}[b]{0.45\textwidth}
\includegraphics[width=\textwidth]{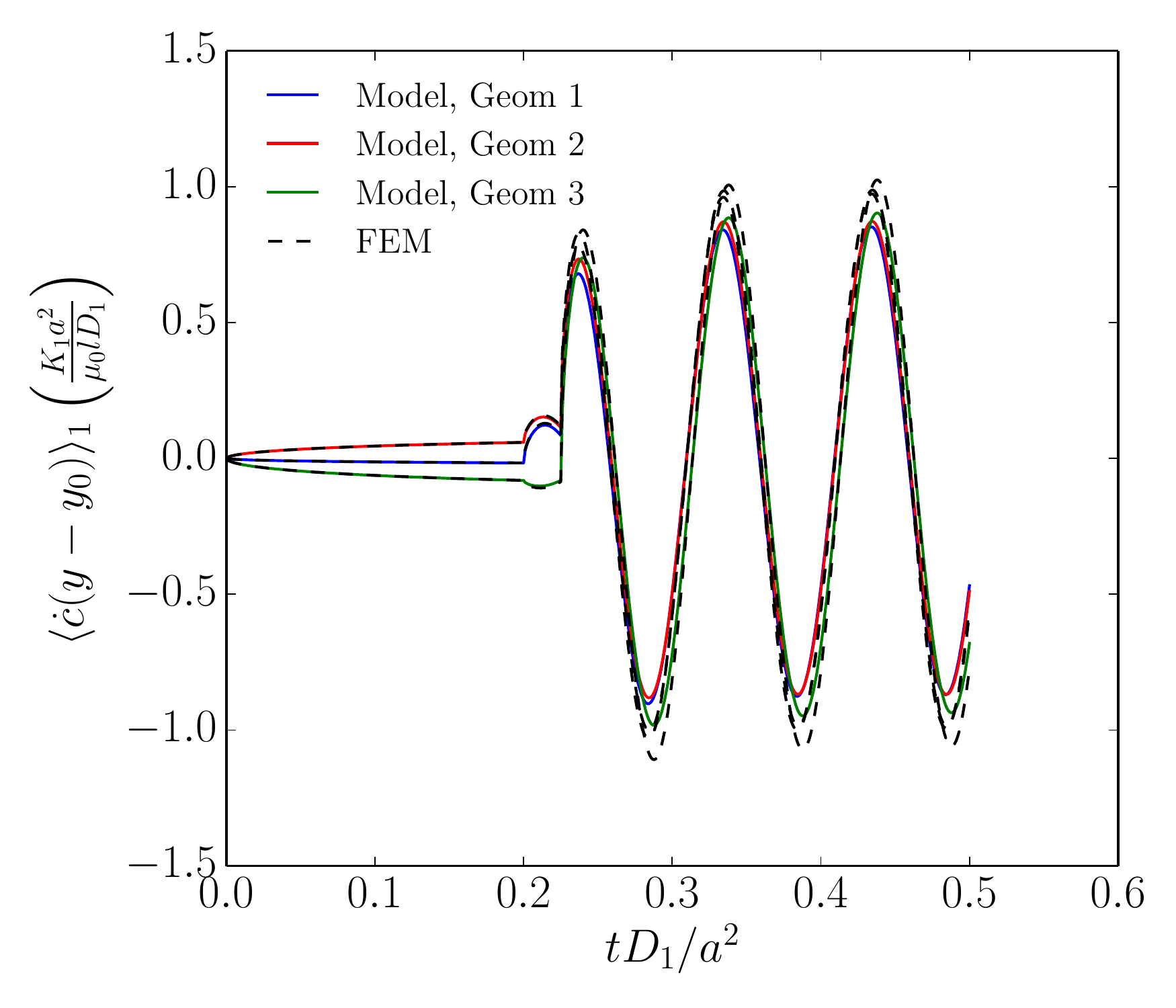}
\caption{}
\end{subfigure}
\caption{Effective behaviour corresponding to the loading conditions (\ref{loading3-1})-(\ref{loading3-3}). (a) Effective concentration, (b) inclusion average concentration, and (c)-(d) macroscopic flux components for the first geometry. (e)-(f) Moments of the inclusion concentration rates for the three considered geometries.}
\label{fig-valid3}
\end{center}
\end{figure}

Predictions of the mean-field model are compared to the reference FE results in Fig. \ref{fig-valid3}. Predictions of the concentration response on Figs (\ref{fig-valid3})(a)-(b) are very good, both at RVE and inclusion level. Both components of the macroscopic flux, Figs (\ref{fig-valid3})(c)-(d), are also very well predicted by the mean-field model. The components of the moment of concentration rate in the inclusion are represented in Figs (\ref{fig-valid3})(e)-(f), and are also well predicted by the model. Like in the previous examples, the effect of the anisotropy in the distribution of inclusions on the moment of the concentration is apparent. However, this term has negligible impact on the macroscopic flux, which is dominated by diffusion through the steady-state matrix. 

\section{Two-scale simulations}\label{sec:macro-simu}
\subsection{Comparison to full-field results}
\begin{figure}
\begin{center}
\includegraphics[width=\textwidth]{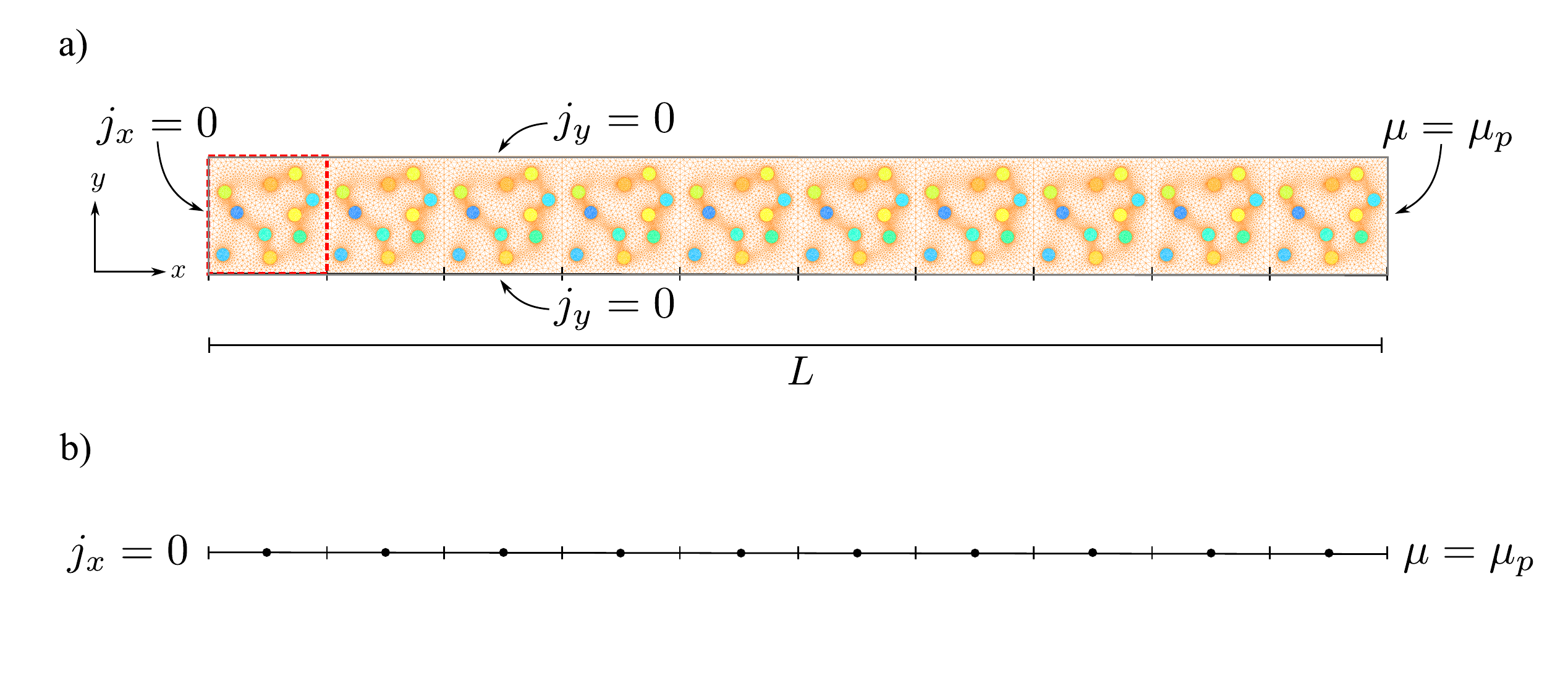}
\caption{a) Schematic of the diffusion boundary-value problem in a composite slab of length $L$ and Finite Element mesh. Details of the microstructure are fully resolved. (b) Two-scale simulation of the same problem using an effective medium. At macroscale, a 1D Finite Element model with first-order elements is used, with element size corresponding to the RVE size. The constitutive behaviour in each element is obtained from the mean-field model.}
\label{fig-macroFEM}
\end{center}
\end{figure}

We now use the mean-field model developed and validated in the three previous sections to solve 2D boundary-value problems at macroscopic scale. We consider the transient diffusion problem through a slab of material of length $L$ in the $x$-direction, as represented in Fig \ref{fig-macroFEM}(a). The microstructure was generated by juxtaposing 10 identical unit cells of size $l$ with random distributions of 10 inclusions in the $x$-directions, so that $L=10l$. The volume fraction of inclusion is $f=0.1$. The material properties are such that $K_2/K_1=6$ and $k_2/k_1=10^5$. The slab is subject to the following boundary conditions:
\begin{equation}
\mu(L,y,t) = \mu_p(t), \quad j_x (0,y,t) = j_y(x,0,t) = j_y(x,L,t) = 0. 
\end{equation}        
The last three boundary conditions correspond to symmetry planes. Full-field solution to this problem is obtained using FEM on a mesh of the fully-resolved microstructure. Effective concentrations and fluxes over each unit cell are then calculated from their definition (\ref{macro_c}) and (\ref{macro_j2}), where volume averages are numerically calculated by carrying out weighted averages over the integration points. 

The same macroscale boundary-value problem is also solved using a two-scale approach. At macroscale, the actual heterogeneous medium is replaced by an equivalent homogeneous medium with behaviour described by the mean-field model, Eqs (\ref{macro_c_integral})-(\ref{macro_flux_integral}). Here, for simplicity, we assumed an isotropic response, that is, we set $\bm s_1=\bm 0$ and $\bm S_1 = (1/12)\bm 1$ in the mean-field model. The effective isotropic conductivity $\bar k$ was estimated using the Hashin-Shtrikman upper bound in 2D, Eq. (\ref{HS_2D}). The problem can then be solved in 1D (Fig. \ref{fig-macroFEM}(b)). 

The 1D diffusion problem is solved numerically using FEM. As is standard, the weak form of the diffusion problem is obtained by multiplying the species conservation equation (\ref{macro_mass_cons}) by a test function $w(x)$ that vanishes on $\partial \Omega_{\mu}$ (here, $x=L$), integrating along the slab, using integration by parts and the boundary condition $j_x(0,t)=0$, to finally obtain:
\begin{equation}
\int_0^L \left( \dot{\bar c} w - \bar j_x \frac{dw}{dx} \right) dx = 0.
\end{equation} 
The weak form is discretised in space using a finite element interpolation of the unknown field $\bar{\mu}$:
\begin{equation}\label{fe_interp}
\bar{\mu} = \sum_{I=1}^{N_n} \bar{\mu}^I \xi^I,
\end{equation}
where the superscript $I=1,\dots,N_n$ refers to the node number, $\bar{\mu}^I$ are the nodal values and $\xi^I$ the shape functions. We adopt a standard Galerkin formulation and use the same shape functions to interpolate the test function $w$. Here, linear shape functions were used. We obtain $N_n$ equations for the $N_n$ nodal values: 
\begin{equation}
\int_0^L \left( \dot{\bar c} \xi^I - \bar j_x \frac{d\xi^I}{dx} \right) dx = 0 \quad\quad (I=1,\ldots,N_n).
\end{equation} 
The integral is evaluated numerically using a simple quadrature method with one integration point. Integration in time is performed using a fully-implicit Euler scheme. At every integration point and time step $t^{(n+1)}$, the macroscopic chemical potential $\bar{\mu}^{(n+1)}$ and $\bar g_x^{(n+1)}$ are calculated from the interpolation (\ref{fe_interp}), and updates $\bar c^{(n+1)}$ and $\bar j^{(n+1)}_x$ (as well as updates of internal variables) are calculated using the scheme outlined in Section \ref{sec-update}. The slab was discretised using 10 elements of size $l$, assuming that each element corresponds to one unit cell in the full-field model (Fig. \ref{fig-macroFEM}(b)). The comparison between the two approaches is presented in the following. 

\begin{figure} 
\begin{center}
\begin{subfigure}[b]{0.5\textwidth}
\includegraphics[width=\textwidth]{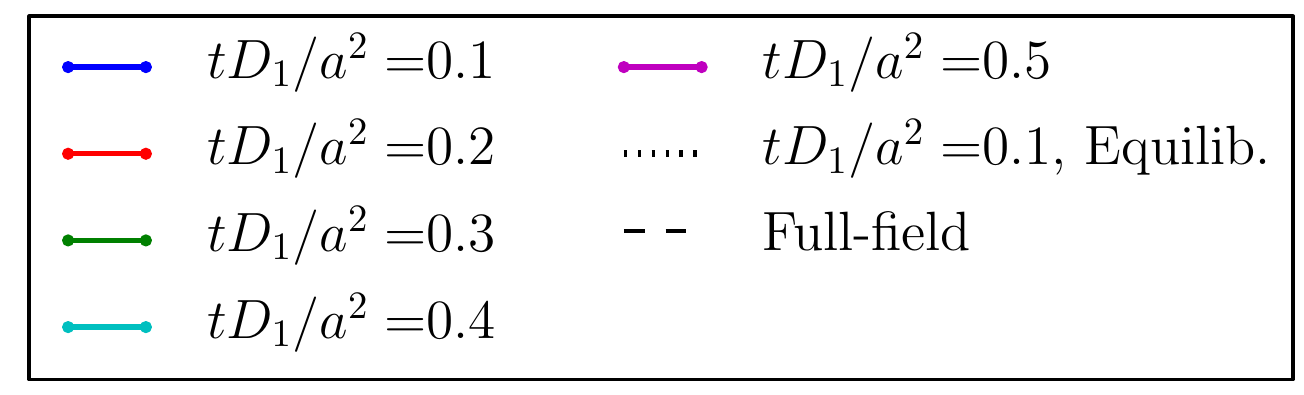}
\end{subfigure}\\
\begin{subfigure}[b]{0.45\textwidth}
\includegraphics[width=\textwidth]{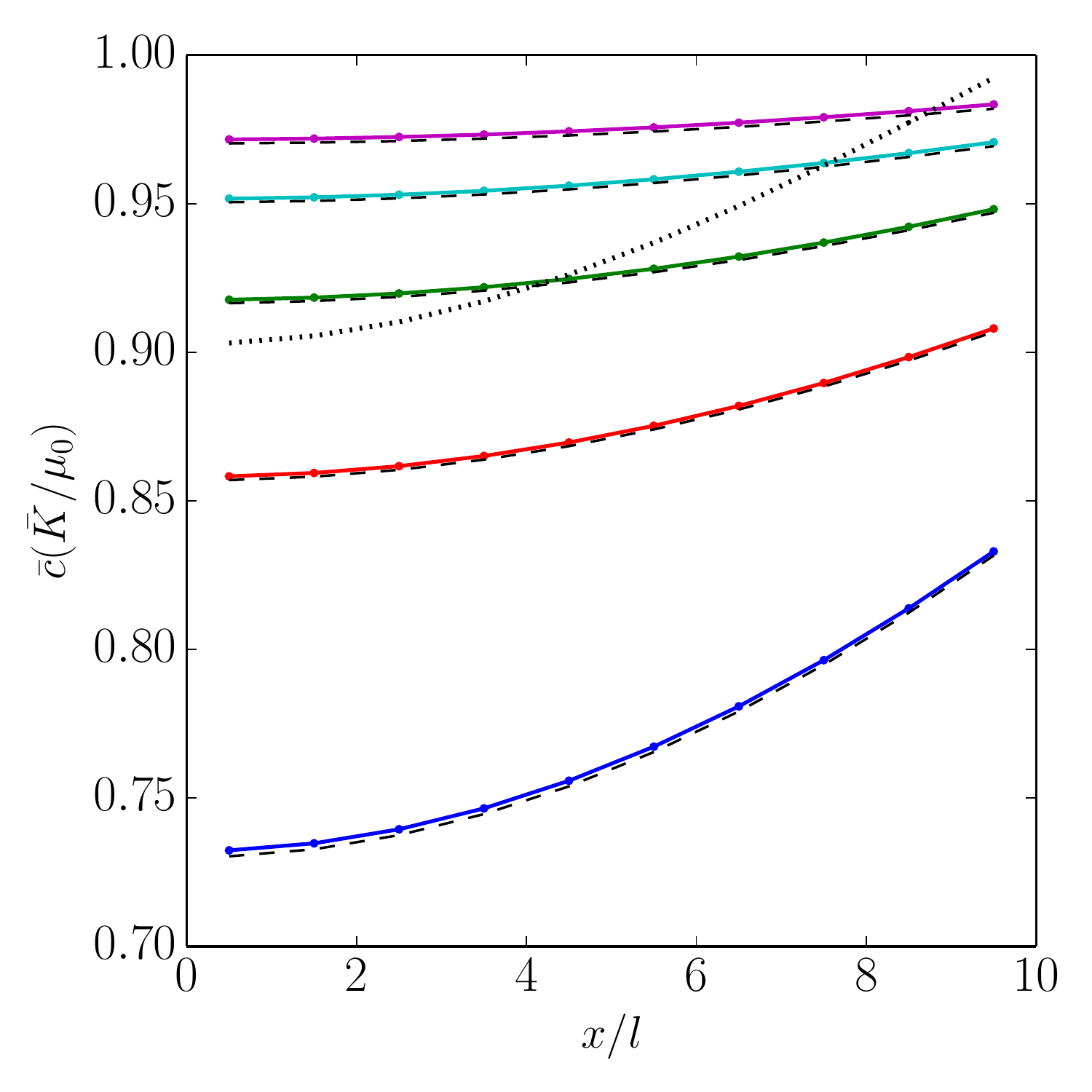}
\caption{}
\end{subfigure}
\begin{subfigure}[b]{0.45\textwidth}
\includegraphics[width=\textwidth]{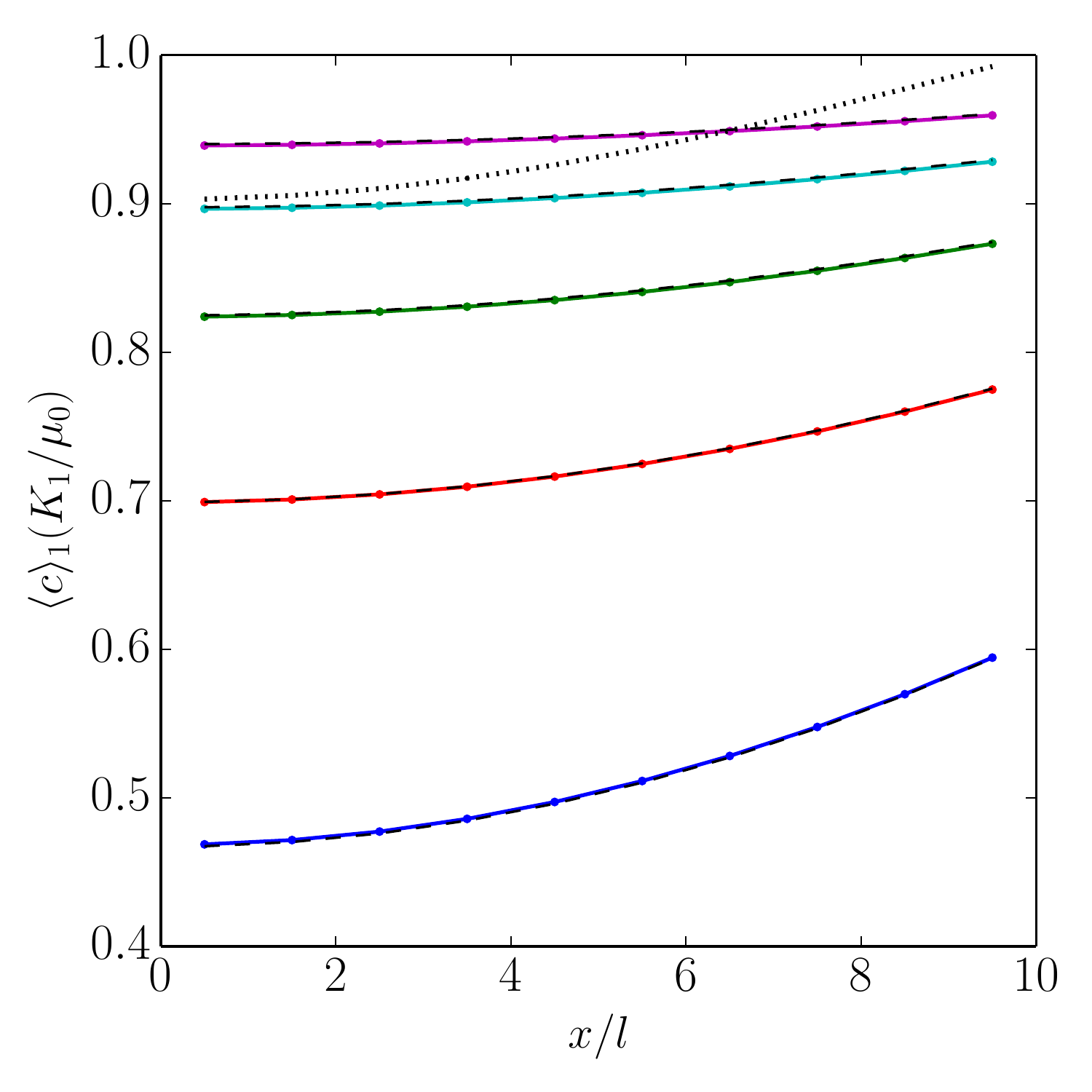}
\caption{}
\end{subfigure}

\caption{Profiles of macroscopic concentration (a) and average inclusion concentration (b) in a composite slab subject to a step load at $x=L$ and natural boundary conditions on the three other boundaries. Black lines are the 2D full-field results and color lines are the results obtained by combining a 1D finite element approach with the mean-field model.}
\label{fig-macro-step}
\end{center}
\end{figure}

As a first example, consider a step load of the following form: 
\begin{equation}\label{mup_step}
\mu_p(t) = \mu_0 H(t),   
\end{equation} 
with $\mu_0$ a constant. Concentration profiles obtained with the full-field and the two-scale approaches are shown in Fig. \ref{fig-macro-step}. Simulation times are normalised by the characteristic time for diffusion in the inclusion, $\tau_1$. The actual data points are located at the integration points, while the lines are guides for the eye.  The agreement is excellent, for both the macroscopic and the inclusion concentration response. The non-Fickian response of the composite is apparent from the concentration response in $x=L$, where the value of chemical potential is applied. Indeed, for Fickian behaviour, one would expect that the concentration is one-to-one related to the (constant) chemical potential. This is not the case here, as the concentration in $x=L$ slowly evolves in time to reach their equilibrium value. The prediction of the two-scale approach using Eq. (\ref{ss}) for the effective concentration is also shown. In that case, the macroscopic concentration at $x=L$ is at equilibrium, however the concentration profile in the slab does no agree with the full-field results. The discrepancy is due to the transient inclusion response, Fig. \ref{fig-macro-step}(b), and has an impact also at macroscale Fig. \ref{fig-macro-step}(a). 

\begin{figure} 
\begin{center}
\begin{subfigure}[b]{0.45\textwidth}
\includegraphics[width=\textwidth]{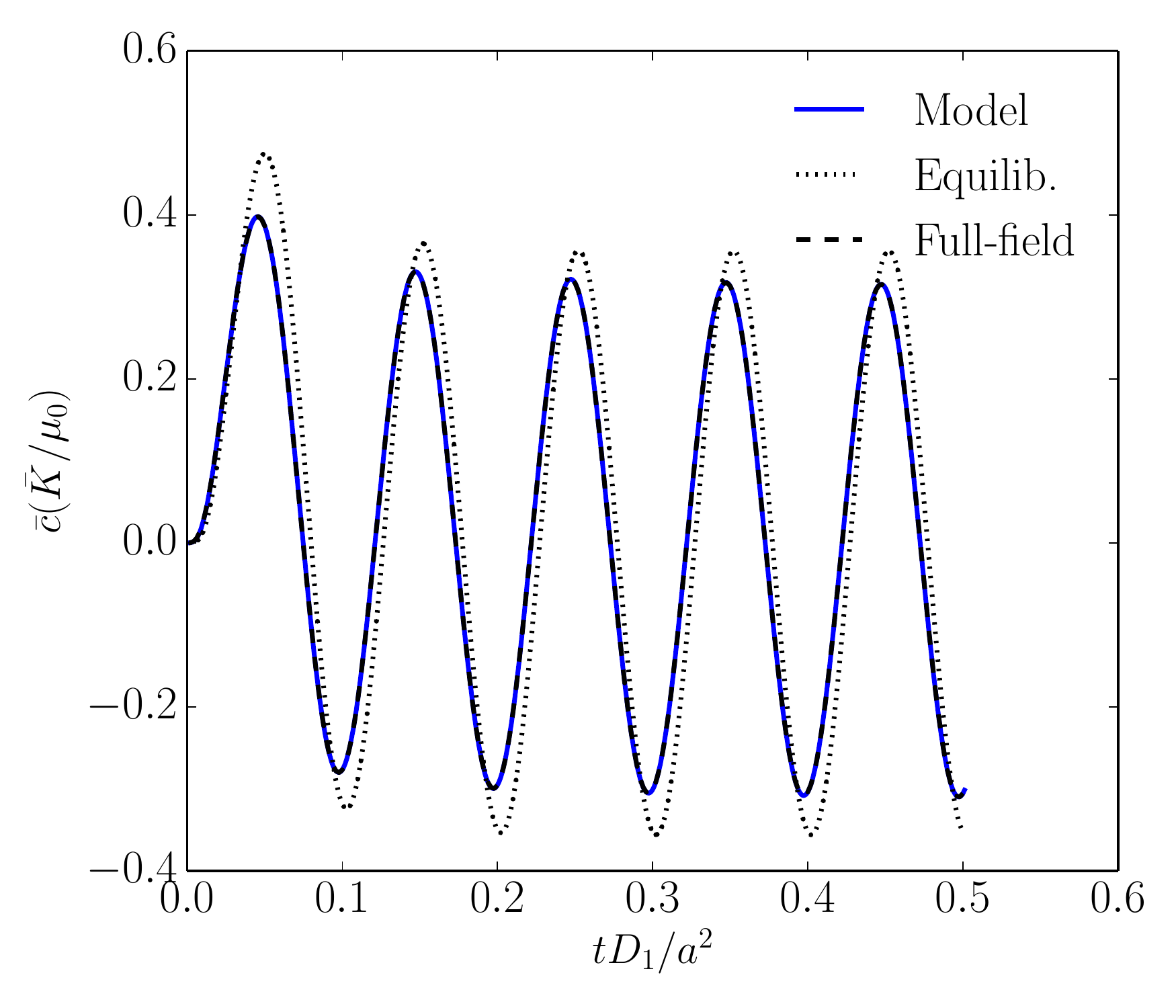}
\caption{}
\end{subfigure}
\begin{subfigure}[b]{0.45\textwidth}
\includegraphics[width=\textwidth]{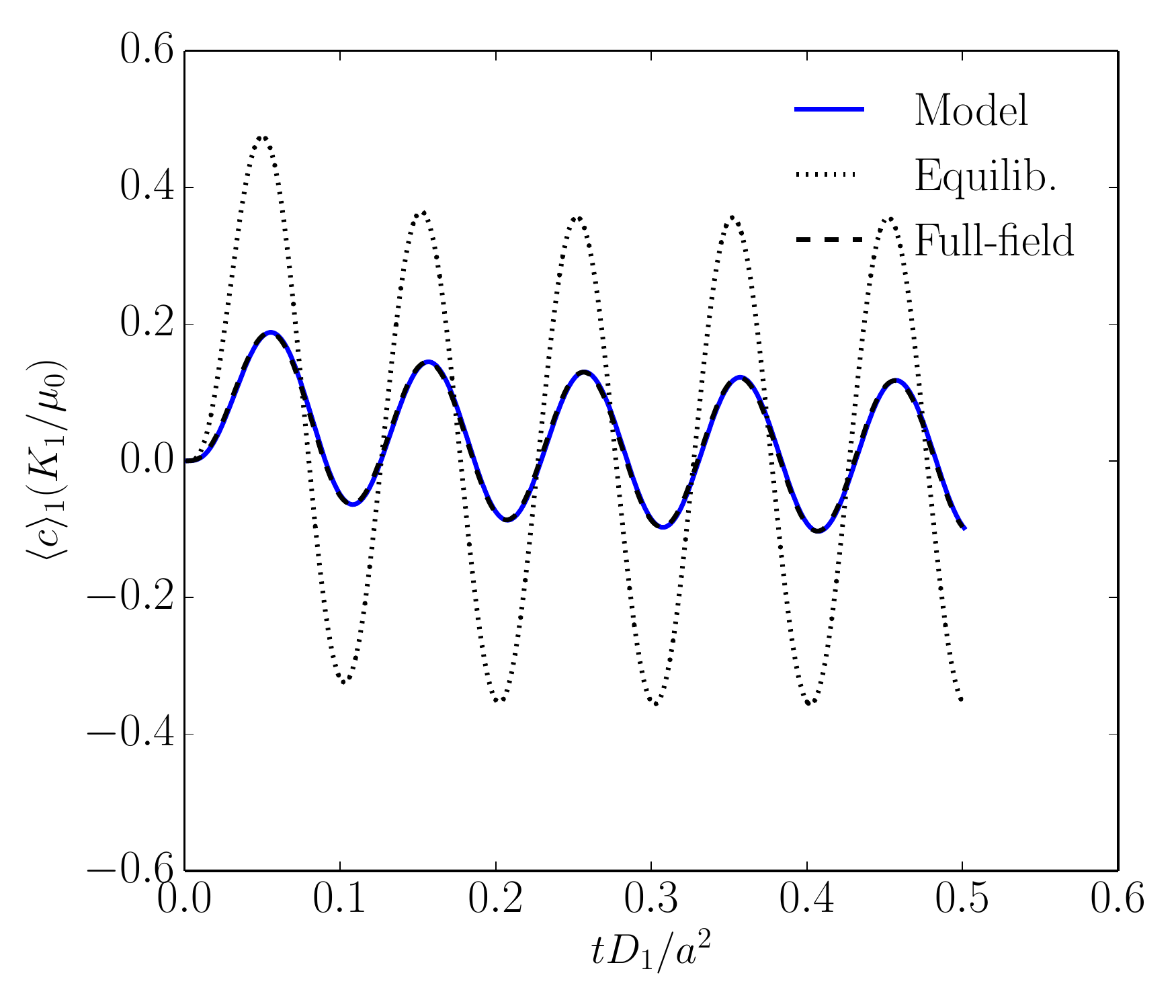}
\caption{}
\end{subfigure}
\caption{Time evolution of the (a) macroscopic and (b) average inclusion response at macroscopic coordinate $x/L=0.05$ in the slab subject to harmonic loading conditions.}
\label{fig-macro-harmonic}
\end{center}
\end{figure}

As a second example, consider the following harmonic loading conditions:  
\begin{equation}
\mu_b(L,t) = \mu_0 \sin(\omega t)  
\end{equation}
with $\omega = 2\pi/T$ and $\mu_0$ is a constant. The macroscopic and average inclusion response in the first unit cell with centre of volume at $x=0.05L$ are represented in Fig. \ref{fig-macro-harmonic}. Full-field predictions correspond to the volume average of the local fields in that unit cell, and mean-field results are obtained for the integration point located at $x=0.05L$. The agreement between full-field and mean-field results is remarkable.        

\begin{figure} 
\begin{center}
\begin{subfigure}[b]{0.45\textwidth}
\includegraphics[width=\textwidth]{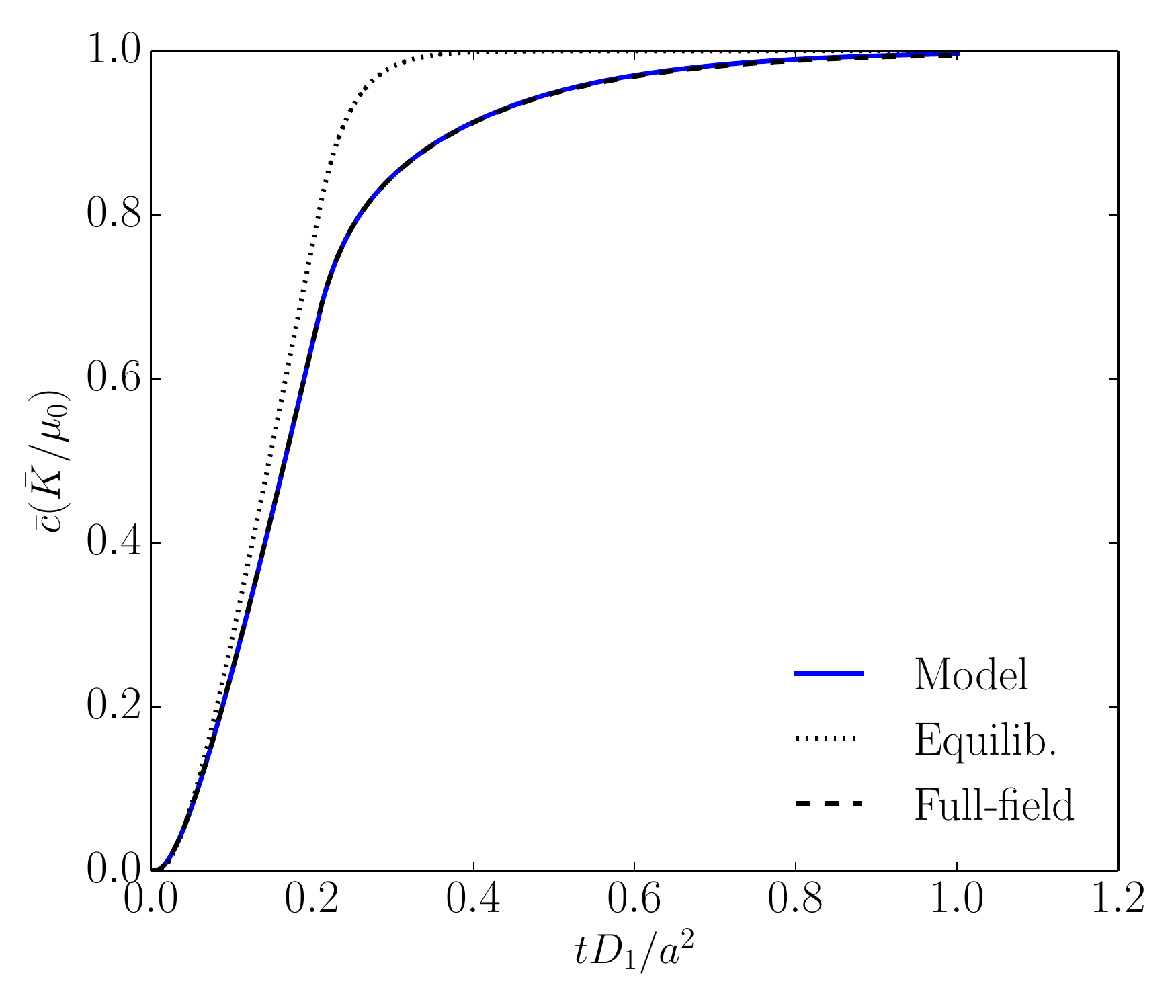}
\caption{}
\end{subfigure}
\begin{subfigure}[b]{0.45\textwidth}
\includegraphics[width=\textwidth]{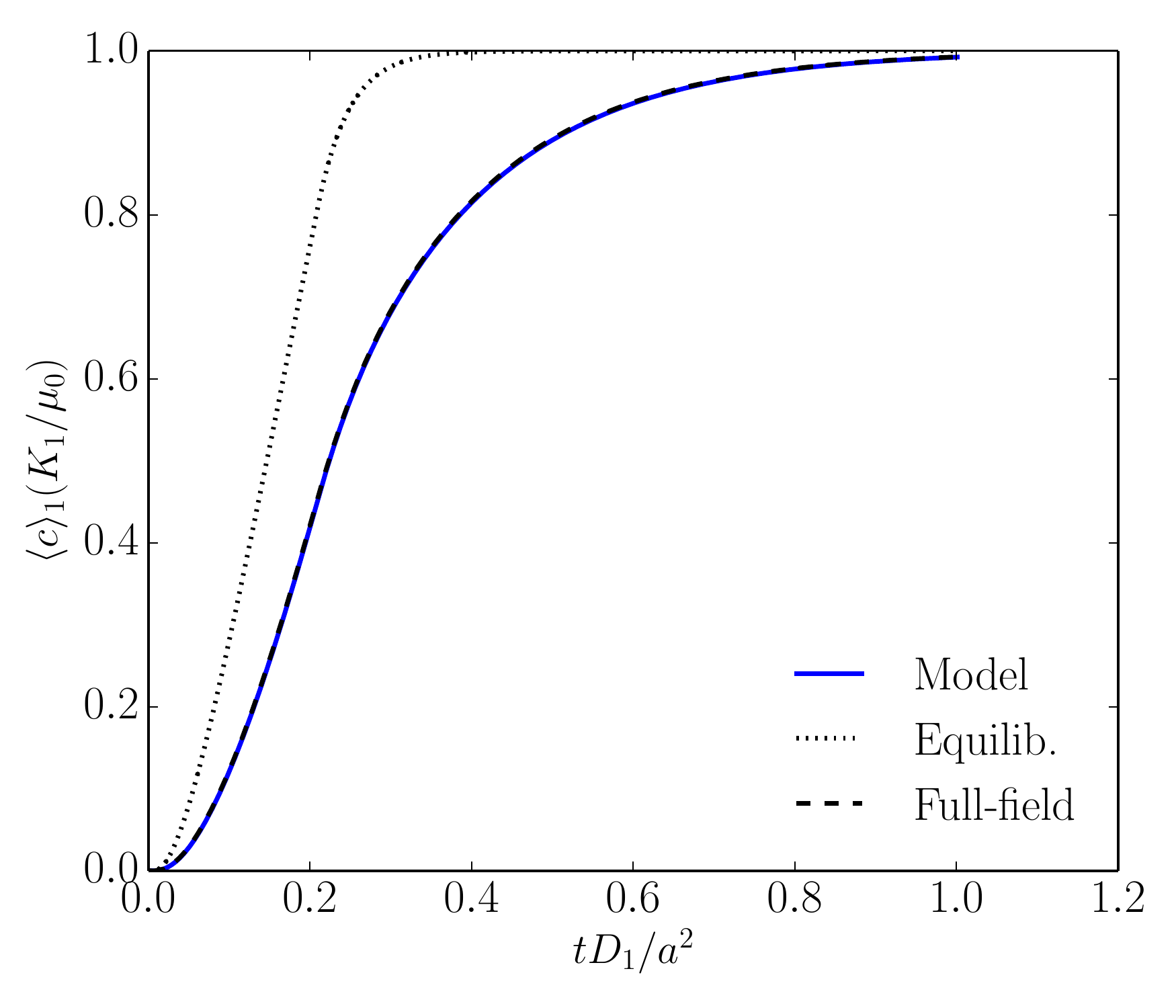}
\caption{}
\end{subfigure}
\caption{Time evolution of the (a) macroscopic and (b) average inclusion response at macroscopic coordinate $x/L=0.05$ in the slab subject to a ramp load.}
\label{fig-macro-ramp}
\end{center}
\end{figure}

Finally, we consider a ramp loading of the form:
\begin{equation}
\bar{\mu}(t) = \left\{ \begin{array}{ll}
\mu_0 \frac{5t}{\tau_1} & \textnormal{if} \ t\leq 0.2\tau_1 \\
\mu_0 & \textnormal{otherwise}
\end{array} \right.
\end{equation}
The macroscopic and average inclusion responses in the unit cells at $x=0.05L$ are represented in Fig. \ref{fig-macro-ramp}. Once again, the agreement between full-field results and mean-field predictions is excellent.

\subsection{Limiting regimes}
The results of the previous section show that the transient behaviour of the composite is non-Fickian, as a result of the slow diffusion in the inclusion phase. The significance of the slow relaxation of the inclusion depends however of the time scale for diffusion throughout the entire system. Let us introduce a characteristic length $\Lambda$, which represents the length scale for diffusion through the matrix over a time scale comparable to the inclusion diffusion time:
\begin{equation}\label{Lambda}
\Lambda = \sqrt{D_2 \tau_1} = \sqrt{\frac{D_2}{D_1}} a. 
\end{equation} 
The length $\Lambda$ is intrinsic to the material and is independent of the characteristic size $L$ of the macroscopic boundary-value problem. The macroscopic length $L$ and intrinsic material length $\Lambda$ can be combined to define a dimensionless parameter $\chi$: 
\begin{equation}\label{chi}
\chi \equiv \frac{\Lambda^2}{L^2} = \frac{a^2}{L^2} \frac{D_2}{D_1}.
\end{equation} 
The dimensionless parameter $\chi$ represents the ratio of the inclusion diffusion time, $a^2/D_1$ to the time for diffusion through the matrix over a distance $L$, $L^2/D_2$.

When the size of the macroscale boundary value problem is much larger than the characteristic length, $L \gg \Lambda$ ($\chi \ll 1$), diffusion in the inclusion phase can be considered at equilibrium relative to macroscale diffusion through the matrix. In this case, the RVE may be considered at steady-state, and the overall behaviour is Fickian. In contrast, when the macroscopic length $L$ becomes comparable or less than the characteristic length $\Lambda$ ($\chi \geq 1$), the transient response in the inclusions at microscale impacts the overall behaviour. In the example of the previous sub-section, $L/\Lambda=0.23$ ($\chi=18.9$), and the transient inclusion response impacted the overall behaviour significantly. 
\begin{remark}
In the limit $L\ll \Lambda$ ($\chi \gg 1$), diffusion through the entire system is very fast (thanks to the presence of fast diffusion paths through the matrix), and the field of macroscopic chemical potential instantaneously reaches its equilibrium value, while the concentration response is transient. One may however question the validity of the model in this case, as it may no longer be possible to define a RVE since the separation of scale hypothesis (\ref{separation_scales}) does no longer hold.   
\end{remark}

The limiting regimes are illustrated in the following example. Consider again the 1D diffusion problem through a slab of length $L$. The slab is subjected to a step load, identical to Eq. (\ref{mup_step}). The problem is numerically solved using the 1D finite element method in combination with the isotropic mean-field model. The material properties are such that $K_2/K_1=1$ and $k_2/k_1=10^5$. The radius of the inclusions (assuming 10 inclusions with volume fraction $f=10 \%$) is $a=0.056l$. The intrinsic length is thus $\Lambda=18l$ ($\chi=314(l/L)^2$). Three values of the macroscopic length $L$ are successively considered: $L=100l$, $L=10l$ and $L=l$. In all cases, the number of finite elements in the simulation is set to 100, which means that the finite element length does no longer represent the RVE size. We have verified that our results are insensitive to the number of elements in the FE discretisation. 

\begin{figure} 
\begin{center}
\begin{subfigure}[b]{0.45\textwidth}
\includegraphics[width=\textwidth]{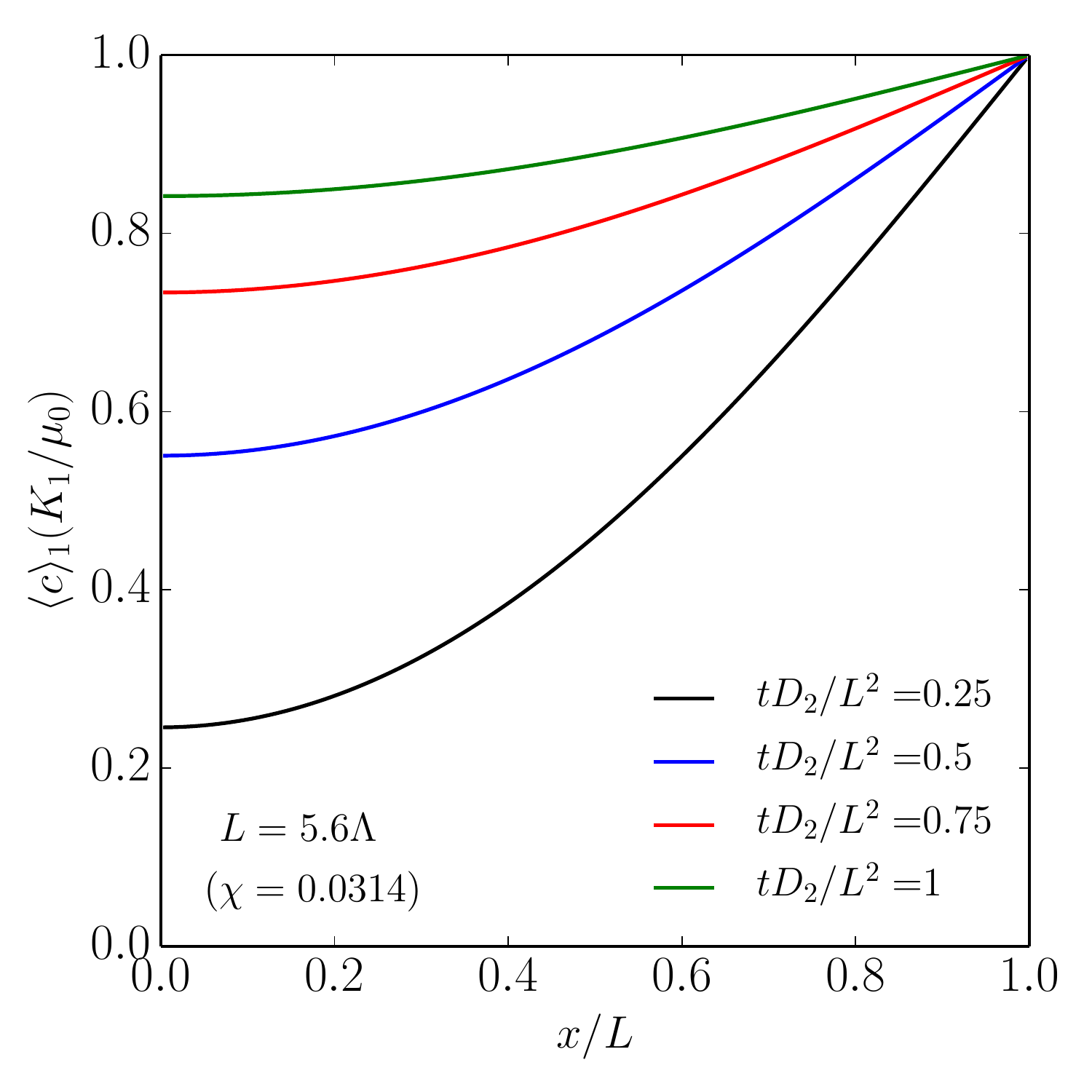}
\caption{}
\end{subfigure}
\begin{subfigure}[b]{0.45\textwidth}
\includegraphics[width=\textwidth]{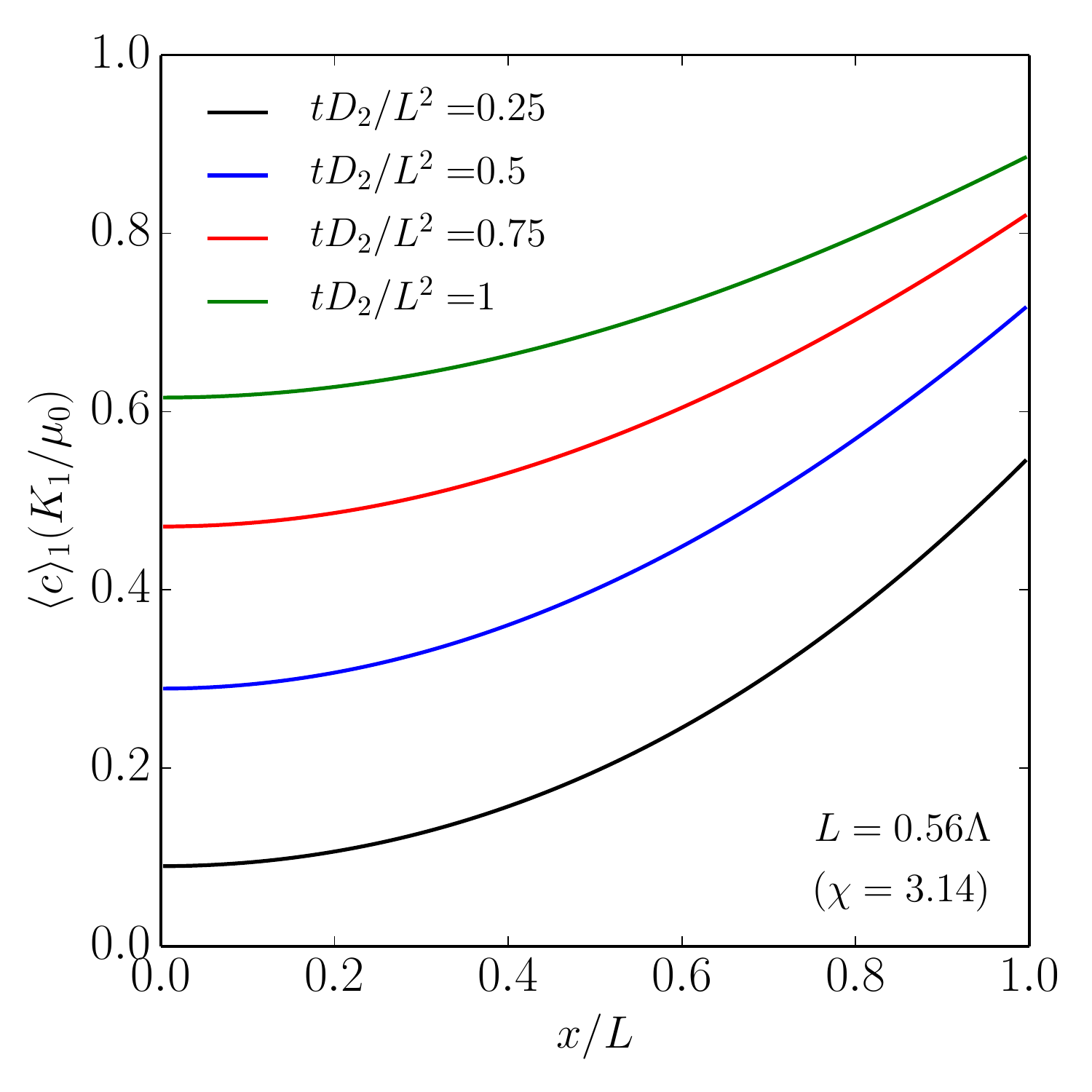}
\caption{}
\end{subfigure}
\begin{subfigure}[b]{0.45\textwidth}
\includegraphics[width=\textwidth]{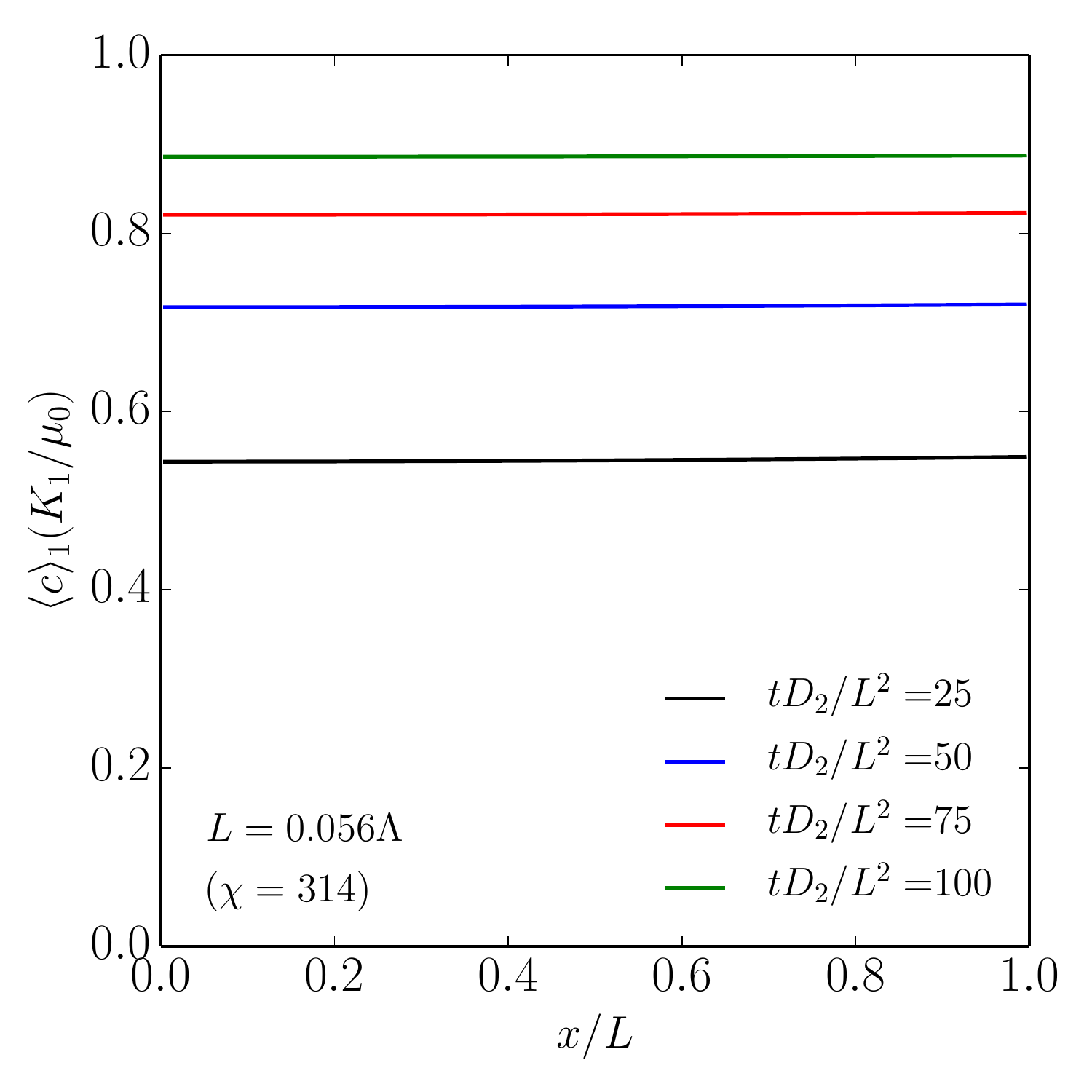}
\caption{}
\end{subfigure}
\begin{subfigure}[b]{0.45\textwidth}
\includegraphics[width=\textwidth]{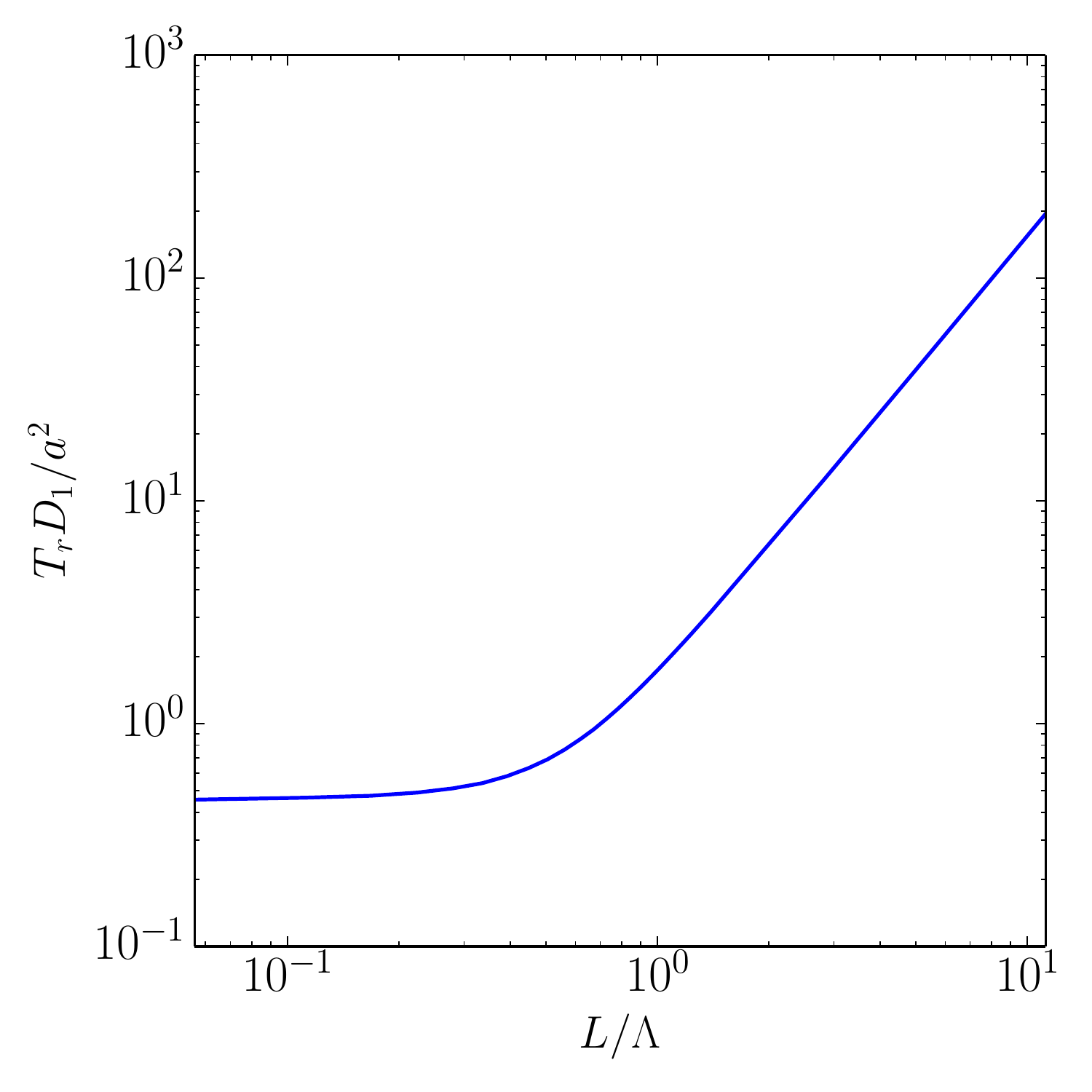}
\caption{}
\end{subfigure}
\caption{(a)-(c) Profiles of average inclusion concentration in slabs for varying ratio of the slab length $L$ to the intrinsic length $\Lambda$. (d) Overall relaxation time $T_r$ as a function of the $L/\Lambda$ ratio.}
\label{fig-Lambda}
\end{center}
\end{figure}

Figure \ref{fig-Lambda} shows the profiles of inclusion concentrations at various simulation times in the three boundary-value problems. In Fig. \ref{fig-Lambda}a, the macroscopic length $L$ is much larger than $\Lambda$, $L=100l=5.6\Lambda$ ($\chi=0.0314$). The overall relaxation of the system is limited by the long-range diffusion through the matrix, rather than the diffusion through the inclusion, due to the large difference in length scales. As a result, the RVE problem is at steady-state, and the overall system behaves in a Fickian manner. This can be seen from the inclusion concentration value in $x=L$, which is at all times equal to its equilibrium value, $\langle c \rangle_1 = \mu_0/K_1$. The case $L=10l=0.56\Lambda$  ($\chi=3.14$) is represented in Fig. \ref{fig-Lambda}b, showing the clear non-Fickian effects, similar to the previous sub-section. Finally, Fig. \ref{fig-Lambda}c shows profile in the case $L=l=0.056\Lambda$ ($\chi=314$). In this case, diffusion through the matrix of the RVE is so fast that the all fields can be considered homogeneous. However, the concentration is not at equilibrium, and gradually relaxes in time.  

The overall relaxation time of the system under a step load is represented in Fig. \ref{fig-Lambda}(d) as a function of the problem size $L$ in a log-log plot. The overall relaxation time is defined as the time needed for the average inclusion response in the first integration point (nearest to $x=0$) to reach 95\% of its equilibrium value. Fickian behaviour is characterised by a relaxation time scaling with $L^2$, and is observed for $L>\Lambda$. When $L<\Lambda$, the behaviour is non-Fickian, and the relaxation time tends to a constant value that coincides with the time for diffusion in an inclusion of radius $a$. The two-scale model predictions are similar to the predictions of phenomenological theories for coupled interdiffusion and viscous flow \citep{brassart2018}, as well as a theory for coupled self-diffusion and viscous flow \citep{li2014}.

\section{Conclusion}
In this work we have developed a mean-field estimate for the transient diffusion response of composites in which there is high contrast in the phase diffusivities. According to Eq. (\ref{macro_c_integral}), the macroscopic concentration is a function of both the macroscopic chemical potential and its gradient, and involves a history-dependent contribution due to transient diffusion in the slow inclusions. The dependence in the macroscopic chemical potential gradient however vanishes in the case of isotropic composites, as expected (Eq. (\ref{macro_c_iso})). The macroscopic flux also depends on both the macroscopic chemical potential and its gradient (Eq. (\ref{macro_flux_integral})), and the dependence in the former vanishes in the isotropic case (Eq. (\ref{macro_flux_iso})). The mean-field model can be written in terms of internal variables for efficient time-integration, enabling two-scale simulations at very low computational cost compared to full-field or FE$^2$ simulations. 

The model has been validated by comparing its predictions to full-field results at both RVE and macroscale, showing excellent agreement. It was found that the macroscopic flux is dominated by the fast transport through the matrix, and is accurately described by the effective conductivity of the composite in steady-state and assuming non-conducting inclusions. On the other hand, the history-dependent contribution of the inclusions to the macroscopic flux is negligible. The transient inclusion response however significantly affect the overall concentration response. In particular, it leads to a non-Fickian behaviour when used to solve macroscale boundary-value problems. 

While the effective transient behaviour of composites was previously investigated using either asymptotic homogenisation, e.g. \cite{auriault1983}, or computational homogenisation, e.g. \cite{larsson2010}, we believe that the proposed approach based on mean-field approximation is original. The main advantage of the approach is that the effective constitutive equations are obtained in closed-form, allowing two-scale simulations at a reasonable cost. Our model also differs in its construction and final expressions from the volume-averaged expressions of the Porous Electrode Theory (\cite{thomas2002,smith2017}), in which transient diffusion in the inclusions is described by a source term in the averaged conservation equations.  In contrast, our approach does not introduce any source term, but instead suggests that the macroscopic chemical potential is history dependent, as a result of the micro-to-macro transition. The latter also provides a justification to phenomenological models which do not rely on the chemical equilibrium assumption at macroscopic scale, see e.g. \citep{brassart2013} (coupled diffusion and elasto-plasticity) and \citep{brassart2016,brassart2018} (coupled diffusion and viscoplasticity). 

The presently-proposed approach has a number of limitations. First, it rigorously applies to cases where the steady-state assumption holds in the matrix, but not in the inclusions. Such cases are found for a range of excitation time scales when the diffusivity contrast between the phases is very large. Second, the model assumes that the macroscopic gradient of chemical potential is not too large over the length scale of a RVE, in order for the first-order homogenisation approach to hold and also for the assumption of uniform chemical potential on the inclusion boundaries to be reasonable. Finally, the model is restricted to linear diffusion behaviour, since it relies on the single inclusion solution for which an exact solution only exists in the linear case. While the present work only considered spherical or circular inclusions, more general inclusion shapes (e.g. ellipsoids) could easily be considered. Strategies for identifying the chemical creep function for arbitrary inclusion shapes have been proposed in \cite{brassart2017}.  

Future work could include the extension to nonlinear diffusion behaviour due to concentration-dependent mobility coefficients and chemical moduli. A first, heuristic approach at the single inclusion level  is provided in \citep{brassart2017}, however more sophisticated techniques could certainly be proposed. Extending the mean-field model to coupled chemo-mechanical problems would also be important to address problems such as diffusion in battery electrodes. The first step to consider would be to revisit the linear diffusion problem in a single inclusion, including chemo-mechanical coupling in two ways: volumetric expansion associated with concentration change and pressure-dependency of the chemical potential. Analytical or semi-analytical solutions can probably be obtained for simple geometries, see e.g. \cite{Hetnarski1964} or \cite{OrtnerWagner2014} for thermo-elasticity. Alternatively, a computational approach for example based on modal analysis could be adopted, see e.g. \cite{brassart2017}. At the composite level, a suitable mean-field assumption would also need to be introduced in order to partition the strain between the phases, see e.g. \cite{nemat1998}. A third, important direction for future work will be to generalise the model to the fully-transient case in order to address smaller diffusivity contrasts.   

\begin{appendices}
\renewcommand*{\thesection}{\appendixname~\Alph{section}}

\section{Estimates of structure tensors for isotropic inclusion distributions}\label{sec:app_A}
Estimates for $\bm s_1$ and $\bm S_1$ in the case of an isotropic distribution of inclusions are obtained by considering ensemble averages for many RVE realisations:   
\begin{eqnarray}
\bm s_1 &=& \frac{1}{l} \oint \left( \frac{1}{N} \sum_{k=1}^N (\bm x_k-\bm x_0) \right) P_N d\bm x_1 d\bm x_2 ... d\bm x_N \label{ensemble_av1}\\
\bm S_1 &=& \frac{1}{l^2} \oint \left( \frac{1}{N} \sum_{k=1}^N (\bm x_k-\bm x_0) \otimes (\bm x_k-\bm x_0) \right) P_N d\bm x_1 d\bm x_2 ... d\bm x_N \label{ensemble_av2}
\end{eqnarray}     
where $P_N$ is the N-particle probability density function (\cite{torquato2002}). For a random distribution of \textit{overlapping} circular (2D) or spherical (3D) inclusions, it is simply given by:
\begin{equation}
P_N = \frac{1}{V^N},
\end{equation}
where $V$ is the RVE area (2D) or volume (3D). The overlapping inclusion assumption is reasonable at a low volume fraction of inclusions. Using the latter expression and carrying out the integration in (\ref{ensemble_av1}) and (\ref{ensemble_av2}), we find:
\begin{eqnarray}
\bm s_1 &=& \bm 0  \label{s1_iso} \\
\bm S_1 &=& \frac{1}{12} \bm 1  \label{S1_iso}
\end{eqnarray}
In this work we assumed that (\ref{S1_iso}) also holds for non-dilute isotropic distributions of non-penetrable inclusions.

\section{Effect of the conductivity contrast on model predictions}\label{sec:app_B}
\begin{figure} 
\begin{center}
\begin{subfigure}[b]{0.45\textwidth}
\includegraphics[width=\textwidth]{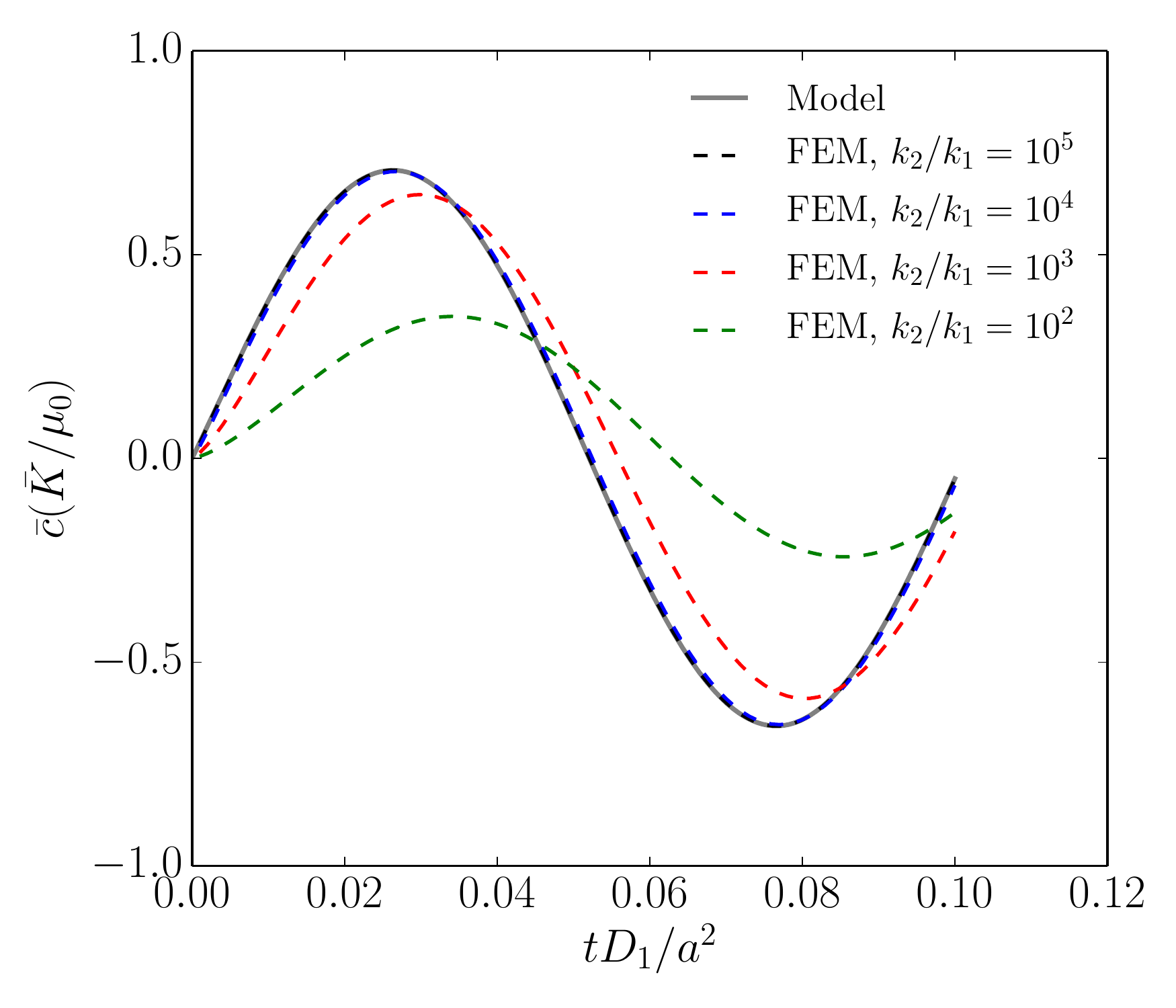}
\caption{}
\end{subfigure}
\begin{subfigure}[b]{0.45\textwidth}
\includegraphics[width=\textwidth]{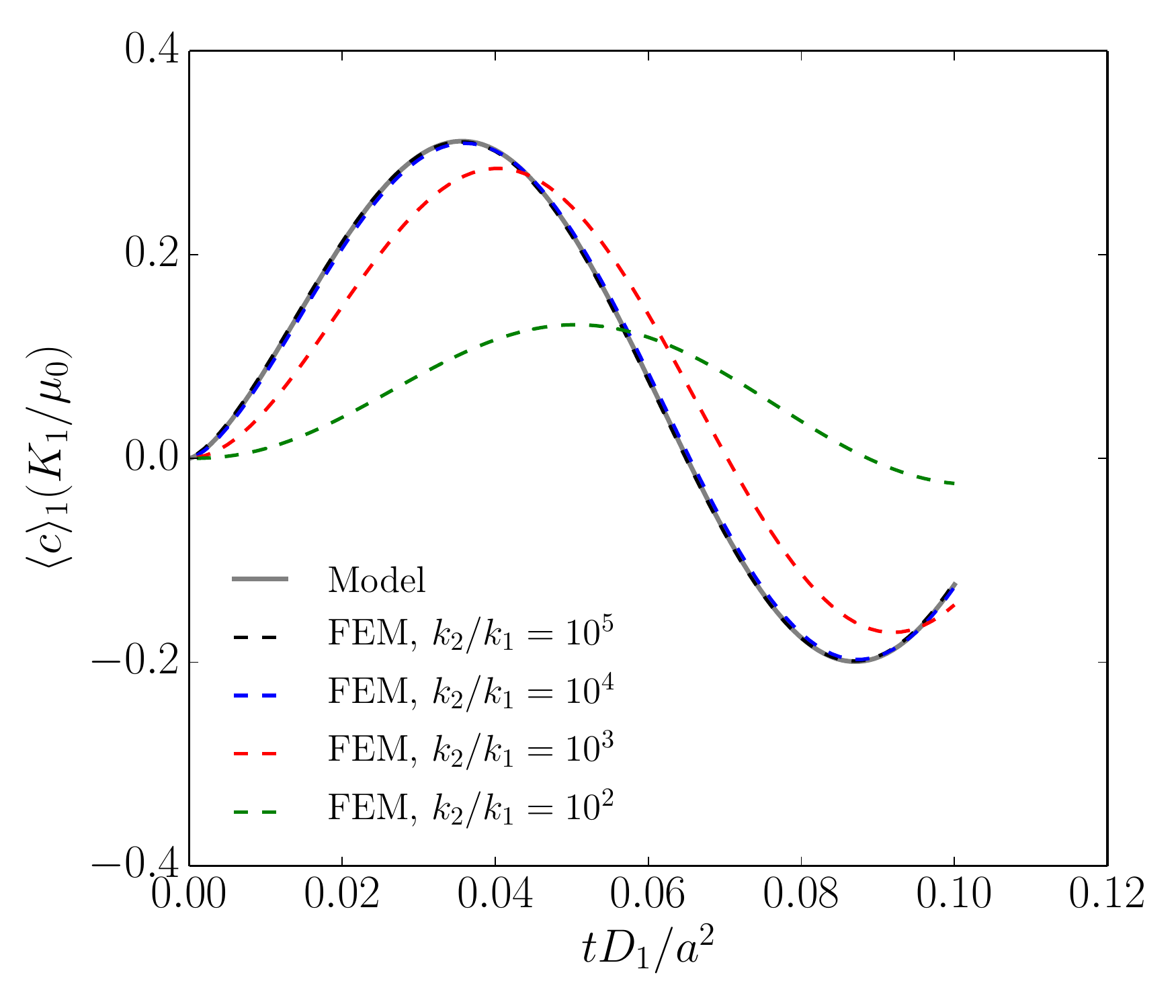}
\caption{}
\end{subfigure}
\caption{Macroscopic concentration and inclusion average concentration corresponding to the loading conditions (\ref{loading1}) applied to the first geometry, for decreasing values of the conductivity contrast $k_2/k_1$.}
\label{fig-appB-1}
\end{center}
\end{figure}

Figure \ref{fig-appB-1} shows the effective concentration response in the first geometry subject to a macroscopically-uniform, time-varying chemical potential, Eq. (\ref{loading1}). FE results are shown for decreasing values of the conductivity contrast $k_2/k_1$. On the other hand, the mean-field model is insensitive to the conductivity contrast as far as the concentration response is concerned. Indeed, the predictions of the average concentration in the matrix and in the inclusion, Eqs (\ref{c_matrix}) and (\ref{c_allinclusions2}), do not depend on the phase conductivities in the absence of a macroscopic chemical potential gradient. While the FE results are practically indistinguishable for contrast values of $10^5$ and $10^4$, significant differences appear for lower values of the contrast. The matrix is then no longer at steady-state, and therefore the assumption of uniform chemical potential within the matrix and on the inclusion boundary does no longer hold. A similar loss of accuracy is found regarding mean-field predictions of moment of concentration rate in the inclusions (not shown).

\begin{figure} 
\begin{center}
\begin{subfigure}[b]{0.45\textwidth}
\includegraphics[width=\textwidth]{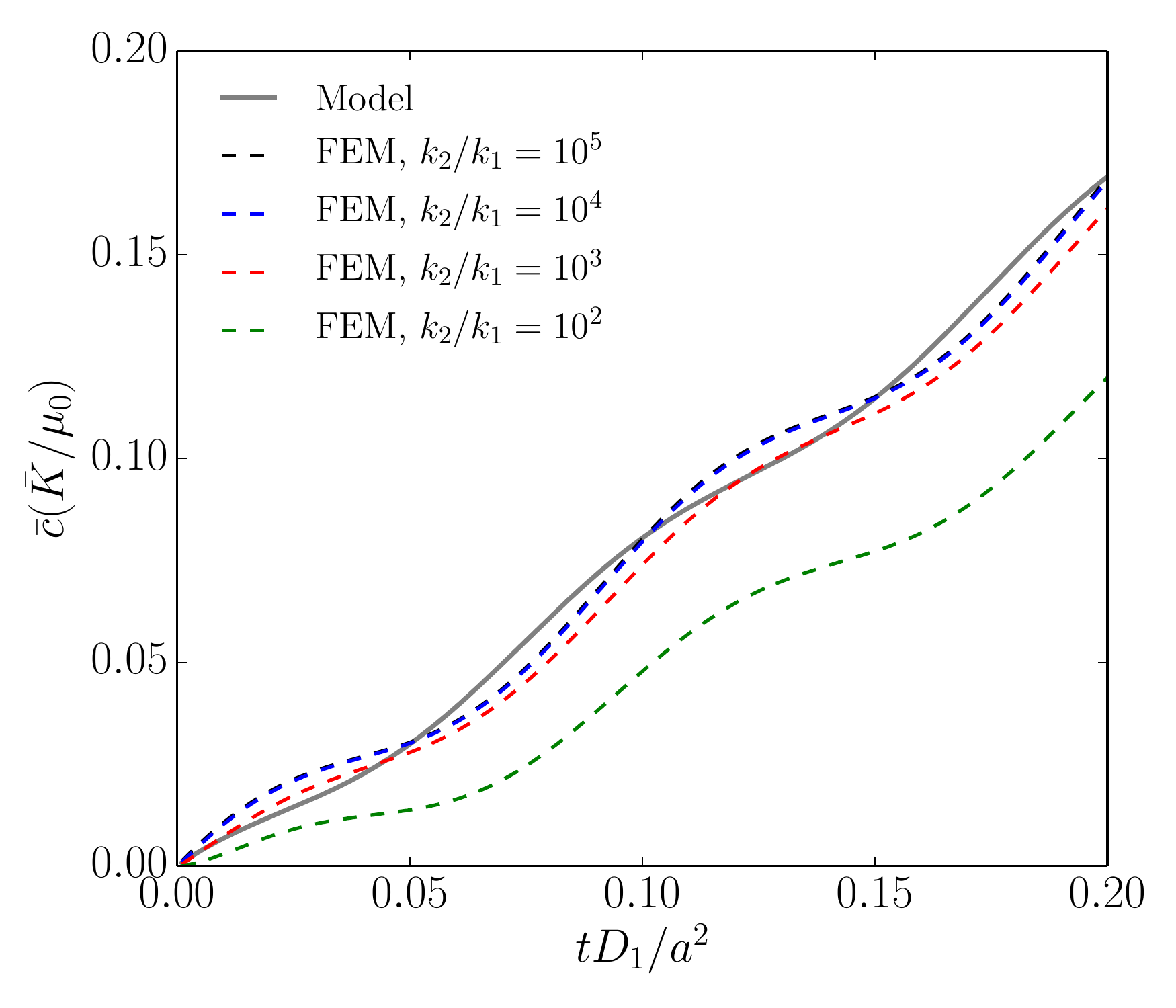}
\caption{}
\end{subfigure}
\begin{subfigure}[b]{0.45\textwidth}
\includegraphics[width=\textwidth]{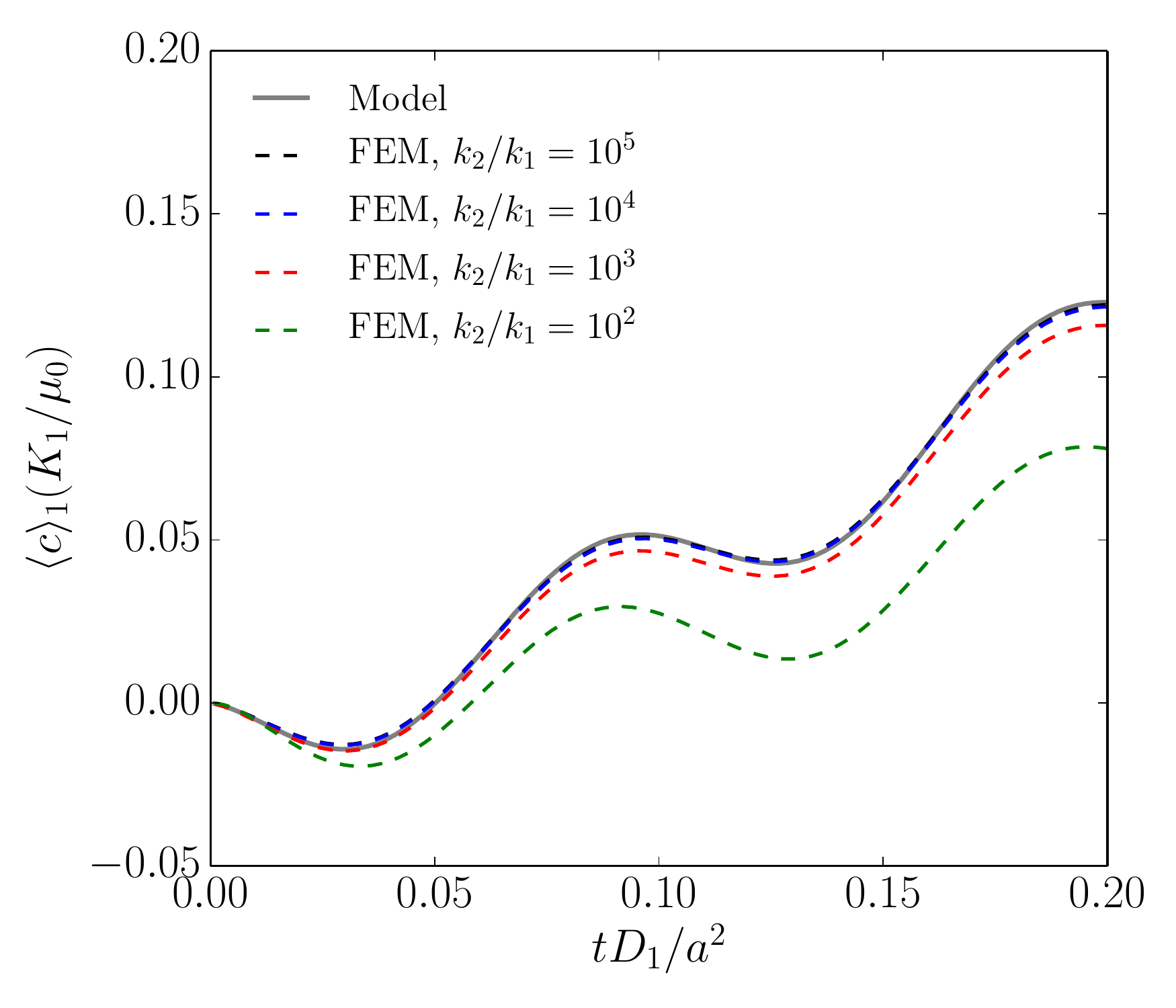}
\caption{}
\end{subfigure}
\begin{subfigure}[b]{0.45\textwidth}
\includegraphics[width=\textwidth]{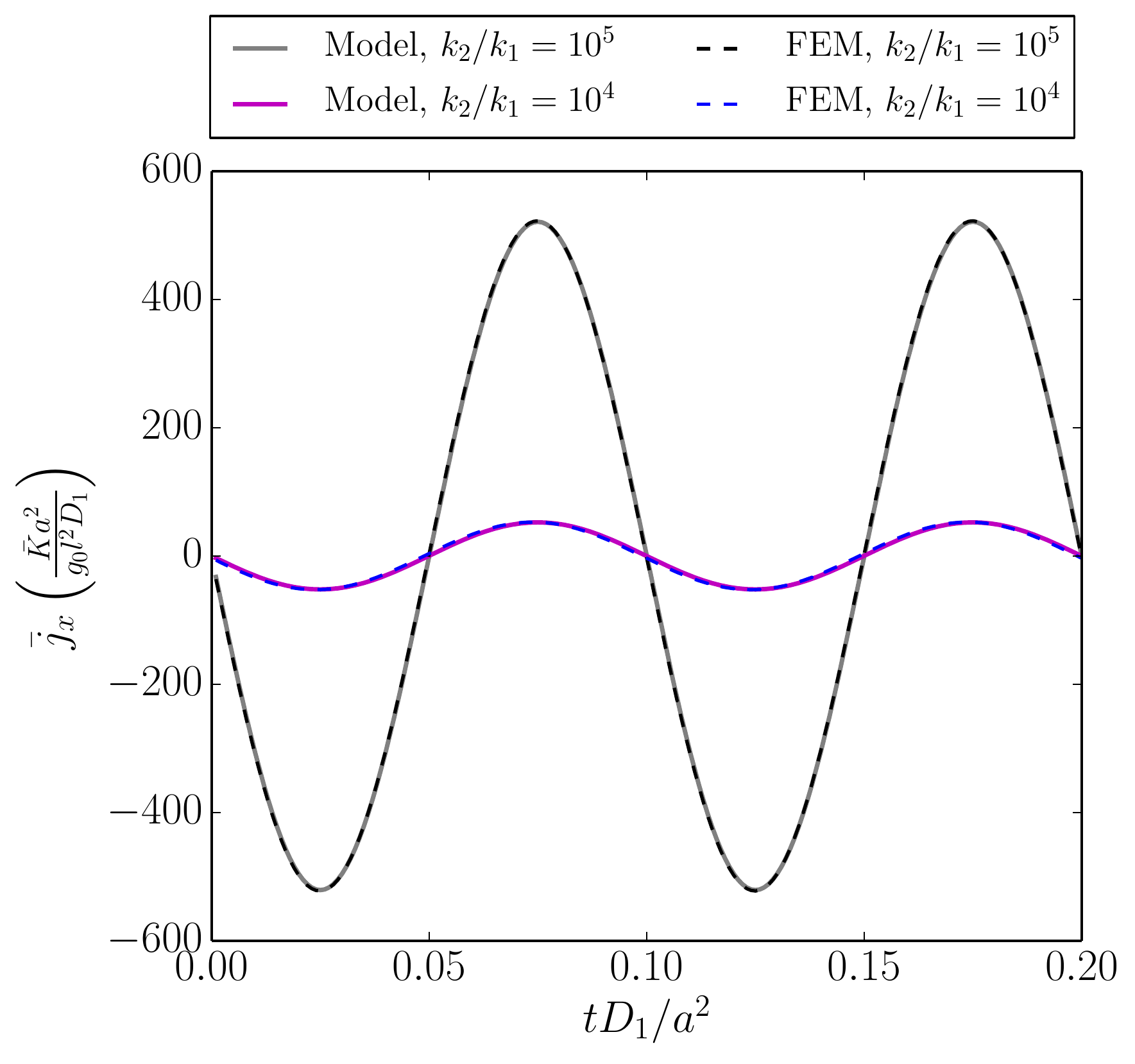}
\caption{}
\end{subfigure}
\begin{subfigure}[b]{0.45\textwidth}
\includegraphics[width=\textwidth]{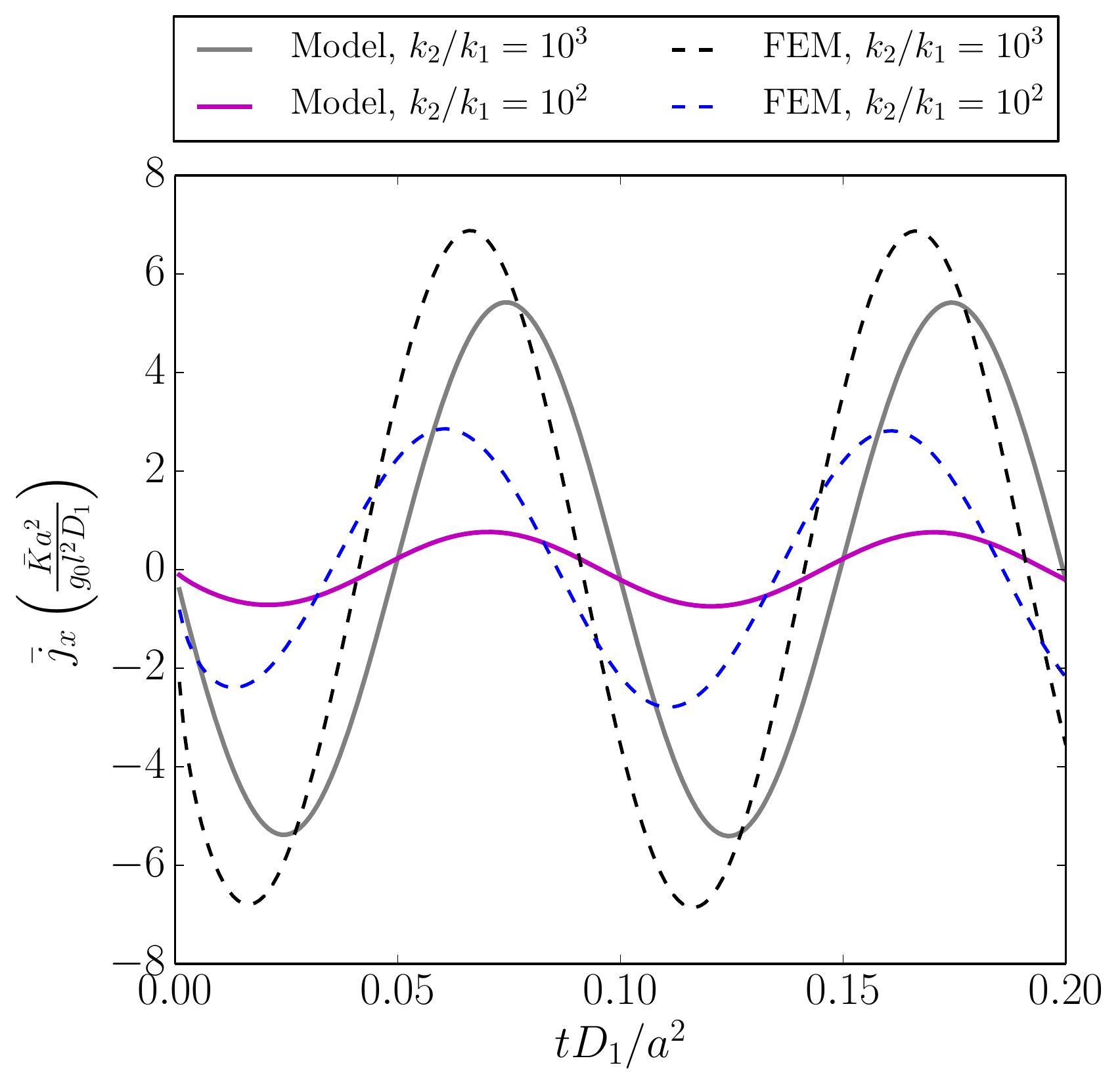}
\caption{}
\end{subfigure}
\caption{Effective behaviour corresponding to the loading conditions (\ref{loading2}) applied to the first geometry, for decreasing values of the conductivity contrast $k_2/k_1$. (a)-(b) Macroscopic and average inclusion concentration. (c)-(d) $x$-component of the macroscopic flux.}
\label{fig-appB-2}
\end{center}
\end{figure}

The case of macroscopically non-uniform, time-varying chemical potential is examined in Fig. \ref{fig-appB-2} for decreasing values of the conductivity contrast. The loading conditions are given in Eq. (\ref{loading2}). Results are shown for the first geometry, but the same conclusions hold for the other two considered geometries. Similar to the previous example, the mean-field model does no longer accurately predict the concentration response for contrasts of $10^3$ and below (Figs \ref{fig-appB-2}(a)-(b)). The fact that the model is less accurate in predicting the matrix average concentration (even at high diffusivity contrast) was previously mentioned in Section \ref{sec-nonunif-mu}. The macroscopic flux component in the x-direction is shown in Figs \ref{fig-appB-2}(c)-(d). Here, we used the Hashin-Shtrikman bound (\ref{HS_2D}) as an analytical estimate of the effective conductivity. The mean-field model is valid for the two larger values of the contrast (\ref{fig-appB-2}(c)). In this case the macroscopic flux is dominated by the average flux in the matrix, and is well predicted by the mean-field model. The mean-field model looses its accuracy at lower values of the contrast (\ref{fig-appB-2}(d)). While Hashin-Shtrikman estimate of the volume average of the flux (\ref{av_flux_macro}) remains accurate for the contrasts $10^3$ and $10^2$ (not shown), the moment of the concentration rates in both the matrix and in the inclusion then also contribute significantly to the macroscopic flux. Among these two contributions, the moment of the concentration rate in the matrix is the most significant and is not captured by the mean-field model, which assumes a steady-state matrix.

\end{appendices}

\end{document}